\documentclass[a4paper,11pt,twoside]{report}
\usepackage{emptypage}
\usepackage[utf8]{inputenc}
\usepackage[italian,english]{babel}
\usepackage{url}
\usepackage[abbreviations = false]{siunitx}
\usepackage{caption}
\usepackage{varioref}
\usepackage{natbib}
\usepackage{tabularx}
\usepackage[framed,numbered,autolinebreaks,useliterate]{mcode}
\newcommand{\ap}{\textsuperscript}
\newcommand{\gr}{\textbf} 
\newcommand{\co}{\textit}
\usepackage{morefloats}
\usepackage{mathtools}
\usepackage{amsmath,amssymb}
\usepackage{mathrsfs}
\usepackage{geometry,marginnote} 
\usepackage{csquotes}
\usepackage{multirow}

\usepackage[pdftex]{graphicx}
\usepackage{lmodern}
\usepackage{xcolor}
\usepackage{microtype}
\usepackage{titlesec}
\usepackage{sectsty}
\definecolor{myblue}{RGB}{0,82,155}
\chapterfont{\color{black}}
\sectionfont{\color{black!88}}
\subsectionfont{\color{black!70}}
\titleformat{\chapter}[display]
  {\vspace{-2.5cm}\normalfont\bfseries\color{black}}
  {\thispagestyle{plain}\filleft\hspace*{-60pt}%
    \rotatebox[origin=c]{90}{%
      \normalfont\color{black}\Large%
        \textls[200]{\textsc{Chapter}}%
  }\hspace{10pt}%
    {\setlength\fboxsep{0pt}%
    \colorbox{white}{\parbox[c][3.5cm][c]{2.8cm}{%
      \centering\color{black}\fontsize{120}{120}\selectfont\thechapter}%
    }}%
  }
  {10pt}
  {\titlerule[4pt]\vskip3pt\Huge\sffamily}
\usepackage{fancyhdr}
\renewcommand{\headrulewidth}{1pt}
\renewcommand{\headrule}{\hbox to\headwidth{%
    \color{black}\leaders\hrule height \headrulewidth\hfill}\vspace{-0.5cm}}

\makeatletter
\def\footrule{{
  \vskip-\footruleskip\vskip-\footrulewidth
  \color{\footrulecolor}
  \hrule\@width\headwidth\@height
  \footrulewidth\vskip\footruleskip
}}
\makeatother
\renewcommand{\footrulewidth}{1pt}
\newcommand{\footrulecolor}{black}
\begin{document}
%-----------------------------------------------------------------------%
%                            	FRONTESPIZIO
%-----------------------------------------------------------------------%
\newpage
\thispagestyle{empty}

\begin{center}
~\vspace*{-0.1cm}
%-------------------------------UNIVERSITA------------------------------%
\huge{\textsc{Politecnico di Torino}}

\vspace{0.5cm}
\Large{Corso di Laurea Magistrale in Ingegneria per l'Ambiente e il Territorio}

\vspace{2.0cm}
\huge{Tesi di Laurea Magistrale}

%-------------------------------TITOLO----------------------------------%
\vspace{2.0cm}
\huge{\textbf{Mathematical modeling of \\ cholera epidemics in South Sudan}}

\vspace{0.8cm}
\Large{Modellazione matematica delle \\ epidemie di colera nel Sudan del Sud}

%-------------------------------LOGO--------------------------------------%
\vspace{0.8cm}
\begin{figure}[h]
\centering
\includegraphics[width=0.20\textwidth]{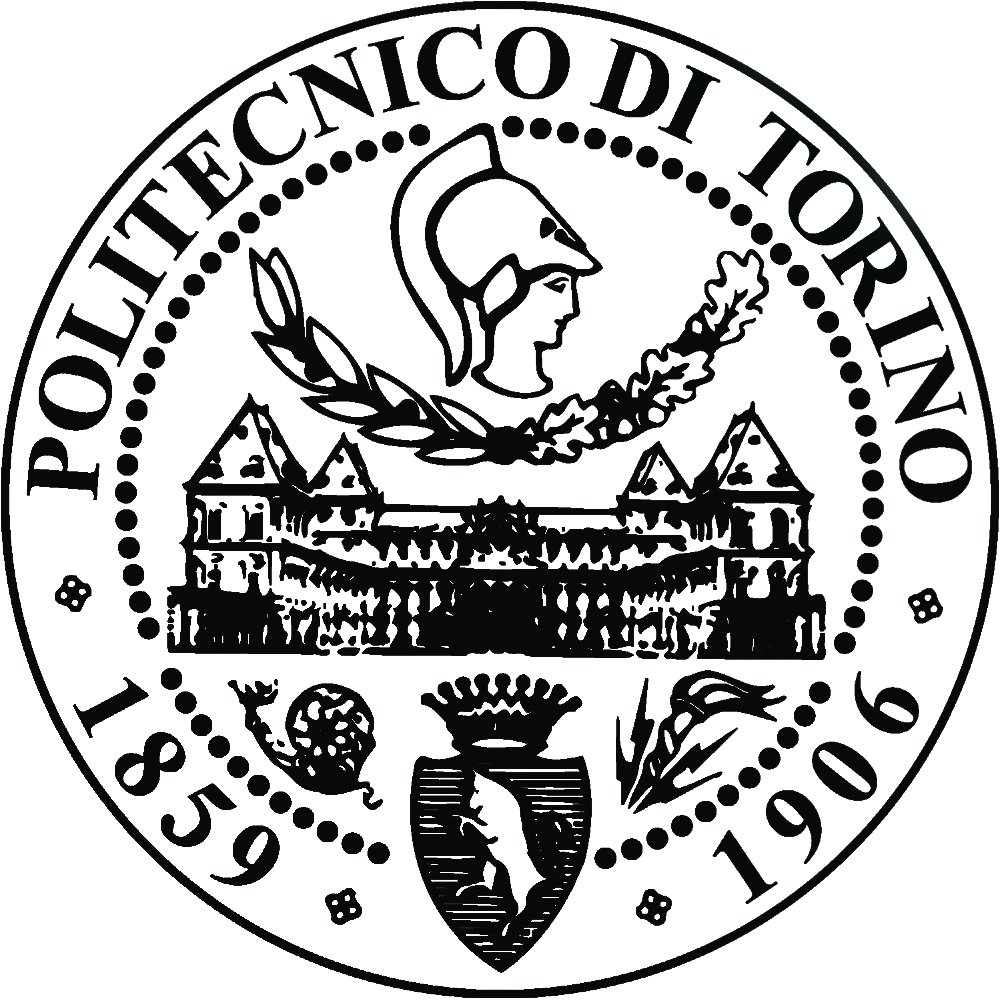}
\end{figure}

%---------------------------RELATORE POLI---------------------------------%
\vspace{0.2cm}
\begin{flushleft}
\Large{\textbf{Relatore:}} \\
\begin{tabular}{l l}
Prof.~Francesco & \textsc{Laio}
\end{tabular}
\end{flushleft}

%---------------------------RELATORI EXT---------------------------------%
\vspace{0.001\textwidth}
\begin{flushleft}
\Large{\textbf{Relatori esterni:~~~~~~~~~~~~~~~~~~~~~~~~~~~~~~~~~~~~~~~~~ Candidata:}} \\

\vspace{0.01\textwidth}
%\begin{tabular}{l l}
~Prof.~~Andrea ~~~~\textsc{Rinaldo}  ~~~~~~~~~~~~~~~~~~~~~~~~~~~~~~~~~~~~ Carla  ~~ \textsc{Sciarra} \\
~Doct.~Damiano ~~\textsc{Pasetto}  ~~~~~~~~~~~~~~~~~~~~~~~~~~~~~~~~~~~~ matricola: 219921
%\end{tabular}
\end{flushleft}

%------------------------------ LOGO EPFL-------------------------------%
\vspace{1.0cm}

\normalsize{\textbf{Supervisori esterni presso \\ \'Ecole Polytechnique F\'ed\'erale de Lausanne:}}

E. Bertuzzo, F. Finger, D. Pasetto, A. Rinaldo

\vspace{0.02cm}
\begin{figure}[h]
\centering
\includegraphics[width=0.2\textwidth]{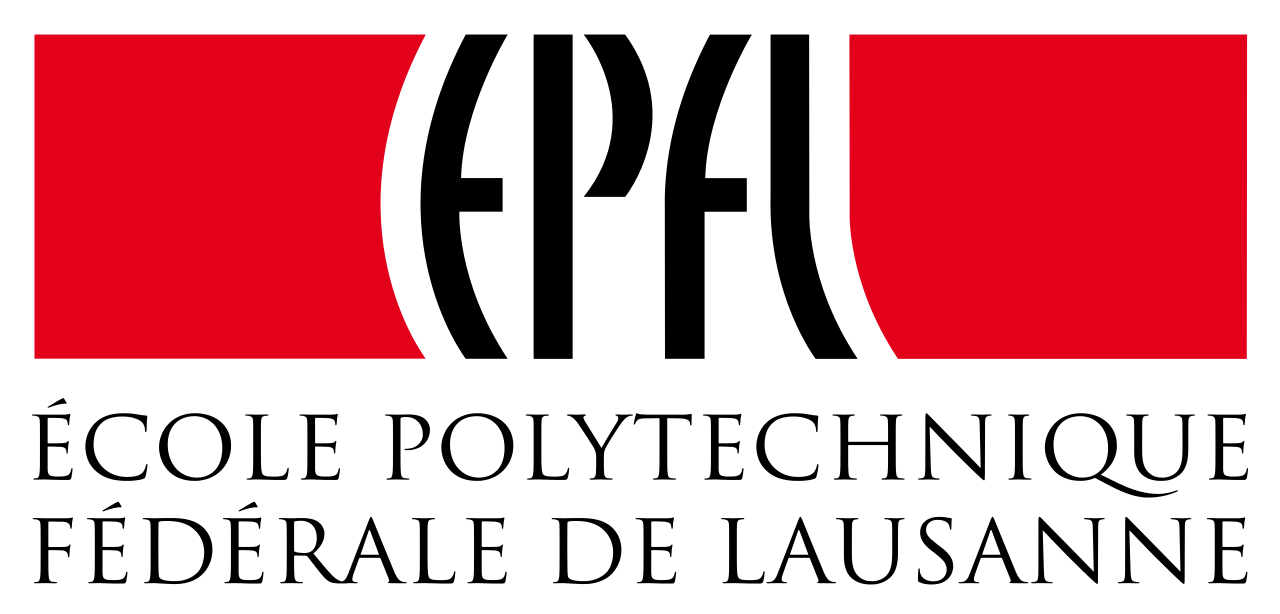}
\end{figure}

\normalsize{\textbf{Anno Accademico 2015-2016}}
\end{center}

\newpage
\thispagestyle{empty}
%~\vspace*{21cm}
\begin{flushleft}
This work is subject to the Creative Commons Licence.\\ 
Questo testo è soggetto alla Creative Commons Licence.
\end{flushleft}

%-----------------------------------------------------------------------%
%                            	  DEDICA
%-----------------------------------------------------------------------%
\newpage
\thispagestyle{empty}
~\vspace*{120pt}
~\vspace*{40pt}
\begin{quote}
from the speech addressed at Board Meeting of the Vaccine Fund, May 2002:

\large
\emph{"\dots We must ensure that children and parents and communities are educated and taught: unless all of our communities understand the importance of immunization, we will not succeed in preventing the millions of deaths that are occurring unnecessarily. And we must ensure that vaccines and health care are accessible and affordable for all families."} 
\begin{flushright}
Nelson Mandela 
\end{flushright}
\end{quote}
%~\vspace*{30pt}
%\begin{quote}
%\large
%\begin{flushright}
%\emph{Dedicated to my family: \\ Mom, Dad, Rita, Amalia, Rafael \\ and to my grandfather Carlo \dots} \\ 
%\end{flushright}
%\end{quote}

\newpage
\thispagestyle{empty}
\null%This page is intentionally left blank.
%\newpage
%\thispagestyle{empty}
%-----------------------------------------------------------------------%
%                          LISTE DEI CONTENUTI
%-----------------------------------------------------------------------%
\chapter*{Abstract}
\thispagestyle{plain}
\setcounter{page}{5}
\addcontentsline{toc}{chapter}{Abstract}
Cholera is one of the main health issue around the world, especially in Africa, where every year thousands of people die due to this enteric disease. International agencies, as the World Health Organization and the United Nation are making big efforts to help people in need and to stop this infection and, even if progresses have been made, a lot of work is still required.

Epidemiological models of cholera outbreaks, like the \textit{SIRB} that is proposed in this thesis, have been developed in the last decades to better understand the routes of transmission of the infection and to provide key tools in elaborating intervention  strategies in case of an emergency.

The mathematical approach used in this thesis work to simulate the epidemics consists of a spatially-explicit model driven by rainfall and human mobility. This methodology, which has already been successful in the reproduction of several cholera outbreaks, has several advantages: as first, the model considers the hydrological network in which the ``cause bacteria" \textit{Vibrio cholerae} can survive and move, which is one of the possible ways the epidemics can spread in infection-free areas; as second, the model accounts for the aggravating effect of the rain, which increases the bacteria concentration in the water through the run-off of excreta and latrines; as last the model considers human mobility, by which the bacteria can be suddenly spread far away by a daily traveler.
\\ \par In this work, we analyze and model the cholera epidemics that affected South Sudan, the newest country in the world, during 2014 and 2015. South Sudan possibly represents one of the most difficult context in which adapt the deterministic mathematical cholera model, due to the unstable social and political situation that clearly affects the fluxes of people and the sanitary conditions, increasing the risk of large outbreaks. Despite the limitation of a static gravity model in describing the chaotic human mobility of South Sudan, the \textit{SIRB} model, calibrated with a data assimilation technique (Ensemble Kalman Filter), retrieves the epidemic dynamics in the counties with the largest number of infected cases, showing the potentiality of the methodology in forecasting future outbreaks.

%% ABSTRACT - IT
\chapter*{Contenuto}
\thispagestyle{plain}
\setcounter{page}{6}
\addcontentsline{toc}{chapter}{Contenuto}
Il colera è una delle più importanti problematiche sanitarie nel mondo, specialmente in Africa, dove ogni anno migliaia di persone muoiono affette da questa malattia enterica. Le agenzie internazionali, come l'Organizzazione Mondiale della Sanità e le Nazioni Unite, lavorano per fermare l'epidemie, e nonostante i progressi, tanto lavoro è ancora necessario.
I modelli epidemiologici del colera, come il \textit{SIRB} che viene utilizzato in questa tesi, sono nati con l'intento di studiare i meccanismi di trasmissione della malattia e di fornire strumenti utili all'elaborazione di strategie in caso di emergenza.

L'approccio matematico utilizzato è un modello spazialmente esplicito in cui le principali forzanti risultano essere la pioggia e la mobitilità umana. La validità di tale metodo è stata comprovata dall'applicazione su altri casi studio. Tra i vantaggi principali del modello proposto, in primis, quello di considerare la rete ambientale in cui il ``batterio causa'' \textit{Vibrio cholerae} può sopravvivere e muoversi, spostando così l'infezione in posti in cui prima non ve n'era presenza; inoltre, il modello tiene conto dell'effetto aggravante delle piogge che possono, tramite lisciviazione degli escrementi e delle latrine, aumentare la concentrazione dei batteri nell'acqua; infine, viene presa in considerazione la mobilità umana, tramite la quale la malattia può essere spostata improvvisamente da un pendolare quotidiano su lunghe distanze.
\\ \par In questo lavoro, sono state analizzate e studiate le epidemie di colera che hanno coinvolto il Sudan del Sud nel biennio 2014-2015. Il Sud Sudan è forse uno dei contesti piú difficili nel quale adattare la formulazione matematica deterministica che è stata proposta, dovuto alle instabili condizioni socio-politiche che evidentemente influenzano i flussi di persone e le condizioni sanitarie del paese, aumentando il rischio che malattie trasmissibili possano diffondersi in tutta la nazione. Nonostante le limitazioni nel descrivere la caotica mobilità umana in Sud Sudan, il modello \textit{SIRB}, calibrato con una tecnica di data assimilation (Ensemble Kalman Filter), riesce a simulare le dinamiche delle epidemie nelle zone dove si regristra il più alto numero di infetti, mostrando quindi la potenzialità della metodologia nel prevedere future epidemie. 

\tableofcontents
\listoffigures
\addcontentsline{toc}{chapter}{List of figures}
\listoftables
\addcontentsline{toc}{chapter}{List of tables}
\addcontentsline{toc}{chapter}{Acronyms}
\thispagestyle{plain}
\chapter*{Acronyms}
\begin{tabular}{l@{\hspace{40pt}}l}
%\textbf{BBC}    &  British Broadcasting Company \\[2pt]
%\textbf{CDC}    & Centers for Disease Control and Prevention \\[2pt]
%\textbf{CIA}    & Central Intelligence Agency \\[2pt]
%\textbf{CPA} & Comprehensive Peace Agreement \\[2pt]
\textbf{DA}     & Data Assimilation \\[2pt]
\textbf{DREAM}  & DiffeRential Evolution Adaptive Metropolis \\[2pt]
\textbf{EnKF}   & Ensemble Kalman Filter \\[2pt]
\textbf{IDPs}   & Internally Displacement Persons \\[2pt]
\textbf{KF}     & Kalman Filter \\[2pt]
\textbf{MC}     & Monte Carlo \\[2pt]
\textbf{MSF}    & Médecins Sans Frontiéres \\[2pt]
\textbf{PoC}    & Protection of Civilians \\[2pt]
\textbf{PDF}    & Probability Density Function \\[2pt]
\textbf{RMSE}   & Root Mean Square Errors \\[2pt]
\textbf{SIRB}   & Susceptible-Infected-Recovered-Bacteria \\[2pt]
%\textbf{SPLM} & Sudan People's Liberation Movement \\[2pt]
\textbf{SS}     & South Sudan \\[2pt]
\textbf{SSMoH}  & South Sudanese Ministry of Health \\[2pt]
\textbf{SSNBS}  & South Sudan National Bureau of Statistics \\[2pt]
\textbf{UN}     & United Nations \\[2pt]
\textbf{UNHCR}  & United Nations High Commissioner for Refugees \\[2pt]
\textbf{UNICEF} & United Nations Children's Emergency Fund \\[2pt]
\textbf{WASH}   & Water, Sanitation and Hygiene \\[2pt]
\textbf{WHO}    & World Health Organization \\[2pt]
\end{tabular}

\newpage
\thispagestyle{empty}
\setcounter{page}{12}
~\vspace*{570pt}
\begin{paragraph}
\noindent
All the maps in this work were processed with the use of ArcGIS 10.3 from data and shapefiles made available from the South Sudan Government, the South Sudan National Bureau of Statistics and the World Health Organization. 
In order to maintain the authenticity of these information, we tried to minimize the number of modifications to the least possible yet adequate to allow us to analyze the new acquired information. Misfits between shapefiles and different spelling of toponyms are possible. 
Results and data of the model have been processed using MATLAB 2015. 
\end{paragraph}
\newpage
\thispagestyle{empty}
%\input{pretext}
%-----------------------------------------------------------------------%
%                          		CAPITOLI
%-----------------------------------------------------------------------%
\pagenumbering{arabic}
%-----------------------------------------------------------------------%
%                          CHAPTER 1: INTRODUCTION
%-----------------------------------------------------------------------%
\chapter{Introduction}
\setcounter{page}{3}
\label{chap1}
%-----------------------------------------------------------------------%
%                           	ON CHOLERA
%-----------------------------------------------------------------------%
\section{Cholera epidemiology}
\label{cholera}
\begin{figure}[!h]
\centering
\includegraphics[width=0.5\textwidth]{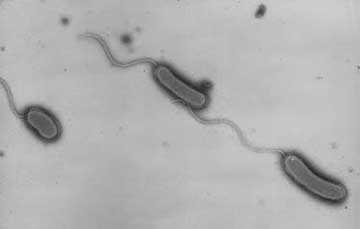}
\caption[\textit{Vibrio Cholerae} bacteria]{\co{Vibrio Cholerae} bacteria. \citep{cholerabacter}}
\label{vibrios}
\end{figure}

Cholera is an enteric disease caused by the bacteria \textit{Vibrio Cholerae}, usually of serogroup $O1$ and $O139$ \citep{Kaper1995.} (Fig.\ref{vibrios}). The infection is subject of studies since the XIX century when the disease spread in the Indian subcontinent, although proofs of a possible cholera infection can be found back in the 5th century BC in Sanskrit \citep{Harris2012}. Nowadays, cholera is a public health issue around the world, above all in developing countries, most of the time occurring in the African continent \citep{Bhattacharya2009}. The bacterium naturally lives in the human intestine but it also survives and reproduces in the aquatic environment, causing spread of the infection thanks to the waterways and the river networks. In fact, it has been shown that the bacteria are found in association with zooplankton and aquatic vegetation, resulting autochtonous in some coastal regions \citep{Colwell1996,Lipp2002}. 

\begin{figure}[!h]
\centering
\includegraphics[width=1\textwidth]{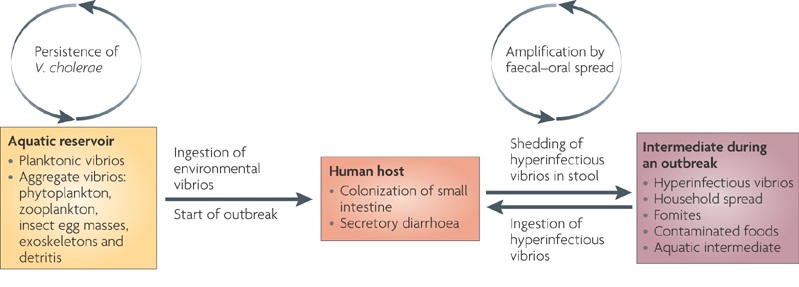}
\caption[Paths and states of cholera epidemics]{Schematic representation of cholera transmission paths and states. Susceptible people can contract the disease by ingestion of environmental vibrios. Symptomatic infection enhances, via excreta, the local concentration of the pathogen, powering the route of transmission with hyperinfectious states and amplifying the faecal-oral spread of the disease \citep{Nelson2009}.}
\label{transmission}
\end{figure}

Transmission of the disease occurs through ingestion of contaminated water or food \citep{Miller1985}. Once ingested, the bacteria \textit{V.Cholerae} colonize the small intestine and elaborate \textit{cholera toxin}, a protein toxin that triggers fluid and electrolyte secretion by intestinal epithelial cells \citep{Nelson2009}.
%The 'environmental' \textit{V.Cholerae} once reintroduced into the human organism, results infectious due to change in genetic and microbiological properties that are not objects of this thesis. When reintroduced, the bacteria colonize the small intestine and elaborate \textit{cholera toxin}, a protein toxin that triggers fluid and electrolyte secretion by intestinal epithelial cells \citep{Nelson2009}. 
Recently, laboratory analysis have suggested that the passage of the bacteria into the gastrointestinal tract can raise in hyperinfectious status (Fig.\ref{transmission}), facilitating the \textit{human-to-human} transmission of the illness \citep{Bertuzzo2008}. Other studies are now highlighting the role of animals in carrying the bacteria and the house environment as a reservoir of bacteria \citep{Kaper1995.}.
\\ \par The infection can be asymptomatic or can affect the human host with watery diarrhea that lasts a few days. Other symptoms can be nausea and muscle cramps. In some cases, severe diarrhea can lead to dehydration and electrolyte imbalance, until death \citep{WHO2010Weekly}. The disease can be defined as \textit{endemic} or \textit{epidemic}. It is endemic in cases in which the infection recurs in time and place, whereas epidemic denotes cholera occurring unpredictably. 
\par The role that environmental factors play in the spreading phenomena of cholera infection is then evident. As the bacteria can move in the aquatic environment, each change in the hydrological cycle may affect the pathogenic concentration in water: rain and its seasonal behavior, droughts and floods, can enhance or reduce the transmission process. 
\par Nevertheless, relevance has to be given to the environmental matrix in which the disease spreads into disease-free regions \citep{Bertuzzo2008}, together with consideration on human mobility and travelers carrying the disease in long-distance journeys. Susceptible people traveling on a daily basis may contract the disease in destination sites and take the disease back to the possibly uninfected communities where they regularly live. At the same time, infected individuals not showing severe symptoms, can carry the illness releasing bacteria via their faeces \citep{Mari2012}. Finally, symptomatic infected individuals locally increase the bacteria concentration, which is then spread along the hydrological network.
%-----------------------------------------------------------------------%
%                           	MODELING
%-----------------------------------------------------------------------%
\section{Modeling the epidemics}
\label{modelinf}
\par Epidemiological models for large-scale diseases as cholera, likewise measles, SARS and other contagious diseases, are relatively recent and are now taking off for their capability to understand the process dynamic and forecasting, analyze intervention scenarios and attack strategies to reduce the spreading phenomena. The first cholera model was proposed by \citet{Capasso1979} to describe the epidemic in Bari (Italy) during 1973. \citet{Codeco2001} extended the model, considering the dynamics of the susceptible population, together with the dynamics of the infected population and the free-living pathogens. Particular progress in the mathematical modeling of cholera epidemics has been made in response to the dramatic Haiti outbreak in 2010, in the attempt of aiding the real-time emergency management in allocating health care resources and evaluating intervention strategies \citep{Bertuzzo2014}. 
\\ \par The model used in this work to study the dynamic of cholera epidemics in South Sudan, has been developed in the Laboratory of Eco-Hydrology at the \'Ecole Polytechnique Fédérale de Lausanne, together with the cooperation of the Polytechnic University of Milan and the University of Padua within the project DYCHO - \textit{Dynamics and Controls of large-scale Cholera Outbreaks} -, supported by the Swiss National Science Foundation. This deterministic model simulates the dynamics of Susceptible, Infected, Recovered individuals, and Bacterial concentrations -- \textit{SIRB model} -- in a spatially-explicit setting of local human communities. In this way the model results can be compare with the real epidemiological data \citep{Bertuzzo2014} collected in different zones of the infected region. With respect to the standard zero-dimensional SIR models, the spatially-explicit framework has the advantages to consider the environmental matrix along which the disease can spread and the river network in transporting and redistributing \textit{V.Cholerae} \citep{Bertuzzo2010}. An additional advantage of the proposed formulation is to include rainfall data and human mobility as drivers of transmission.
\\ \par A first formulation of this model has been used to simulate the 2000 cholera epidemic in KwaZulu-Natal province in South Africa \citep{Bertuzzo2008,Mari2012}. Other applications of the spatially explicit and rainfall-driven model include  cholera epidemics in Haiti in October 2010 \citep{Bertuzzo2011,Bertuzzo2012,Bertuzzo2014,Rinaldo2014,Mari2015,pasetto2016} and in Lake Kivu Region, Democratic Republic of Congo, between 2004-2011 \citep{Finger2014}. A last conception of the model has been applied for 2005 cholera epidemic in Senegal by \citet{Finger2016}, using mobile phone data to fully catch the influence of mass gatherings in the spreading of the disease.
%-----------------------------------------------------------------------%
%                           	CASE STUDY
%-----------------------------------------------------------------------%
\section{South Sudan epidemics}
\label{intrss}
In this thesis we study the epidemics that affected South Sudan in 2014 and 2015. Epidemiological records have been made available by the South Sudan Minister of Health - \textit{SSMoH}. In recent years, 4 major cholera outbreaks have occurred in South Sudan. The unstable social conditions of the Nation, together with poverty and lack of sanitation, enhanced the risk until the outbreak of a new infection in 2014.

\begin{figure}
\centering
\includegraphics[width=1\textwidth]{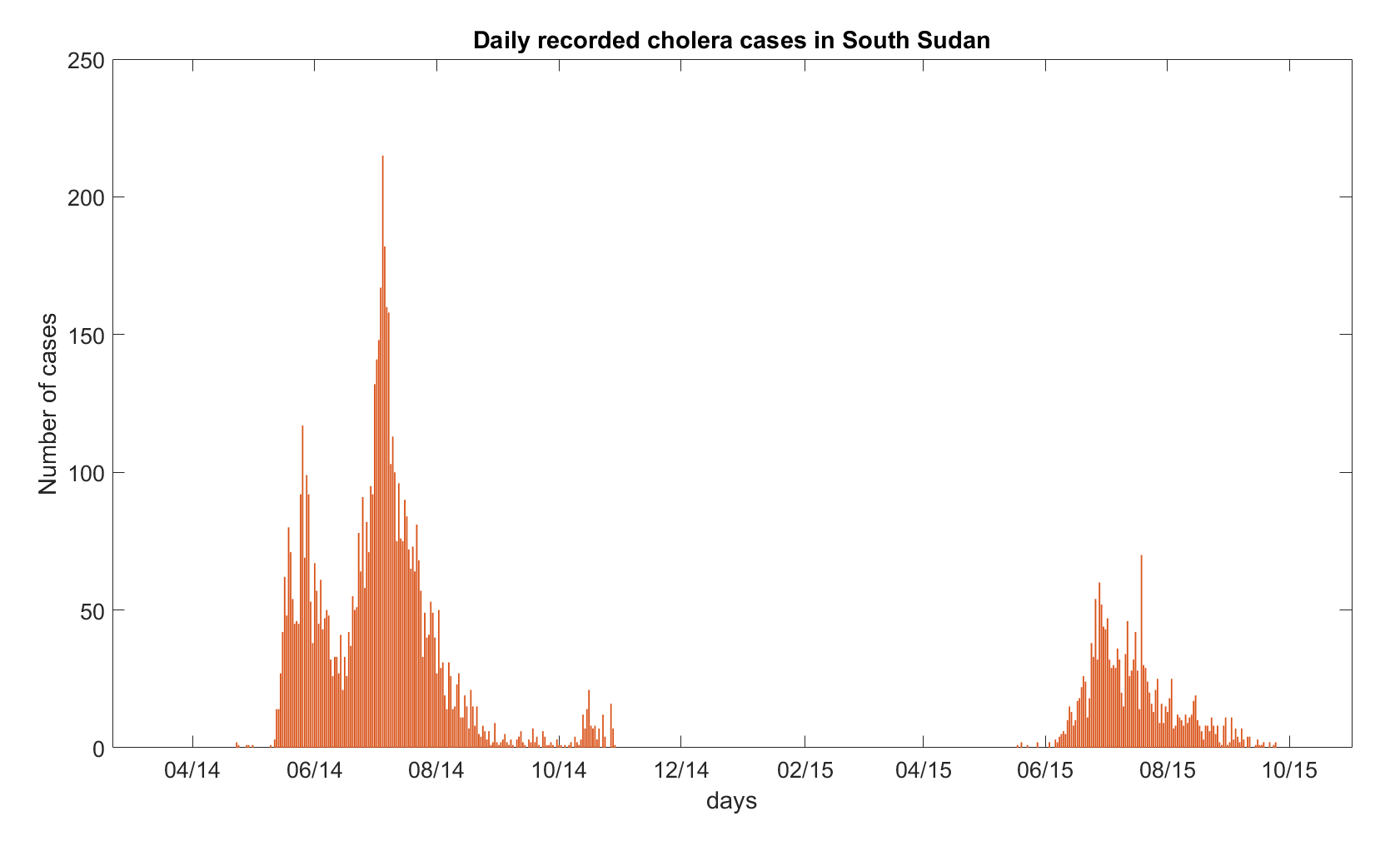}
\caption[Time series of suspected cholera cases]{Time series of all suspected cholera cases registered by SSMoH and WHO all over South Sudan territory.}
\label{cholera_cases}
\end{figure}

Despite the oral cholera vaccination campaign into the Protection of Civilians camps at the beginning of 2014, whose population-level effectiveness was highly affected by mass population displacements \citep{Azman2016}, the first cholera case was confirmed in April 23, 2014 in the capital Juba. Four weeks later, the SSMoH declared the outbreak. In total, 6,269 suspected cases were recorded, including 156 deaths. The epidemic ended in October 2014, not evolving in endemic. Unfortunately, the country passed through a similar condition during the following year. The 2015 epidemic, if compared to the previous year, affected less the country in terms of number of cases and duration (Fig. \ref{cholera_cases}). First case showed up in May 18, 2015 and the totally recorded suspected cases were 1,575, with 46 deaths. The epidemic ended in September 2015.

%-----------------------------------------------------------------------%
%                           OBJECTIVES AND STRUCTURE
%-----------------------------------------------------------------------%
\section{Objectives and structure}
This thesis has three aims. First, with our work we would like to contribute to the understanding of the South Sudan epidemic dynamics, by analyzing the context in which the disease spreads. The second task of the thesis consists in the setup of the spatially-distributed cholera model for the SS epidemics, objective particularly challenging due to the critical country conditions, the internal civil wars and the scarcity of mobility data. Different advanced calibration procedures have  been proposed to assess the probability distribution of the model parameters governing the epidemic processes. The third and final goal is to analyze the results of the simulations in order to suggest future directions for further understanding of cholera dynamics in South Sudan, and possibly extending the discussion to general large-scale epidemics.
\\ \par The report is organized as follows:
\begin{description}
	\par \item [Chapter 2: \textit{Background of South Sudan cholera epidemics},] delineation of the area affected by the epidemics. The geography, hydrology and climate of the country will be shortly described as well as the time and space characteristics of 2014-2015 epidemics.
	\par \item[Chapter 3: \textit{Epidemiological model},] details on the mathematical structure of the rainfall-driven, spatially-explicit model of cholera epidemics. This chapter contains as well a brief yet concise digression on the \textit{Ensemble Kalman Filter}, the data assimilation technique here adopted for the calibration of the model against the epidemiological data.
    \par \item[Chapter 4: \textit{Model setup and epidemiological data},] chapter dedicated to the description of the model setup and input data (rainfall and population). Particular attention is dedicated to the domain discretization, based on the available epidemiological records.
	\par \item[Chapter 5: \textit{Results and discussion},] presents and discusses the model responses associated to different calibration procedures for both years of analysis. 
    \par \item[Chapter 6: \textit{Conclusions},] drives toward main conclusions of this thesis work, highlighting the benefits of using the spatially-explicit model for the simulation of the epidemics and suggesting some possible model improvements for future works.
\end{description}
%-----------------------------------------------------------------------%
%                          	CAPITOLO 2 South Sudan
%-----------------------------------------------------------------------%
\chapter{Background of South Sudan cholera epidemics}
\label{chap2}
%-----------------------------------------------------------------------%
%                          			HISTORY
%-----------------------------------------------------------------------%
Originally belonging to Sudan, the Republic of South Sudan is the newest nation in the world. 
The territory of Sudan, referring to both Northern and Southern parts, has always been land of great interest by conquerors due to the presence of the Nile and, recently, due to oil reservoirs. After years of Arab domination and British-Egyptian rule, Sudan gained its independence in the second half of the XX century. From then onward two civil wars (1955-1972; 1983-2005), due to heterogeneity in ethnicity and in religion between the two sudanese parts, affected the entire population, leading to more than 2 millions deaths and 4 millions refugees. In 2005, at the end of the Second Sudanese Civil War, the so-called `Comprehensive Peace Agreement' among the two factions established the terms of autonomy of the Southern Sudan. The country obtained independence from the Northern part thanks to a referendum on self-determination in July 2011, becoming the 54\ap{th} country of the African Continent \citep{treccani,Collins2015,keyindicators}.

\begin{figure}[t]
\centering
\includegraphics[width=1\textwidth]{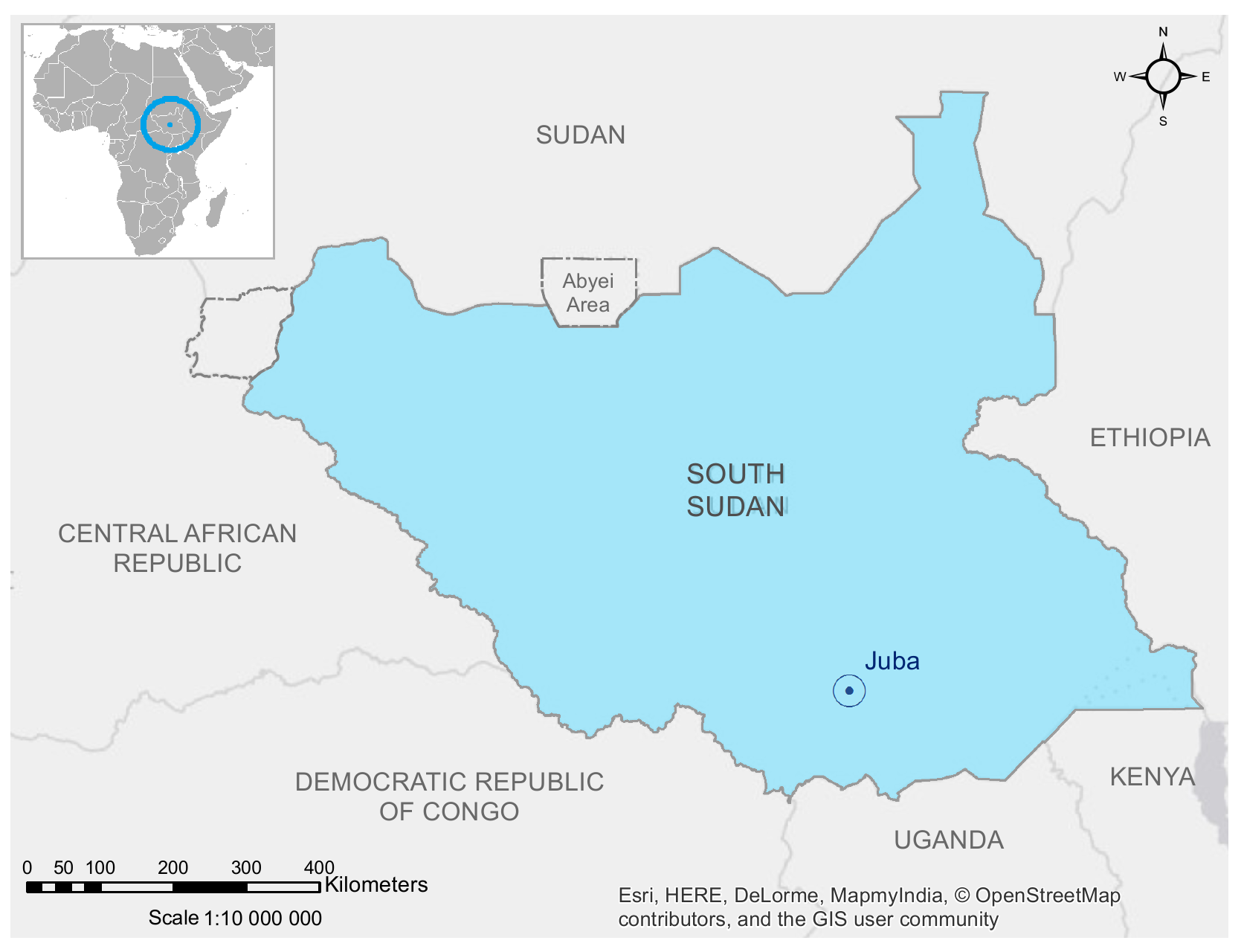}
\caption[South Sudan geography]{Position of South Sudan territory and its capital Juba in the African continent. Dashed lines show disputed areas and undefined borders. Data provided by SSMoH and \citet{openstreetmap}.}
\label{position}
\end{figure}
%-----------------------------------------------------------------------%
%                         GEOGRAPHY AND LAND ADMIN
%-----------------------------------------------------------------------%
\section{Geography and land administration}
\subsection{Geography, topography and hydrography}
\label{geogr}
%-----------------------------------------------------------------------%
%                           	GEOGRAPHY
%-----------------------------------------------------------------------%
\par South Sudan is a landlocked country situated in Central Africa. It shares its borders with Sudan to the north, Ethiopia to the east, Kenya, Uganda and the Democratic Republic of Congo to the south, and Central African Republic to the west. The capital city is Juba, which is also the most populated one, located in the south east. The total extension of the country is about 644,329 km\ap{2} \citep{CIA2015}. Due to political and governance conflicts, borders are actually not well defined, as instance the Abyei area and the north-western borders. For the purpose of this thesis, we do not take into account the disputed areas. Figure \ref{position} shows South Sudan placement and borders.

%-----------------------------------------------------------------------%
%                           	TOPOGRAPHY
%-----------------------------------------------------------------------%
\par
\begin{figure}[t]
\centering
\includegraphics[width=1\textwidth]{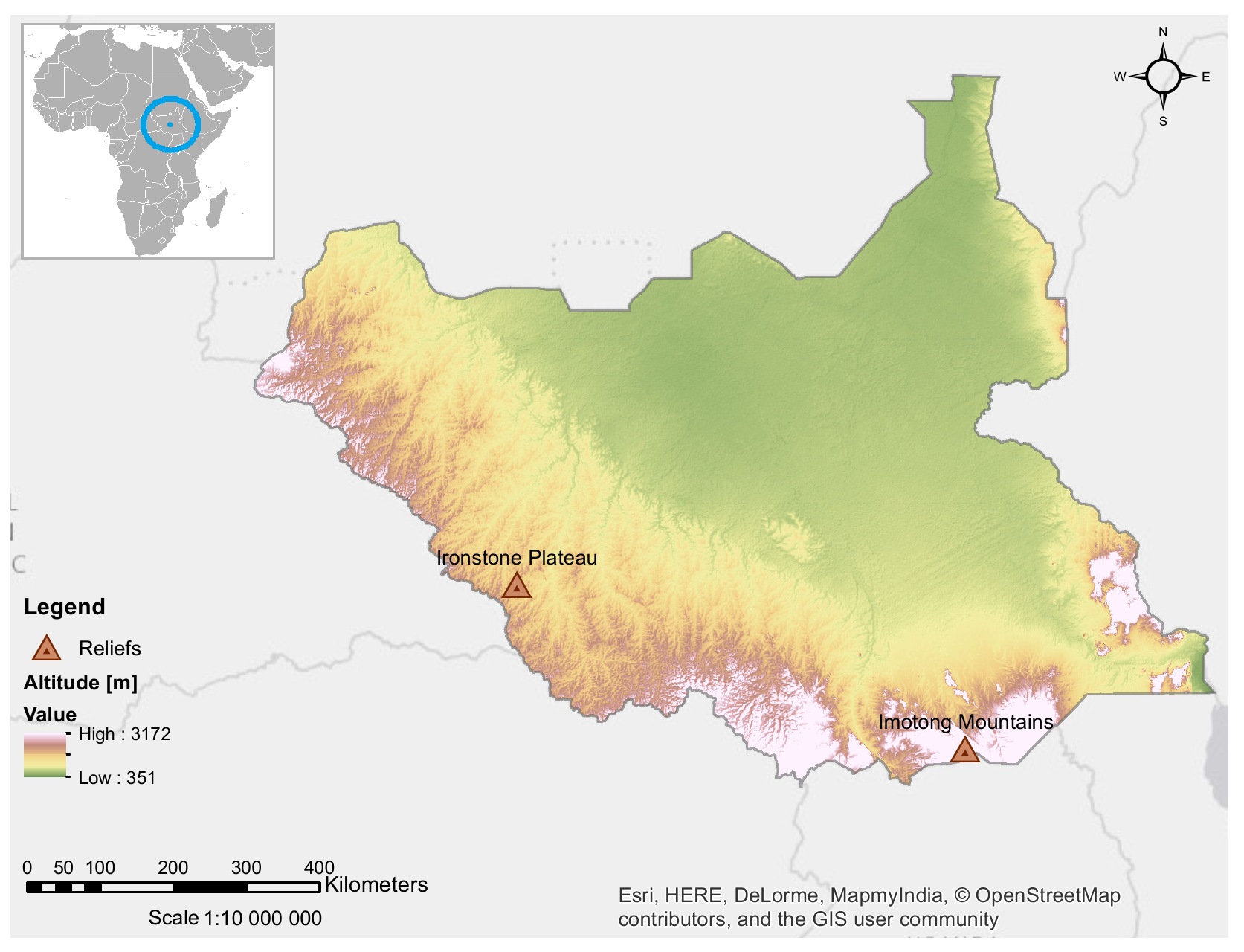}
\caption[South Sudan topography]{Digital Elevation Model of South Sudan provided by SSMoH and position of the upland areas. Resolution 90 meters. Data provided by SSMoH and \citet{openstreetmap}.}
\label{elevation}
\end{figure}

 The country is predominantly flat. There are only two upland areas: 
\begin{itemize}
\item the \textit{Imotong Mountains}, situated on the Uganda borders, characterized by high peaks. Here the Mount Kinyeti, the highest point in South Sudan, raises up to 3,172 meters \citep{Collins2015,GlobalSecurity2016}
\item the \textit{Ironstone Plateau}, at the feet of the `Nile Congo watershed' on the west side borders, whose peaks' elevation is among 800 and 1,700 meters. \citep{Collins2015}
\end{itemize}
Figure \ref{elevation} shows the topography of the area under study.

%-----------------------------------------------------------------------%
%                           	HYDROGRAPHY
%-----------------------------------------------------------------------%
\par
\begin{figure}[t]
\centering
\includegraphics[width=1\textwidth]{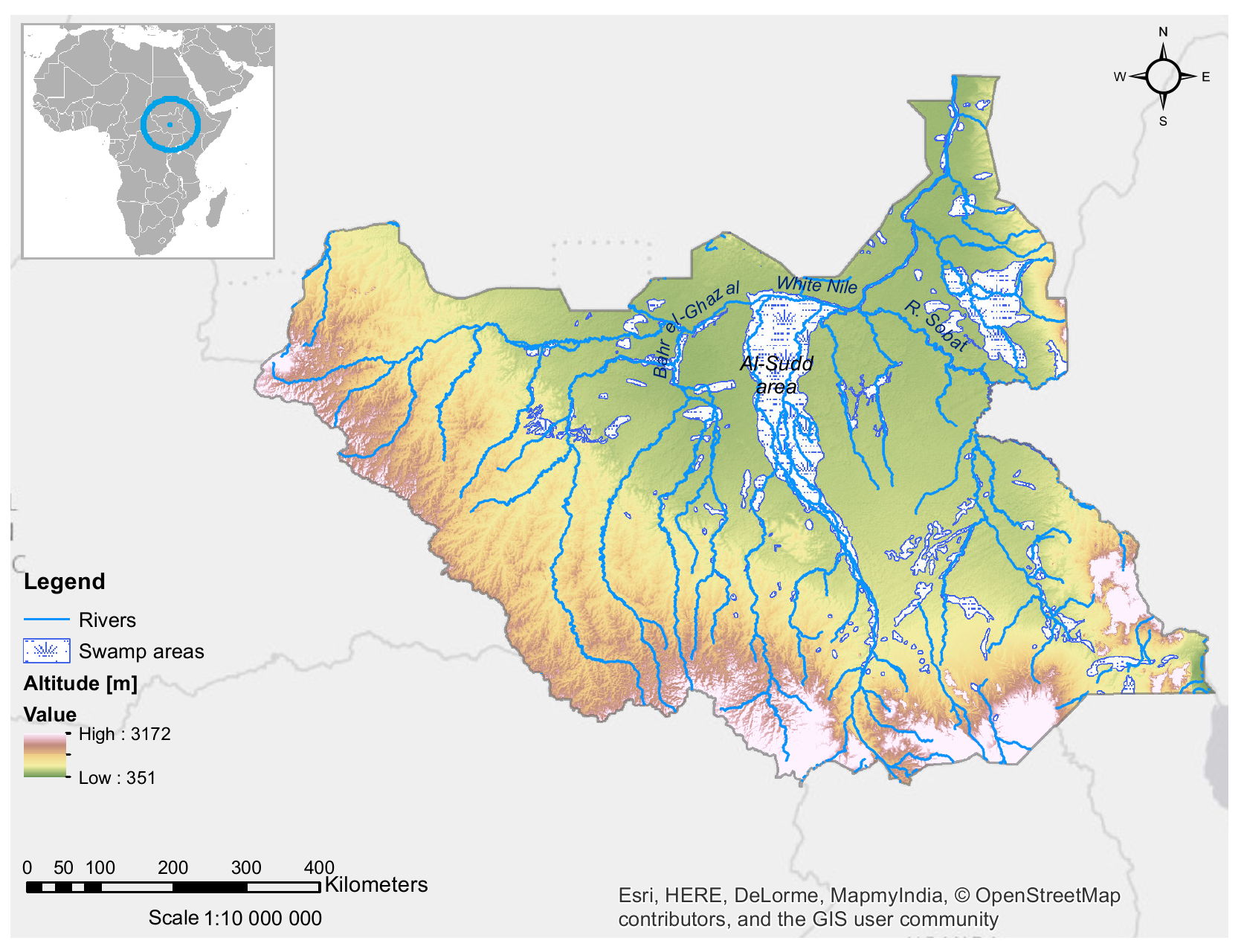}
\caption[South Sudan hydrography]{Hydrographic network of South Sudan. Al - Sudd swamp area is highlighted. Data provided by SSMoH and \citet{openstreetmap}.}
\label{hydrography}
\end{figure}

 At the hearth of the nation a clay plain is present, where the Mountain Nile flows from south to north. This area, named \textit{Al-Sudd}, is characterized by swamps,lagoons, side channels, and several lakes created by the Mountain Nile and its tributaries (mainly the Sobat River from Ethiopia and the Bahr Al-Ghazal River from the west) \citep{CIA2015,Collins2015} (Fig.\ref{hydrography}). In this area, the Nile becomes the White Nile thanks to the debit of the other streams.\par The Al-Sudd swamp is the main drainage area, covering more than 100,000 km\ap{2}, almost the 15\% of the country's total area. The label `Sudd' is the local term defining the vegetation that covers the entire region \citep{Collins2015}.

In 1978 the Jonglei Canal was planned to bypass Al-Sudd swamp and provide a straight well-defined channel for the Nile to flow. The building of the structure stopped due to the political instability and during the rain season , this part of the country is not viable using means of land transport \citep{treccani,keyindicators}.
%-----------------------------------------------------------------------%
%                           	ADMINISTRATION
%-----------------------------------------------------------------------%
\newpage
\subsection{Administrative division}
\label{admin}
\begin{figure}[t]
\centering
\includegraphics[width=1\textwidth]{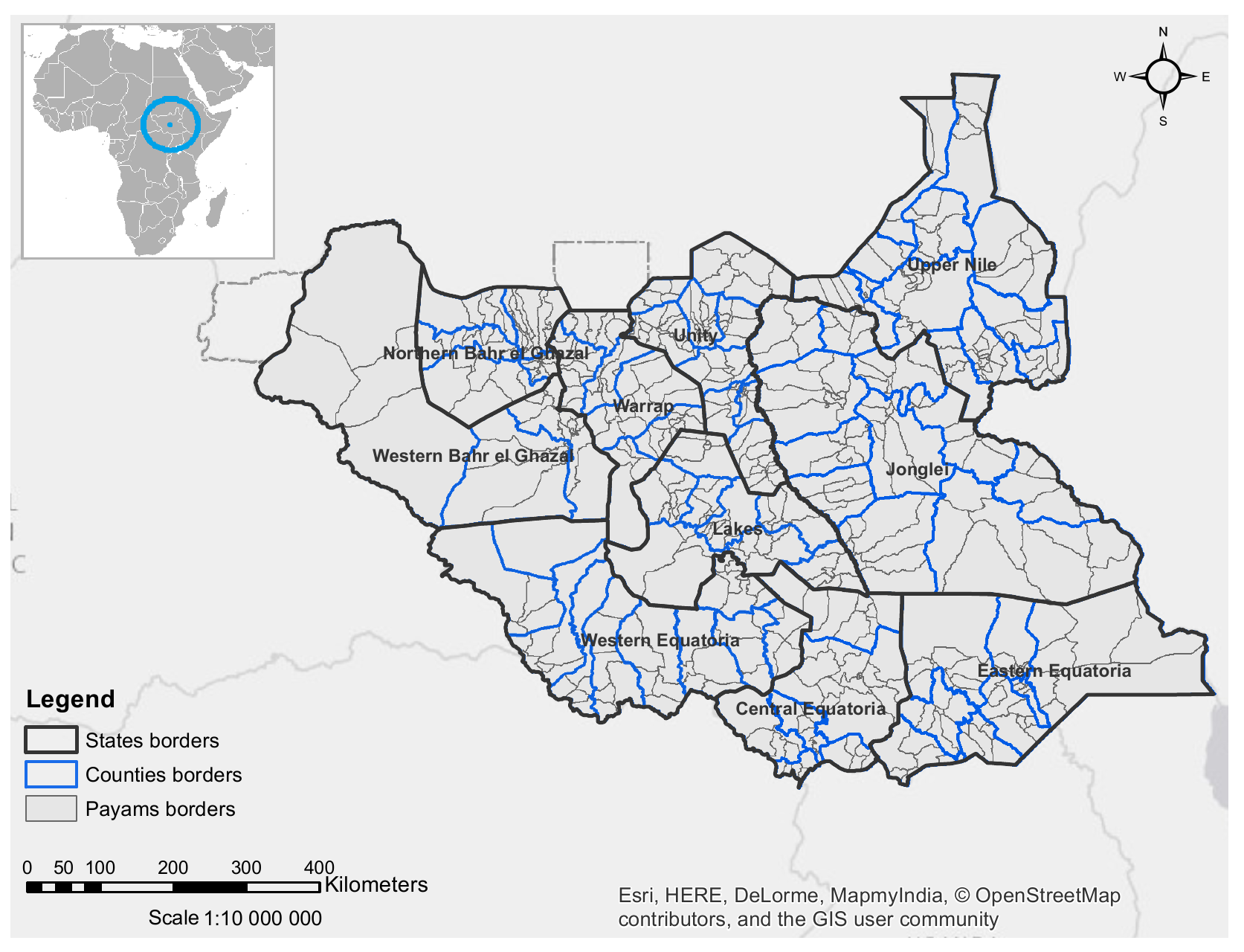}
\caption[South Sudan administrative division]{Administrative division of South Sudan up to November 2015. The map shows the 10-states division, together with the organization in counties - thick dark grey line - and Payams - thin black line.}
\label{administration}
\end{figure}

For administrative purposes, the nation is divided in ten states (Fig.\ref{administration}), following the three historical regions of Sudan: \textit{Bahr el Ghazal, Equatoria,} and \textit{Greater Upper Nile}. Each state is then divided in \textit{counties}, themselves divided in \textit{payams}, which are aggregates of villages. The minimum number of inhabitants required per payam is 25,000. They can be further subdivided into a variable number of \textit{Bomas} \citep{Grawert2010}.

A new subdivision has been proposed by President Salva Kiir in December 2015 for establishing 28 states, whose borders are defined following ethnicity and the historical three regions \citep{SudanTribune2015}. For the aim of this thesis, and following all the official reports found, we will consider the structure showed in Figure \ref{administration}, i.e. before December 2015. 
%-----------------------------------------------------------------------%
%                           	CLIMATE
%-----------------------------------------------------------------------%
\newpage
\section{Climate}
\label{clima}
\begin{figure}[t]
\centering
\includegraphics[width=1\textwidth]{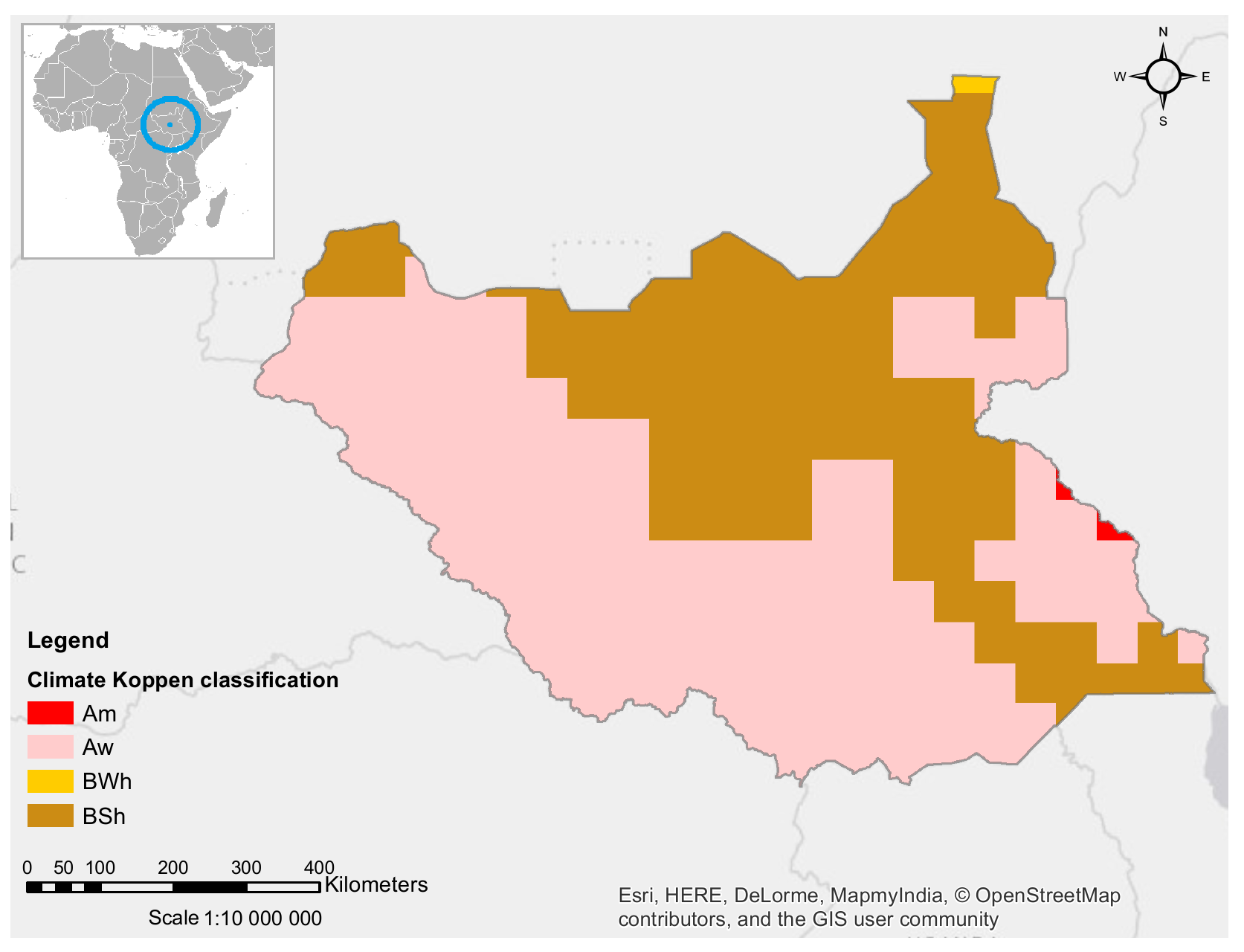}
\caption[South Sudan K\"{o}ppen-Geiger climate classification]{K\"{o}ppen-Geiger classification on the region of South Sudan from observed data between 1976 and 2000. Regular 0.5 degree lat/lon grid, about 56 km at these latitudes \citep{koppenclimate}.}
\label{climate}
\end{figure}

The southern sudanese territory is vast, so that diverse climates are present along it. Analyzing the country in latitude, according to the K\"{o}ppen-Geiger Climate classification based on temperature and precipitation (Fig.\vref{climate}), we can distinguish dry, arid and semiarid climates in the north (\ang{12}N, Group B) and, wet savanna climate in the south (\ang{3}N, Group A) \citep{Collins2015,koppenclimate}. Specifically, a small north-eastern part on the borders with Sudan is classified as \textit{BWh}, presenting warm desert climate conditions; while on the west side of the same border, classification states a warm semi-arid climate \textit{BSh}.

\par The majority of the country is although influenced by the annual fluctuations of the \textit{Inter-Tropical Convergence Zone}, determining a tropical savanna climate group (\textit{Aw}) \citep{GlobalSecurity2016}. Cool and dry winds from north-east blow at the beginning of the year, meeting after few months the moist southwesterlies winds. The wet season starts roughly in April and lasts until November, though its length is variable. Rainfalls result to be heavier in the southern upland areas, due to orographic phenomena. In the tropical savanna region, temperature has not spatial variability, and its values increase going towards the end of the dry season and reaching more than \SI{35}{\degreeCelsius} in March. July is the coldest month, when temperature falls down to \SI{20}{\degreeCelsius} or less \citep{WorldAtlas2016,GlobalSecurity2016}.
\par The meteorological situation has effects on the social and economical southern sudanese aspects and on the environment. As instance, the tropical forest that grows during the wet season is cut and burnt during the dry season, compromising the ecosystems. Roads, that in some cases only consist in rough tracks or not even that, become impassable during the rainfalls.
%-----------------------------------------------------------------------%
%                       SOCIAL AND POLITCAL ASPECTS
%-----------------------------------------------------------------------%
\section{Economic, social, political and demographic aspects}
\label{social}
%Possible SUBSECTION
The Republic of South Sudan is one of the poorest and less developed country in the world, appearing at the 169\ap{th} place in the Human Development Index scale \citep{keyindicators}. \par Subsequently to the establishment of the secession, conflicts were born in the area of Abyei, on the borders with Sudan, as a result of disputes for the control of oil reservoirs. Moreover, the government was not able to manage a country with so many internal fractures because of ethnicity. Corruption and conflicts of interests rule the country. The political party that brought the nation to the independence, the Sudan People's Liberation Movement (SPLM), is now divided and fighting for power, wasting money in army. At the same time, violent rebels fight against the government. In December 2013, violence broke out in the streets of the capital Juba, giving start to another internal war \citep{Kelly2016} that keeps going \citep{bbc}.
%\par The long history of wars has long deprived of an adequate education system, although a good education seems a top priority for southern sudanese people. English is the official language of the nation and is also the learning language, despite the numerous idioms spoken in the territory. More than the 80\% of the population is estimated to be illiterate. South Sudan has proportionately fewer girls going to school than any other country in the world. According to the \textit{United Nations Children's Emergency Fund} - UNICEF, fewer than one percent of girls complete primary education \citep{UNMISS2010,keyindicators}.
%\par Despite of the incredible economic resource of oil and the abundance of natural resources, more than half the population lives on less than \$1 per day. The country's economy depends on oil, from which the government derives almost 98\% of its budget revenues, and on agriculture and mining \citep{CIA2015}.
\par The lack of infrastructures that characterizes the country complicates the relationships with the surrounding countries. Roads are mainly unpaved; a single-track railroad connects the city of Weu in Western Bahr el Ghazal to Babonosa in Sudan. International flight connections are present from Juba and Malakal. Internal flight connections, supported by UN are scheduled \citep{keyindicators,UNMISS2010}. 
%-----------------------------------------------------------------------%
%                           	POPULATION
%-----------------------------------------------------------------------%
\subsection{Population}
\label{popolazione}
At the time of writing, the exact value of the population of South Sudan is unknown. The last census was organized in 2008 during the Sudanese administration, when the population reached by the census in the southern part was 8,260,490 \citep{SSYearbook10}. Estimation made by \citet{popUNprospect} states 12,340,000; while \citet{CIA2015} attests 12,042,910 people in July 2015. \citet{general_projectionSSNBS} projected for 2015 11,000,000 people.
\par The nation has the highest population growth rate in the world, registering an annual increment of 4\% \citep{CIA2015}.
\par The 2008 census data show that the population is very young, with half of it below 30 years old. Life expectancy is very low compared to values in neighboring countries. Population is uneven between the ten states, being Jonglei the most inhabited one (1,443,500 people in 2008) while Western Bahr El Ghazal the least one (358,692 in 2008) (Fig. \ref{administration} for geo-political references). The 83\% of the population lives in rural areas \citep{SSYearbook10}.
\par Moreover, distribution and density of the population are affected by refugees movements. South Sudan hosts both external refugees from Sudan, as a consequence to the long War in Darfur, and internal refugees, subsequently to the outbreak of violence in December 2013.
At this regard \citet{2015WHOUpdates} reported: 
\begin{quote}
\emph{``The humanitarian situation in South Sudan has deteriorated since the outbreak of violence on 15 December 2013. Total of 195,416 persons has been displaced to camps from Bor, Bentiu and Malakal in Jonglei, Unity and Upper Nile states.''}
\end{quote}
People move and establish in arranged settlements (defined as Internally Displacement Persons IDP) or to PoC - Protection of Civilians sites, as to say, sites where civilians seek protection and refuge at existing United Nations bases when fighting starts\citep{UNMISS2010}.\par Since the beginning of new fights, south-sudanese people are additionally moving abroad seeking for protection: \citet{UNHCR2016}, in the last update on June 3, 2016 reports 844,406 displaced persons crossing into Ethiopia, Kenya, Sudan and Uganda. The number of refugees that moved into the nation is 272,261 .\par Migrations are affected, besides frequency and intensity of conflicts, by the seasonality of rain, whose direct consequence is on the impossibility to cross the country in the mud, as described previously.
%-----------------------------------------------------------------------%
%                          WATER HEALTH AND SANITATION
%-----------------------------------------------------------------------%
\subsection{Water and sanitation}
\label{sanitation}
\begin{figure}
\centering
\includegraphics[width=0.65\textwidth]{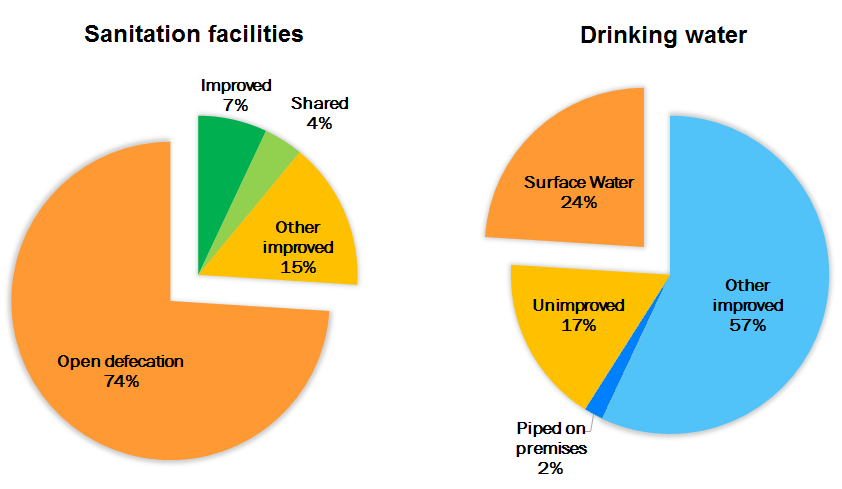}
\caption[Water, sanitation and hygiene data]{Water, sanitation and hygiene conditions in South Sudan. Color scale following UN definitions on WASH. Data from \citet{WHO/UNICEF2015JointAssessment}.}
\label{wash}
\end{figure}

According to \citet{WHO/UNICEF2015JointAssessment}, the 59\% of the population has access to improved sources of drinking water (considering as improved `Piped on premises' and `Other improved', definition by UN, see reference), i.e. people use water that received some kind of treatment. The 24\% instead uses surface water (see Fig. \ref{wash}). In urban areas the percentage of people with access to improved systems is 67\%, while is 57\% in rural areas. Many inhabitants have to walk for more than 30 minutes to collect drinking water \citep{keyindicators}. \par As regards sanitation, only the 7\% of the population uses improved sanitation facilities. In urban areas, the 16\% has access to improved sanitation facilities and the 10\% shares them. The 74\% of the peoples still practice open defecation. These conditions enhance the risk of water-driven infectious diseases, as cholera, to spread. 

\par Conflicts and social instability worse the situation. \cite{2015WHOUpdates} in its report on December 2013 added:
\begin{quote}
\emph{``Poor water, hygiene and sanitation conditions in the camps for the internally displaced people increases risk of communicable diseases.''}
\end{quote}
%-----------------------------------------------------------------------%
%                           	EPIDEMICS
%-----------------------------------------------------------------------%
\section{Cholera epidemics}
\label{colerass}
\begin{figure}[t]
\centering
\includegraphics[width=0.6\textwidth]{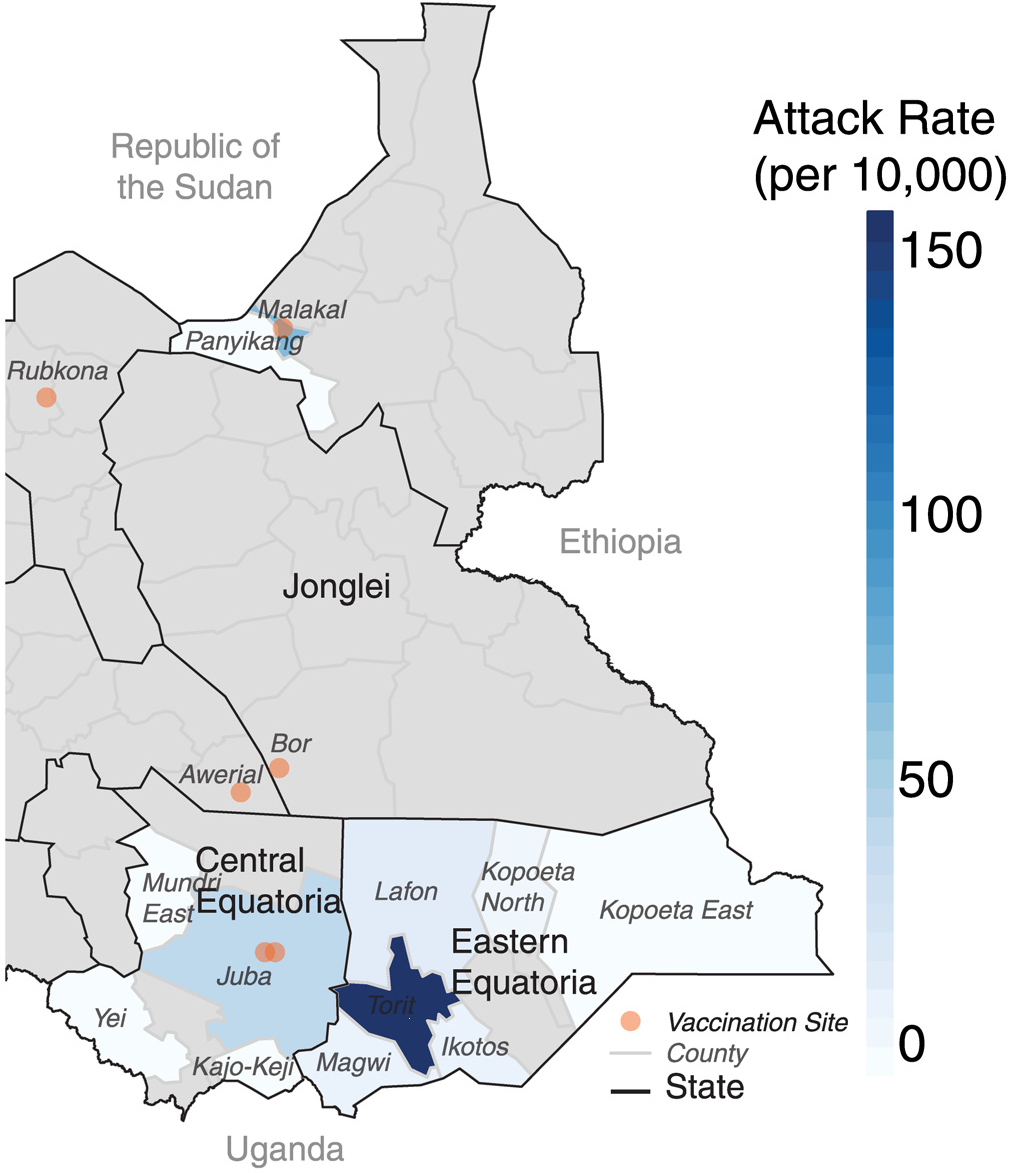}
\caption[Vaccination sites and epidemic attack rate for 2014]{County-level attack rates in the 2014 cholera epidemic with respect to the vaccination locations (orange dots). Grey areas represent counties with no suspected cholera cases. Figure from \citet{Abubakar2015}.}
\label{attackrate}
\end{figure}

Cholera epidemics affected the country back in 2006, when the number of cases totaled 19,777; in 2007 were 22,412; in 2008, cases totaled 27,017; and in 2009 cases totaled 48,035 \citep{Ujjiga2015}. In 2012 an outbreak has been avoided, thanks to a preventive mass vaccination campaign decided on a risk assessment for the potential impact of cholera into refugee camps \citep{vaccinationprev}.
As WHO anticipated, the fear of a new outbreak moved the South Sudanese Ministry of Health - SSMoH - to action. As reported by \citet{Abubakar2015} and \citet{Azman2016}, at the beginning of 2014 the SSMoH requested the intervention of WHO to vaccinate 163,000 IDPs in six small camps throughout the country using the global oral vaccine stockpile. The six refugee camps are located in Awerial, in Lakes state, two sites in Juba, a camp in Rubkona, Unity state, Bor in Jonglei state and Malakal in Upper Nile (Fig.\ref{attackrate}). 

\par A clinic-based cholera surveillance system to record basic patient data and laboratory results has been implemented by SSMoH and WHO. The collected epidemiological data consist in the \textit{cholera suspected cases}, which definition states:\\ \emph{<<anyone with acute watery diarrhea diagnosed by a clinician. Suspected cases were considered confirmed if the patient had a culture-positive fecal sample>>} \citep{Azman2016}. 
\\ \par Our analysis will include all the suspected cases among the health facilities in 2014 and 2015, data available thanks to the Ministry of Health releasing epidemiological records. 

\subsection{Year 2014}
Despite the vaccination campaigns whose response varied in space and among the population \citep{Azman2016}, the first case of cholera was confirmed during the vaccination period in the area of the capital Juba on April 23, 2014. The origin of this infection are unknown, although cholera had been reported in the neighboring area of Uganda during previous weeks \citep{Abubakar2015}. In a month, the officials declared a cholera outbreak. In 2014, 6,269 suspected cholera cases were recorded, in which appear 156 deaths. The epidemic last until October 29, 2014. Cases were recorded inside and outside the camps.

\begin{figure}[t]
\centering
\includegraphics[width=1\textwidth]{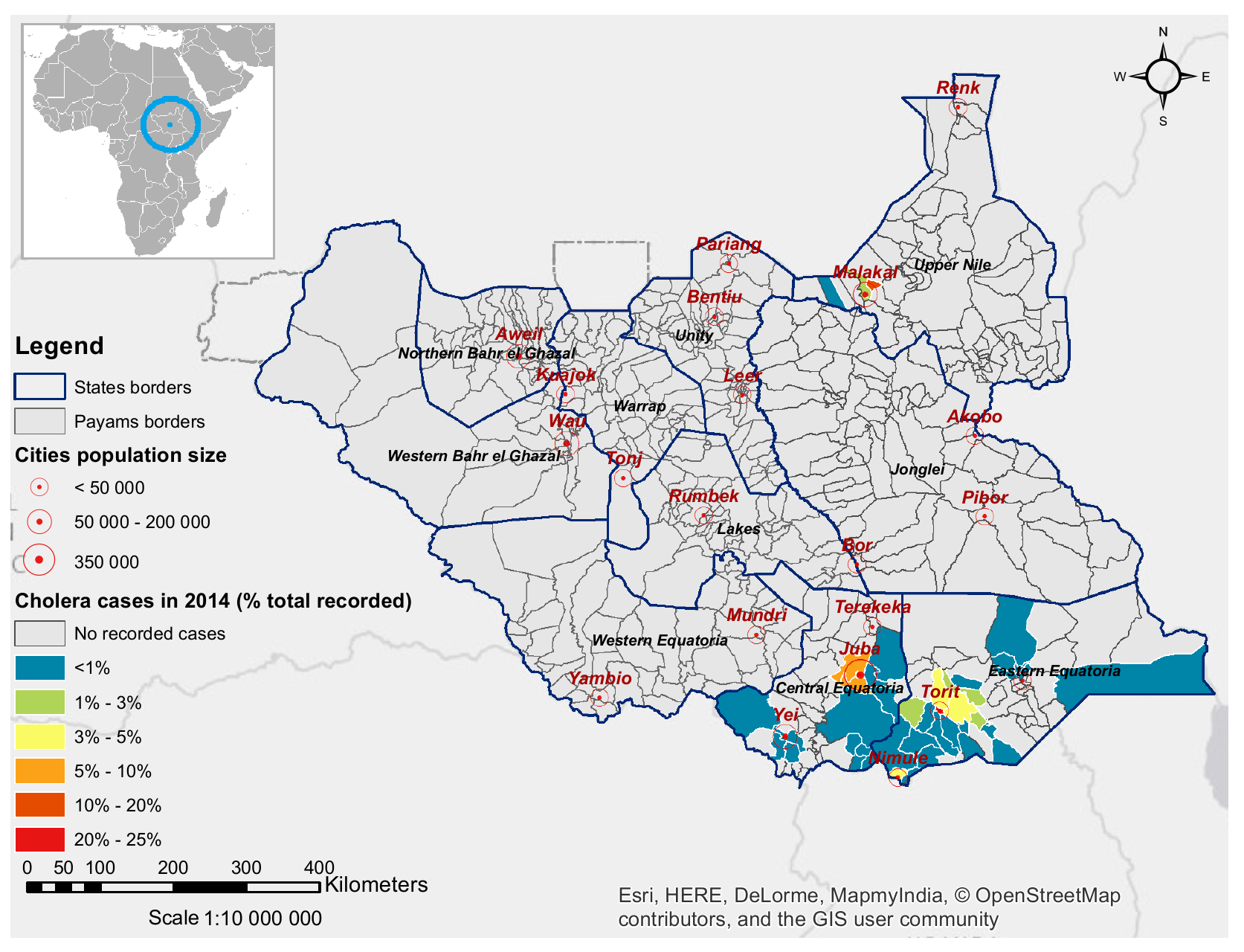}
\caption[Cholera epidemic: map for 2014]{Percentage of the total cases occurred in each payam during the epidemic of 2014. The population sizes of the main cities is based on the census data \citep{census2008tables} and are representative of the population distribution.}
\label{map2014}
\end{figure}

As \cite{Abubakar2015} reported, most of the cases occurred outside the camps. The attack rate of the infection (Fig. \ref{attackrate}), defined as the total number of recorded cases among the population considered, was diverse between areas all over the country due to differences in baseline health care infrastructures. As figure \ref{map2014} shows, during year 2014, the epidemic involved the states of Central Equatoria, Eastern Equatoria and Upper Nile (administrative division Fig. \ref{administration}). In the state of Western Equatoria, only three cases were recorded in the payam near to the city Mundri. As it will be better explained in Chapter 4, these cases do not affect the proposed modeling approach, and will be neglected in the following.

Figures \ref{attackrate} and \ref{map2014} show that most of the cases were recorded around the capital area of Juba and in a Payam on the border with Sudan, near Malakal. The infection jumped to Malakal presumably from an infected traveler \citep{Abubakar2015}, over a large area with no confirmed cholera. Juba and Malakal counties were the only two places that experienced cases either within the population targeted for vaccination or the surrounding areas \citep{Abubakar2015}. 

\subsection{Year 2015}
The continuous conflicts, displacement of people and other disease outbreaks as measles and tuberculosis, had worsened the sanitation condition, overall between people living in the PoC sites. Although the adoption of further prevention measures \citep{WHO2015}, cholera cases were reported by WHO in Eastern Equatoria State (\ref{administration}) with 43 cases and three deaths between 11 - 19 February 2015. A new vaccination campaign started, in spite of which, another cholera outbreak had been declared in May 2015, with total number of suspected cases 1,757 including 46 deaths. The epidemic died out on September 24, 2015 \citep{2015WHOUpdates}. \citet{WHO2015} supported Oral Cholera Vaccination campaigns in Bentiu and Juba PoCs, targeting more than 100,000 people. Following the reports, other two were planned in Malakal and Juba.   

During year 2015, the disease showed up in the two states of Central Equatoria and Jonglei (Fig. \ref{administration}). The regions hit more severely from the epidemic were Bor South, payam of Bor, and Juba.

\begin{figure}[h!]
\centering
\includegraphics[width=1\textwidth]{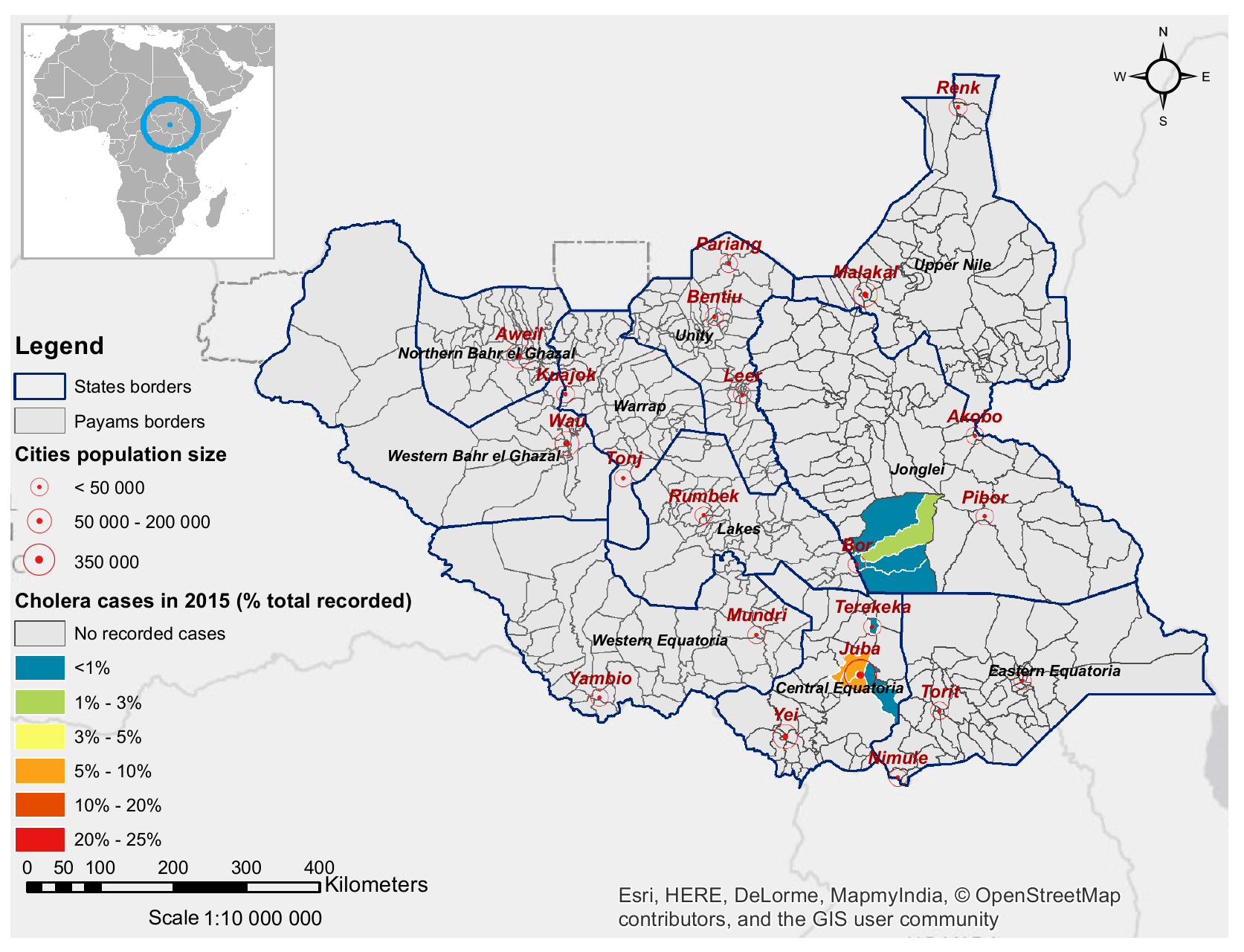}
\caption[Cholera epidemic: map for 2015]{Percentage of the total cases occurred in each payam during the epidemic of 2015. The population sizes of the main cities is based on the census data \citet{census2008tables} and are representative of the population distribution.}
\label{map2015}
\end{figure}

As Figures \ref{map2014} and \ref{map2015} show, the disease hit the nation differently in places and numbers during the two years; notwithstanding the capital area of Juba, which is also the most populated, was always deeply involved. 

%-----------------------------------------------------------------------%
%                          	 CHAPTER 3: MODEL
%-----------------------------------------------------------------------%
\chapter{Cholera epidemiological model}
\label{chap3}
Epidemiological models were born to face emergencies arising from lethal diseases and infections. As soon as epidemiological data started to be recorded, the scientific community committed itself to analyze the dynamics of pathogenic diseases, trying to identify the processes that enhance the infection spread and, moreover, to find a feasible mathematical approach to simulate the outbreak. \\ During the last years, the progresses in the computational power have brought to the numerical resolution of mathematical models that are even able to predict epidemics in time and space, , and which reliability has been demonstrated for the Haitian cholera epidemics (see e.g. \citep{Andrews2011,Righetto2011,Bertuzzo2014,Rinaldo2014} or for influenza (see e.g., \citet{influenzafore}) and even HIV/AIDS (see e.g., \citet{hivfore}).
%-----------------------------------------------------------------------%
%                          EPIDEMIC MODELING
%-----------------------------------------------------------------------%
\section{Model assumptions}
Epidemiological models are based on two main hypotheses \citep{Barabasi}:
\begin{itemize}
	\item \textit{Homogeneous mixing hypothesis} also called \co{fully mixed} or \co{mass-action approximation}, according to which each individual has the same probability to get in contact with an infected person. In this way we do not need to know the exact contact network between people;
    \item \co{Compartmentalization hypothesis} that classifies each individual according to the state of the disease they are at.
\end{itemize} 

Going into details on \co{compartmentalization}, during a cholera infectious process it is possible to distinguish three states, or classes, which each individual can be in:
\begin{description}
	\item{\gr{Susceptible \co{S}}}: state in which individuals are in healthy conditions, still not in contact with the pathogen and therefore in risk of infection;
    \item{\gr{Infected \co{I}}}: state in which individuals contracted the infection and can possible transmit it;
    \item{\gr{Recovered \co{R}}}: individuals that for a certain time are immune to the infection and do not contribute to the transmission.
\end{description}
Each pathogen has its own characteristic and by looking at the process of the related disease, we can also define other states:
\begin{description}
	\item{\co{Immune}} state in which individuals cannot become infected during a variable period of time after recovering. Measles immunity period is life-long, while cholera immunity response, some months or years, can vary with age of individuals and other factors \citep{Leung2012}.
    \item{\co{Asymptomatic and symptomatic}} states, that differentiate people according to their reaction to the pathogen: individuals getting in contact with it can develop, or not, evident infectious state. Asymptomatic infected do not contribute to the transmission process. Examples are hepatitis of type A and, as already explained in Sect. \vref{cholera}, cholera;
    \item{\co{Removed}} state that counts for people died due to the disease, possible event in cases of cholera, influenza and measles if not treated at all or in time.
\end{description}

Each state defines a compartment. Individuals can change their state during the infectious process, therefore can move from a compartment to another. At the beginning of an infection, in absence of previous epidemics and vaccination campaigns, all the population is susceptible. When an infected individual enters the system, all the individuals in the susceptible state can contract the illness and follow the path to the infected state, in case of a symptomatic individual or to the recovered state, in case of an asymptomatic individual.

The mathematical approach chosen to simulate the dynamics of the diseases, seen as changes in the state of individuals, hence movements of individuals between compartments, depends on the characteristics of the pathogen, its related disease and environmental factors. All the models are developed on specific requirements of any \textit{base model}, i.e. of a mathematical base structure used to describe the infectious process. The most frequently used base models in epidemiology are \co{Susceptible - Infected} (\co{SI}) model,\co{Susceptible - Infected - Susceptible} (\co{SIS}) model and, \co{Susceptible - Infected - Recovered} (\co{SIR}) model.

In order to understand the modeling approach that has been used in this work, we give a short explanation of the SIR basic model. 
%-----------------------------------------------------------------------%
%                          	   SIR MODEL
%-----------------------------------------------------------------------%
\subsection{SIR model} 
As \citet{KermackMc} defined it, a SIR model is the basic model in case of epidemic dynamics that include the recovery process and provide temporary immunity to the recovered individuals, as the case of \co{V. cholerae} inducted disease. Immunity implies that a recovered individual is not susceptible to the disease for a variable period of time. Immunity also represents a way to limit the spreading of the disease and the number of infected individuals.

Let us define \textit{N} as the number of individuals in a population or in a community, and \co{t} be the time. The state variable $S(t)$, $I(t)$, $R(t)$ represent respectively the number of Susceptible, Infected and Recovered individuals at time \textit{t} of an infectious process. 

The main idea of the \textit{SIR} model is that each individual is a node of the \textit{contract} network. The disease can move from node to node via mechanisms of transmission that are typical of the pathogen considered and that link nodes between them. Each node has \co{k} edges and we define $\beta$ as the rate of transmission, i.e. the number of individuals that contract the infection despite the total number of nodes linked to the infected one \citep{Barabasi}. 

%Let assume that the epidemics start at $t=$0 with a certain number of infected individuals, $I_0$, i.e. $I(0)=I_0$. If no vaccination campaign and no other cholera epidemics occurred before $t=$0, it is reasonable to assume that a certain part of the remaining population is asymptomatic (and, thus of class $R$), while the other part consists of susceptible individuals. Here we assume that $R(0)$ is proportional to $I(0)$, $R(0)=frac{1-sigma}{sigma} I(0)$.

%In our example, at each time step following \co{t=0}, supposing the infection to start in somehow, the number of susceptible individuals changes with a balance between the new infected and the possible recovered people that gain immunity, so holds:

\begin{equation}
\frac{dS}{dt} = \beta k I(t)[1 - R(t) – I(t)]
\end{equation}
\\ In the same way, the number of people in infected state changes with similar dynamics. The parameter $\gamma$ in the following equation, accounts for the number of people that gain immunity and recover from the illness in time as a rate of recover:
\begin{equation}
\frac{dI}{dt} = -\gamma I(t)+\beta k I(t)[1 - R(t) – I(t)]
\end{equation}
\\Always in the same way we can define the differential equation for the recover state as:
\begin{equation}
\frac{dR}{dt} = \gamma I(t)
\end{equation}
\\The three differential equation here, describe and predict the transition in time of individuals from the healthy state S to the infected I, and the recovered immune state R in a simple contract network in which each node is an individual.

%-----------------------------------------------------------------------%
%                          	   SIRB MODEL
%-----------------------------------------------------------------------%
\section{SIRB Model}
\label{sirb}
Based on the SIR model described in the previous section, the \textit{SIRB} model used in this thesis is built on the previous \citep{Bertuzzo2008,Bertuzzo2010,Bertuzzo2011} and most recent \citep{Rinaldo2014,Mari2015} spatially-explicit epidemiological models for cholera and proposed in \citet{Bertuzzo2012,Bertuzzo2014,Mari2015} for 2010 Haiti epidemic. 
In this mathematical framework, the simple contact and contract network of a SIR model made of individuals, is replaced by a network of human communities distributed in the area under study, in which each node represents a community; in this way, the model embeds explicitly the structure of the contract network that enhance the spread of pathogens by reason of human mobility and hydrological connectivity between these. In each community holds the homogeneous mixing hypothesis. 
\\The model is defined as SIRB due to the fact that transmission of cholera can happen, as explained in \vref{cholera}, via contact with the concentration of the bacteria in the environment. Therefore, together with the dynamic of susceptible, infected and recovered individuals, we need to understand more about the local environmental concentration \textit{B} of \textit{V. Cholera}. This kind of mathematical approach allows the user to simulate in each node the pathogenic concentration and its variation in time and space; it additionally allows to simulate the decrease of exposure to bacteria as consequence of intervention strategies and population awareness.
Moreover, by considering the characteristic of the \co{V. cholerae} induced disease, the set of differential equations distinguishes the \textit{symptomatic} and \textit{asymptomatic} individuals in the Infected and Recovered states. 

\par
\begin{figure}[!h]
\centering
\includegraphics[width=0.5\textwidth]{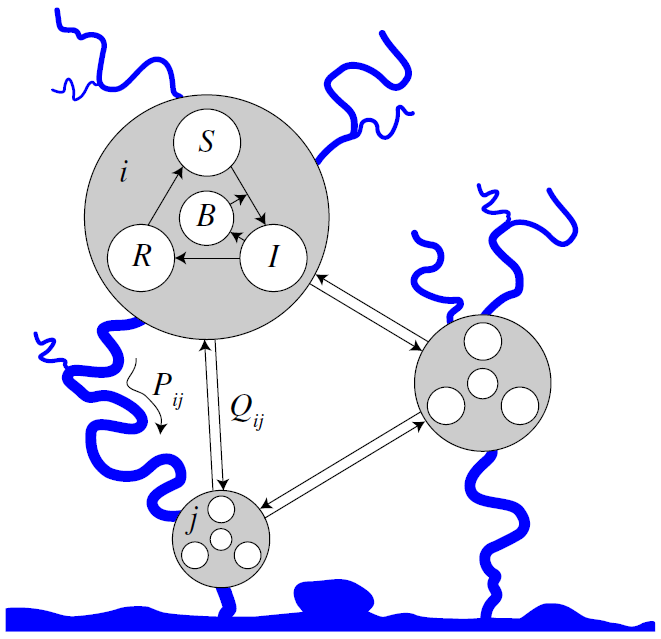}
\caption[Schematic representation of the SIRB model]{Schematic representation of the SIRB model \citep{Bertuzzo2014}. The number of Susceptible, Infected and Recovered, together with the concentration of Bacteria, change within the communities \textit{i}, here represented by gray circles. The epidemic can move to other communities \textit{j} via human mobility, with probability $Q_{ij}$, and via hydrological transport, with chance $P_{ij}$.}
\label{model}
\end{figure}

 Let $S_i(t)$, $I_i(t)$, $R_i(t)$ be the local amount of susceptible, symptomatic infected and recovered individuals and $B_i(t)$ the local concentration of the \textit{V. Cholera} at time \co{t} in each node \co{i} of the network. Cholera transmission dynamics can be described by the following set of coupled differential equation:

%----Susceptible----%
\begin{equation}
\frac{dS_i}{dt}=\mu(H_i-S_i)-F_i(t)S_i+\rho R_i
\label{S}
\end{equation}
%----Infected----%
\begin{equation}
\frac{dI_i}{dt}=\sigma F_i(t)S_i-(\gamma+\mu+\alpha)I_i 
\label{I}
\end{equation}
%----Recovered----%
\begin{equation}
\frac{dR_i}{dt}=(1-\sigma)F_i(t)S_i+\gamma I_i-(\rho+\mu)R_i 
\label{R}
\end{equation}
%----Bacteria---%
\begin{equation}
\frac{dB_i}{dt}=-\mu_B B_i+\frac{p}{W_i}\left[1+\lambda J_i(t)\right] I_i - l\left(B_i-\sum_{j=1}^n P_{ij} \frac{W_j}{W_i}B_j\right)
\label{B}
\end{equation}
In which the value
\begin{equation}
F_i(t)=\beta_i(t)\left[(1-m)\frac {B_i}{K+B_i}+m\sum_{j=1}^nQ_{ij}\frac{B_j}{K+B_j}\right]
\label{F}
\end{equation}
defines the force of the infection. Details on each equation and on the model follow (see Fig. \ref{model} for a schematic representation of the links among the different stages.)

%-------------------SUSCEPTIBLE--------------------%
\paragraph{Susceptible}

The dynamic of susceptible people is described by eq. \eqref{S}. The population size in each node \co{i} is defined by $H_i$ and it is assumed to be at demographic equilibrium. The human mortality rate is defined by the parameter $\mu$ expressed in $d^{-1}$, hence the product $\mu H_i$ is a constant recruitment rate. The parameter $\rho$, measured in $d^{-1}$, defines the rate at which recovered individuals lose their immunity and therefore are susceptible again. The force of infection $F_i(t)$ defines the rate at which susceptible individuals become infected due to ingestion of contaminated water. In this way, the local abundance of susceptible people changes according to basic dynamic of population, defined by the first term of the equation, and to the dynamic of the compartments $I_i$ and $R_i$, with modification of the total infected $F_i(t)$$S_i(t)$ and recovered people from symptomatic and non infected that lose immunity at rate $\rho$$R_i(t)$. 

Through the assumption of demographic equilibrium we can rewrite the equation \eqref{S} simply as:

\begin{equation}
S_i=H_i-I_i-R_i
\label{Scost}
\end{equation}
valid for each node $i$.

%-------------------Force of infection--------------------%
\paragraph{Force of the infection}

The force of infection, eq. \eqref{F}, contains the reason for the infection to pass between individuals and communities, i.e. human mobility and the probability of contact with contaminated water.
The fraction $B_i/(K+B_i)$ defines the probability that an individual in node \co{i} becomes infected due to exposure to a local concentration $B_i$ of \co{V. Cholera}. \co{K} is named \co{half-saturation} constant and expresses the concentration of \co{V. cholerae} in water that yields 50\% chance of catching cholera, assuming that the only route for infection is the ingestion of contaminated water from non-treated sources \citep{Codeco2001}. 

%----Mobility----%
Due to human mobility, susceptible individuals can move from their residing node \co{i} to a destination node \co{j}, where they can be exposed to the local concentration $B_j$ and yield $B_j/(K+B_j)$ probability of get infected in \co{j}. The chance for an individual to travel is given by the value of the parameter \co{m} assumed, in this formulation, to be node-independent. Therefore the term $(1-m)$ defines the chance to remain in the native location \co{i} and be exposed to the concentration $B_i$. 

\par The human mobility is modeled through a \co{gravity model} \citep{Erlander1990}. The \co{gravity model of migration} is a mathematical formulation in transportation derived from Newton's law of gravity and is used to predict the degree of interaction between two sites.  From Newton's third law of mechanics, we can think of bodies as ``locations", and masses as ``importance". In this context, the gravity model is used to compute the \co{connection probability} between nodes, defined as the probability that an individual resident in node \co{i} reaches \co{j} as destination, bringing the risk of infection with him. 

The connection probability $Q_{ij}$ is computed through the following formula :
\begin{equation}
Q_{ij}=\frac {H_j e^{\frac{-d_{ij}}{D}}}{\sum_{k\neq i}^n H_k e^{\frac{-d_{ik}}{D}}}
\label{mobility}
\end{equation} 
The ``importance", or the attractiveness of each location, is the population size $H_i$, while the deterrence factor, as in Newton's, depends on the distance $d_{ij}$ between the nodes. It is represented by an exponential kernel dependent on the shape factor D, measured in $km$.

The formulation of the gravity model states that the more the attractiveness of a place, in this case the population size, the higher would be the probability of choosing that as destination. This probability decreases when the distance among two nodes increases. The shape factor \co{D} controls the importance of distance as a deterrence factor: the higher this value, the less distance affect the connection probability.
	
    %----Exposure rate beta----%

The parameter $\beta_i(t)$ in eq.\eqref{F}, measured in $d^{-1}$, represents the maximum exposure rate, as to say the maximum frequency at which individuals are exposed to the local concentration $B_i$. This value can change in time and space, and specifically, it decreases with the increment of population awareness of cholera transmission risk factors and, with intervention strategies against the disease. We assume the maximum local exposure rate $\beta_i(t)$ to decrease proportionally to the local cumulative attack rate, defined as
\[
\frac{C_i}{H_i}
\]
where $C_i$ stands for the cumulative reported cases and of course depends on time. We describe the variation in time of $\beta_i(t)$ through the following exponential function:
\begin{equation}
\beta_{i}(t)=\beta_{0} e^{-\frac{C_i}{H_i \psi}}
\label{expbeta}
\end{equation}
in which $\beta_{0}$ is the value of the exposure at the beginning of the epidemic, measured in $d^{-1}$. The dimensionless parameter $\psi$ represents the rate at which the value of local exposure $\beta_i(t)$ decreases with awareness of population. The mathematical approach in eq.\eqref{expbeta} assumes that population awareness of cholera risk is higher in regions hit more severely by the epidemics. In this way, as shown for the Haiti epidemic \citep{DeRochars2011}, it is possible to imitate the health response targeted to the most-at-risk communities, as it was even in South Sudan \citep{Azman2016} as explained in Sect.\vref{colerass}. Moreover, in this way it is possible to account for directly possible changes in behavior of individuals, in response to information campaigns.

This formulation of the ``force of infection" represents the main engine in the epidemic dynamics for people to move from one state to another.

The three classes \textit{I,R} and \textit{B} are described as follows.
%-------------------INFECTED--------------------%
\paragraph{Infected}

The equation \eqref{I} describes dynamic of infected people. Symptomatic and asymptomatic individuals are discerned using the dimensionless parameter $\sigma$, representing the fraction of infected individuals that develop symptoms and entering in the symptomatic infected $I_i$ class. The fraction $\sigma$ depends on the dose of bacteria ingested. For simplicity, we assume this dose to be constant in space. Symptomatic infected people can recover from cholera at a rate $\gamma$, measured in $d^{-1}$, or die due to cholera, at rate $\alpha$, or due to other causes at rate $\mu$ as seen for susceptibles in eq. \eqref{S}. In case of death or recovery, individuals move from this class. The product $\sigma$$F_i(t)$$S_i$ defines the number of new symptomatic infected in time, while the second subtractive term removes from the class individuals that died or recovered. 

%-------------------RECOVERED--------------------%
\paragraph{Recovered}

Recovery state dynamics are described by equation \eqref{R}. The fraction of asymptomatic individuals $(1-\sigma)$ recovers much more rapidly from the disease, in around one day \citep{Nelson2009}, and their contribution to the \co{Vibrio cholerae} environmental concentration is lower. In fact, as \citep{Kaper1995.,Nelson2009} stated, they shed bacteria at a lower rate (1,000 times for asymptomatic against $>10^4$ for symptomatic) and the distribution of asymptomatic patients does not strongly affect the local quantity of V. cholerae that is shed for subsequent transmission. Therefore, it is not needed to consider them in the transmission process. However it is fundamental to account for them into the recovery class, as they develop temporary immunity that can contribute in the process of disrupting outbreaks. Furthermore, they act on the susceptible compartment, eq.\eqref{S}, loosing their immunity at a rate $\rho$ or dying at a rate $\mu$, as symptomatic infected do too.
The product $(1-\sigma)$$F_i(t)$$S_i$ counts in this class the number of recovered asymptomatic infected in time; the term $\gamma$$I_i(t)$ adds the number of new recovered individuals from symptomatic infection in time. The last term removes from this class the recovered individuals that died or lost their immunity.

%-------------------CONCENTRATION--------------------%
\paragraph{Environmental concentration}

Changes in time and space of the environmental concentration of \co{Vibrio cholerae} are defined by equation \eqref{B}. This concentration is measured in cells per $m^3$. This equation accounts for all the factors that affect the local amount of free-living vibrios in the water reservoir, as to say, population dynamics of bacteria, rainfall enhancement effect and hydrologic connectivity that disperses them. In this formulation we assume that there is not a local reservoir of bacteria and that the bacteria is not indigenous, hence the concentration is directly affected by the new-born infection.
Regarding population dynamics of bacteria, it is assumed that the death rate of \co{V. cholerae} in the environment exceeds the birth rate. This excess appears in eq.\eqref{B} as the net mortality rate $\mu_{\beta}$, expressed in $d^{-1}$, that reduces the local concentration $B_i$ in time. 

\paragraph{} 
Symptomatic infected individuals, as they are supposed to be non-mobile, contribute exclusively to the local abundance of bacteria at a rate defined as $p/W_i$, where \co{p} is the rate at which one person excretes bacteria, that reach and contaminate the local reservoir of volume $W_i$. $p$ is measured in cells per day per person. $W_i$ is measured in $m^3$. This volume is assumed to be proportional to the population size $H_i$, as to say $W_i$=c$H_i$ as in \citet{Rinaldo2014}. The constant value $c$ is the water consumption per capita, measured in $m^3$ per person. 
In addiction, rainfall induced runoff increases this local abundance due to washout of open-air defecation sites and to overflow of latrines \citep{Rinaldo2014}. The structure of the model accounts for these phenomena via the coefficient $\lambda$, measured in $d$ $mm^{-1}$. The contamination rate \co{p} is therefore increased by additive terms via $\lambda$ and the rainfall intensity $J_i(t)$ \citep{Righetto2011,Rinaldo2014}. 

The hydrologic dispersal parameter \co{l}, measured in $d^{-1}$, represents the rate at which the pathogen can travel from node \co{i} to node \co{j} with probability $P_{ij}$, decreasing the local abundance $B_i$. $P_{ij}$ takes unitary value in cases in which \co{j} is the unique downstream node between the neighborhood of \co{i}, and zero otherwise. The summation term
\begin{equation}
\sum_{j=1}^n P_{ij} \frac{W_j}{W_i}B_j
\end{equation}
is the hydrologic transportation, that, similarly to the mobility probability seen in the force of infection \eqref{F}, defines the probability that the bacteria move from the reservoir of volume $W_i$ in the node \co{i} having local concentration $B_i$, to the reservoir of volume $W_j$ in \textit{j}. 

%-----------------------------------------------------------------------%
%                          	   MODEL PARAMETERS
%-----------------------------------------------------------------------%
\section{Model parameters and calibration}
\label{calibration}
The just-described mathematical framework wants to simulate the behavior of the four state variables $S_i(t)$, $I_i(t)$, $R_i(t)$ and $B_i(t)$ in each node of the network in the most possible realistic way. 

Whatever, approaches of this kind require a lot of assumptions to reduce the natural complexity of environmental processes, and most of the times these are not even sufficient to allow to estimate via analytic methods the values of the variables \citep{Baratti,Vrugt2016}. Therefore, errors and uncertainties lie in all the process of modeling: these lie in measurements of the environmental process, in the mathematical approach chosen, in forcing and, as we are going to see, parameters. All models are characterized by \co{parameters}, as to say, numerical values bringing information about the environmental mechanisms under study. We call \co{calibration} the processes by which we define parameters values for an optimal performance of the model, granting the proposed mathematical structure to reproduce as best the observed real system. Parameters can assume a unique value during the time of simulations, or change during this period. Many research studies have been dedicated in estimating epidemiological and hydraulic parameters, thus some of the model parameters can be found in literature. However, most of the parameters are problem dependent and require calibration. 

Various calibration method exists: parameters can be evaluated through experimental methods, as for the estimation of "dispersion coefficients" in an aquifer \citep{sethi}, or using numerical methods, as the ones used in this thesis for epidemic modeling.

Table \vref{pars1} resumes the parameters in the SIRB model in order of appearance in the equations \eqref{S},\eqref{I},\eqref{R},\eqref{B},\eqref{F},\eqref{mobility},\eqref{expbeta} and recalls their meaning, together with method of evaluation required.

%-------------------Table parameter--------------------%

\begin{table}[htbp]
\centering 
\caption{Parameters of the model}
\label{pars1}
\begin{tabularx}{\textwidth}{c c X c}\hline
\multicolumn{1}{c}{\textbf{Parameters}} & \textbf{Units} & \textbf{Description} & \textbf{Evaluation}\\
\hline
$\mu$ & $d^{-1}$ & Human mortality rate  & Literature\\
$\rho$ & $d^{-1}$ & Loss of immunity rate  & Calibration\\
$\beta_{0}$ & $d^{-1}$ & Exposure rate at the beginning of the epidemic  & Calibration\\
$\psi$ & $-$ & Rate of decrease of the exposure $\beta_{i}(t)$  & Calibration\\
\co{m} & $-$ & Probability of travel, node independent  & Calibration\\
\co{D} & $km$ & Shape factor of the exponential kernel in gravity model  & Calibration\\
\co{K} & $d^{-1}$ & Half saturation constant & Calibration\\
$\sigma$ & $-$ & Fraction of symptomatic infected  & Calibration\\
$\gamma$ & $d^{-1}$ & Recovery rate  & Literature\\
$\alpha$ & $d^{-1}$ & Human mortality rate due to cholera  & Literature\\
$\mu_{\beta}$ & $d^{-1}$ & Bacterial net mortality rate & Calibration\\
\co{p/$W_i$} & $d^{-1}$ & Rate of excreting bacteria per each symptomatic infected, contaminating the water volume $W_i$  & Calibration\\
\co{l} & $d^{-1}$ & Hydrological dispersion & Calibration\\
$\lambda$ & \textit{d} $mm^{-1}$ & Rainfall enhancement effect & Calibration\\
\hline\end{tabularx}
\end{table}

As the table \vref{pars1} shows, most of the parameters should be calibrated. The human mortality rate $\mu$, can be easily found in statistics of the country and census. The recovery rate $\gamma$ and the human mortality rate due to cholera $\alpha$, can be evaluated using epidemiological records. 
\par We can reduce the number of parameter to estimate by introducing the \co{dimensionless bacterial concentration} 
\[
B_i^{*}=\frac{B_i}{K}
\]
where $K$, we recall, is the half-saturation constant from eq.\eqref{F}. This new quantity allows to group three model parameters, as to say:
\begin{itemize}
\item \co{p}, the rate of excreting bacteria by one infected individual;
\item \co{c}, the volume per capita of water reservoir $W_i$  in which the bacteria sheds;
\item \co{K}, the half saturation constant;
\end{itemize} 
in a unique ratio defined as
\begin{equation}
\theta=\frac{p}{cK}
\label{teta}
\end{equation}
The ratio $\theta$ has an interesting meaning: it resumes all the parameters related to water, contamination and sanitation: an higher value of $\theta$, means that the rate of excreting bacteria is higher, and its consequential water pollution worse. We can use this parameter to understand the sanitation conditions and the resulting contamination of the environment.

Moreover, to simplify the implementation of the algorithm, we define $k$ as the reciprocal of the parameter $\psi$:
\begin{equation}
k=\frac{1}{\psi}
\label{kappa}
\end{equation}

in which $\psi$, we recall from eq.\eqref{expbeta}, is the rate of decrease of the local exposure $\beta_i(t)$ due to population awareness. 

\begin{table}[htbp]
\centering 
\caption{Reduced parameters of the model}
\label{pars2}
\begin{tabularx}{\textwidth}{c c X c}\hline
\multicolumn{1}{c}{\textbf{Parameters}} & \textbf{Units} & \textbf{Description} & \textbf{Evaluation}\\
\hline
$\mu$ & $d^{-1}$ & Human mortality rate  & Literature\\
$\rho$ & $d^{-1}$ & Loss of immunity rate  & Calibration\\
$\beta_{0}$ & $d^{-1}$ & Exposure rate at the beginning of the epidemic  & Calibration\\
\co{k} & $-$ & Decrease of the exposure $\beta_{i}(t)$  & Calibration\\
\co{m} & $-$ & Probability of travel, node independent  & Calibration\\
\co{D} & $km$ & Shape factor of the exponential kernel  & Calibration\\
$\theta$ & $d^{-1}$ & Sanitation conditions  & Calibration\\
$\sigma$ & $-$ & Fraction of symptomatic infected  & Calibration\\
$\gamma$ & $d^{-1}$ & Recovery rate  & Literature\\
$\alpha$ & $d^{-1}$ & Human mortality rate due to cholera  & Literature\\
$\mu_{\beta}$ & $d^{-1}$ & Bacteria mortality rate & Calibration\\
\co{l} & $d^{-1}$ & Hydrological dispersion & Calibration\\
$\lambda$ & \textit{d} $mm^{-1}$ & Rainfall enhancement effect & Calibration\\
\hline\end{tabularx}
\end{table}
Table \vref{pars2} shows the final reduced set of parameters; ten parameters have to be calibrated. 
\\ \par The evaluation of the parameters via numerical methods can be both \co{deterministic} or \co{probabilistic}. To define parameters in a deterministic way implies to seek and find unique values for parameters that fit well the observation data. These values represent a ``local optimum" in the domains of definition, meaning that can exist other parameter combinations with an equivalent fit. Moreover, due to the intrinsic and epistemic errors that characterizes hydrological models and the observation data, a unique solution is not representative of these uncertainties and there is no reason to believe that only one set of parameters is true \citep{oncalibuncertain}.

Therefore, a probabilistic approach should be preferred, where all the possible parameters combinations and values in the domains that retrieve a satisfactory fit are considered. This probabilistic approach aims at the estimation of the posterior probability density function (\co{pdf}) of the parameters, that helps in optimizing the performance of the model.
\\ \par In order to get information about these pdfs, \co{statistical inference} is used. This means that we deduce properties of an underlying distribution by exploiting the available observation data \citep{statinf}, the epidemiological model, and the prior distribution of the parameters. Therefore, it is necessary to start from some hypothesis on the pdfs and test them with data.

There are two theory in statistics for inference \citep{Ewens2005}: \co{classical} or \co{frequentist} methodology and the \co{Bayesian} approach. We will briefly give some details on the Bayesian approach in next section in order to understand the calibration method that has been used.

%-----------------------------------------------------------------------%
%                          	   Bayesan approach
%-----------------------------------------------------------------------%
\subsection{The Bayesian approach}
\label{bayesian}
The objective of the Bayesian approach is to compute the probability density function of the parameters conditioned to the available information, i.e., the observed data \citep{Ewens2005}. This is called \textit{posterior} pdf. Bayes rule rewrites the posterior pdf as the normalized product of the likelihood function (measuring the fit of the model with the data) and the starting hypothetical parameter pdf named as \co{prior}.
\\ \par We can clarify this idea thanks to an example \citep{ross}. Let us take the context of a criminal investigation. There is the 60\% of chance \co{a priori} that a certain suspect is guilty. Acquiring a \co{new} piece of evidence, the inspector in charge gets to know that the criminal has a certain characteristic \co{C}, that is also shared with the 20\% of the population. How does the probability that the suspect is guilty change with this new information? Or in other terms, which is the \co{posterior} probability that the suspect is guilty given the \textit{new} piece of evidence?

Let $G$ define the event that the suspect is guilty. At the beginning, the prior probability of the event $P(G)$ is 0.6. Therefore, $G^{c}$ is the event that the suspect is innocent, whose prior probability is $P(G^c)=0.4$. Let $C$ be the event that he possesses the characteristic of the criminal. 

We define the probability of posses the characteristic of the criminal, being guilty, as the conditional probability $P(C \mid G)$, and it has the unitary value, since we know thanks to the new information, that the real guilty has this characteristic. 
We define the probability of posses the characteristic of the criminal, being innocent, as the conditional probability $P(C \mid G^c)$, equal to the percentage of the population that has this characteristic and that is not suspected.

We can define the \co{posterior} probability $P(G \mid C)$ by using Bayes theorem:

\begin{equation}
\label{bayes}
P(A \mid B) = \frac{P(A)P(B \mid A)}{P(B)} 
\end{equation}

that applied to our example gives:
\[
\label{example}
\begin{split}
P(G \mid C) = \frac{P(GC)}{P(C)} = \frac{P(G)P(C \mid G)}{P(G)P(C \mid G)+P(G^c)P(C \mid G^c)} 
        \\= \frac{(0.6)(1)}{(0.6)(1)+(0.4)(0.2)} = 0.8824
\end{split}
\]

The example shows that, using the Bayesian theorem in eq.\eqref{bayes}, the chance for the suspect to be guilty increases from 0.6 to 0.88 thanks to the acquisition of new information.
\\ \par Similarly to the example, we can apply the Bayesian theorem for the posterior pdfs of the parameter of the model. 
Let $\Theta$ be the set of parameters of the model, in our case:
\begin{center}
$\Theta = \{\theta, l, m, D, \lambda, \beta_0, \mu_{\beta}, \rho, \sigma, k\}$
\end{center}
We are interested in finding the posterior distribution of the parameters $P(\Theta \mid Y_{obs})$ to optimize the model. The variable $Y_{obs}$ is the vector of the system observations which are related to the model outputs and, thus, to the parameters. In our case, $Y_{obs}$ is a matrix that contains the recorded suspected cholera cases in time and space. 
Similarly to the example, using the Bayes theorem eq. \eqref{bayes} on conditional probabilities, we can define the posterior as:

\begin{equation}
\label{postformul}
P(\Theta \mid Y_{obs}) \propto P(\Theta) \mathscr{L}(\Theta \mid Y_{obs})
\end{equation}
in which $P(\Theta)$ is the prior distribution of the set $\Theta$. The term $\mathscr{L}(\Theta \mid Y_{obs})$ is called \textit{Likelihood function}. The likelihood is defined as the probability of the observed outcomes dependently by the values of the parameters, so that:
\begin{center}
$\mathscr{L}(\Theta \mid Y_{obs})$ = $P(Y_{obs} \mid \Theta)$
\end{center}
It is important to highlight that the likelihood does not express a distribution of probability yet a probability, dependent on the parameters values. In other words, it gives a measure of the goodness of both the calibration method and the model. Thanks to the likelihood function in Bayesan formalism, the posterior distribution of the parameters of the model can be derived by conditioning the spatio-temporal behavior of the model on measurements of the observed system response \citep{Vrugt2016}.

Probabilistic approaches as the Bayesian one, are preferred because of their ability to handle parameter, state variable and model output uncertainties \citep{Vrugt2016}.
However, the solution of Bayesian approaches is not an easy task and in most of the cases an analytical solution does not exist. This is the reason why numerical methods are wide-spreading. Most methods are based on Monte Carlo methods and Markov Chain processes, as \textit{Iterative Simulations} \citep{Gelman1992}, the \textit{Differential Evolution Markov Chain} DE-MC \citep{terbraak} and the \textit{DiffeRential Evolution Adaptive Metropolis} - DREAM method \citep{Vrugt2012}. These methods, once assumed the shape of the prior and the likelihood function \citep{oncalibuncertain,Vrugt2016}, sample possible parameter realizations from the posterior distribution by simulating the response of the model and optimizing the likelihood function in the Bayesian problem. 

During the first tries and analysis, the DREAM method was used for this work to explore the parameter space and to evaluate the posterior distribution adopting the $DREAM_{ZS}$ variant of the algorithm, as in \citep{Bertuzzo2014} (for further details \citep{Vrugt2016}). In our test case, this approach did not succeeded to accurately fit the dynamics of this case study.

Another typology of calibration methods are based on \textit{Data Assimilation} (\textit{DA}). The main idea of this technique is to include real system observations into the model, in order to correct immediately its response and quantifying simulation uncertainties \citep{Damianotesi}. DA technique can be extended to infer the distribution of the model parameters in a dynamical way. This is useful when the parameter distribution can change in time: DA sequentially, i.e. at each time step, track the parameters using the collected data \citep{pasetto2016}. The technique of the \textit{Ensemble Kalman Filter}, here used to calibrate the spatially explicit SIRB model, belongs to this class. Section \vref{enskf} goes into details of this approach as applied to our cholera model.
%-----------------------------------------------------------------------%
%                          	   ENSAMBLE KF
%-----------------------------------------------------------------------%
\subsection{Ensemble Kalman Filter}
\label{enskf}
The Ensemble Kalman Filter, or \textit{EnKF} was first proposed by \citet{evensen} as an improvement of the so-called Kalman Filter, or \textit{KF}, \citep{Kalman1960}, with the aim of applying the algorithm to non-linear systems and to reduce the computational costs. The Kalman Filter (\textit{KF}) method, in fact, solves the Bayesian problem and finds the posterior pdfs of the state variables in linear problems. Assuming all the pdfs to be Gaussian distributed, KF analytically computes the mean values and the variances of the state variables, since the method provides formulas for advancing them at each time step, when real observations are incorporated into the model. One possibility to overcome the linearity assumption consists in estimating the mean and the variances linearizing the equations and computing the Jacobian matrix at each time step, as performed in the Extended Kalman Filter (\textit{EKF}). It is though evident that the cost of the computation of the Jacobian is high.

EnKF solves the Bayesian problem using a Monte Carlo method, a simple computational algorithm that uses repeated random sampling to approximate the probability distribution of the variables. For the advantages that it brings, EnKF finds large use in hydrological modeling (see e.g., \citep{Camporese2009}) and epidemic modeling (see e.g., \citep{enskftb}).

In the following, we explain how EnKF applies to our cholera model.

%-------------- Data assimilation method--------------%
Let us define $\mathbf{x}_k$, resuming the output of our model at each time step $k$ \citep{pasetto2016}, where $k$ is the epidemiological week (from Sunday to Saturday), as:

\[
\mathbf{x}_k = \{(I_i, R_i, B_i, C_i), ~ with ~ i = 1, \dots,n \}
\]
in which $I_i$ is the number of new infected in each node $i$, $R_i$ the number of recovered, $B_i$ bacteria concentration per $m^{3}$ and $C_i$ the cumulative number of weekly infected in the node. 

The mathematical framework that gives as output $\mathbf{x}_k$ can be written as:

\begin{equation}
\label{spacestate}
\mathbf{x}_k = \mathcal{F} \left(\mathbf{x}_{k-1}, \mathbf{\Theta}, J(t), t_{k-1}, t_{k}\right)
\end{equation}
in which $\mathbf{x}_k$ $\in$ $\mathbb{R}^{4n}$, being $n$ the number of nodes considered; $J(t)$ is the daily rainfall, input of the model and $\mathcal{F}$ is the non-linear operator that solves the system of equations \eqref{S} - \eqref{B} from time $t_{k-1}$ to $t_k$, i.e. from one week to another. 

We can describe the relationship between the weekly recorded cases $\mathbf{y}_k$ $\in$ $\mathbb{R}^{m}$, aggregated in a number of domains $m$, and the state system $\mathbf{x}_k$ using the following equation:

\begin{equation}
\label{observed}
\mathbf{y}_k = \mathbf{H}_k\mathbf{x}_k + \mathbf{\xi}_k
\end{equation}
where $\mathbf{H}_k$ is the so-called \textit{observation function} relating the weekly real observed cases to the weekly simulated ones. The role of the observation function is to upscale the results of the simulation in each node $i=1,\dots,n~$ to the spatial level at which the recorded cases have been aggregated, therefore $\mathbf{H}_k$ $\in$ $\mathbb{R}^{m x n}$. The value $\mathbf{H}_k\mathbf{x}_k$ and $\mathbf{y}_k$ would coincide whether the simulations and the observations were errors-free. The vector $\mathbf{\xi}_k$ $\in$ $\mathbb{R}^{m}$ represents the measurements error, whose components are modeled as independent Gaussian random variables with mean equal to zero and standard deviation $\sigma_{\xi}$, i.e.: 

\begin{equation}
\label{errordistr}
\mathbf{\xi}_{k} ~\sim \mathscr{N} \left(0, R_k\right), ~ i=1,\dots,n
\end{equation}
being $R$ the covariance matrix of the measurements errors, that due to the hypothesis on independent errors, is a  diagonal matrix filled with $\sigma^{2}_{\xi,k}$.

At each new observation acquisition, using the Bayes formula eq.\eqref{bayes}, we can rewrite the posterior distribution $p^{a}$ of the state as:

\begin{equation}
\label{analysis}
p^{a}\left(\mathbf{x}_{k} \mid \mathbf{y}_{1},\dots,\mathbf{y}_{k}\right) = c ~ p^{f}\left(\mathbf{x}_{k}, \mid \mathbf{y}_{1},\dots,\mathbf{y}_{k-1}\right) \mathscr{L}\left( \mathbf{y}_{k} \mid \mathbf{x}_{k}\right)
\end{equation}
similarly to what we have stated for the parameters in equation \eqref{postformul}. This step is define as \textit{analysis step} or \textit{update}. $p^{f}$ represents the forecast pdf at time $t_k$, that was computed using the previous assimilation $p^{a} \left(\mathbf{x}_{k-1}, \mathbf{\Theta}_{k-1} \mid \mathbf{y}_{1},\dots,\mathbf{y}_{k-1}\right)$ at which were acquired the data $\mathbf{y}_{k-1}$. In other words, $p^{f}$ represents the response of the model at each time step before the new acquired observation $\mathbf{y}_{k}$ at $t_k$ changes the pdf to $p^{a}$.

Using the EnKF method, $p^{f}$ at time $k$ for the states $\mathbf{x}_{k}$ is approximated by the ensemble of $N$ random samples taken from the initial distribution, hence:

\[
p^{f}_k(x) ~\sim \left[ \mathtt{x}^{f}_1, \dots, \mathtt{x}^{f}_N \right]
\]
and the update of the states happens for each realization $j$ $\in$ $N$ using the following equation:

\begin{equation}
\mathbf{x}^{a,j}_{k} = \mathbf{x}^{f,j}_{k} + \mathbf{K}^{f,j}_k\left(\mathbf{y}^{j}_{k} - \mathbf{H}_k \mathbf{x}^{f,j}_{k}\right)
\label{enkfstates}
\end{equation}
in which the operator $\mathbf{K}^{f,j}_k$ is called "Kalman Gain" and contains the observation function $H_k$, the covariance matrix of the measurements error $R_k$ and the covariance matrix $P_k$ of the ensemble (for further details see e.g. \citep{Kalman1960,Mandel2009}):

\[
\mathbf{K}^{f}_k = P_k ~ H^{T}_k ~ \left(H_k ~ P_k ~ H^{T}_k + R_k\right)^{-1}
\]

The vector $\mathbf{y}^{j}_{k}$, using \eqref{observed}, $\mathbf{y}^{j}_{k} = \mathbf{y}_{k} + \mathbf{\xi}^{j}_{k}$  represents the random perturbations of the observed measurements $\mathbf{y}_{k}$, introduced to correctly estimate the variance of the updated variables. 

Whether we are interested in computing the update of the parameters $\Theta$, together with the states, the method allows to treat them as state variables, and therefore to change their forecast pdfs. In this case the update is computed using the "state augmentation" as: 

\begin{equation}
{\mathbf{x}^{a,j}_{k} \choose \mathbf{\Theta}^{a,j}_{k}} = {\mathbf{x}^{f,j}_{k} \choose \mathbf{\Theta}^{f,j}_{k}} + \mathbf{K}^{f,j}_k \left(\mathbf{y}^{j}_{k} - \mathbf{Hx}^{f,j}_{k}\right)
\label{enkfaug}
\end{equation}
similarly to what we have stated through eq. \eqref{enkfstates}.

\begin{figure}[h]
\centering
\includegraphics[width=1\textwidth]{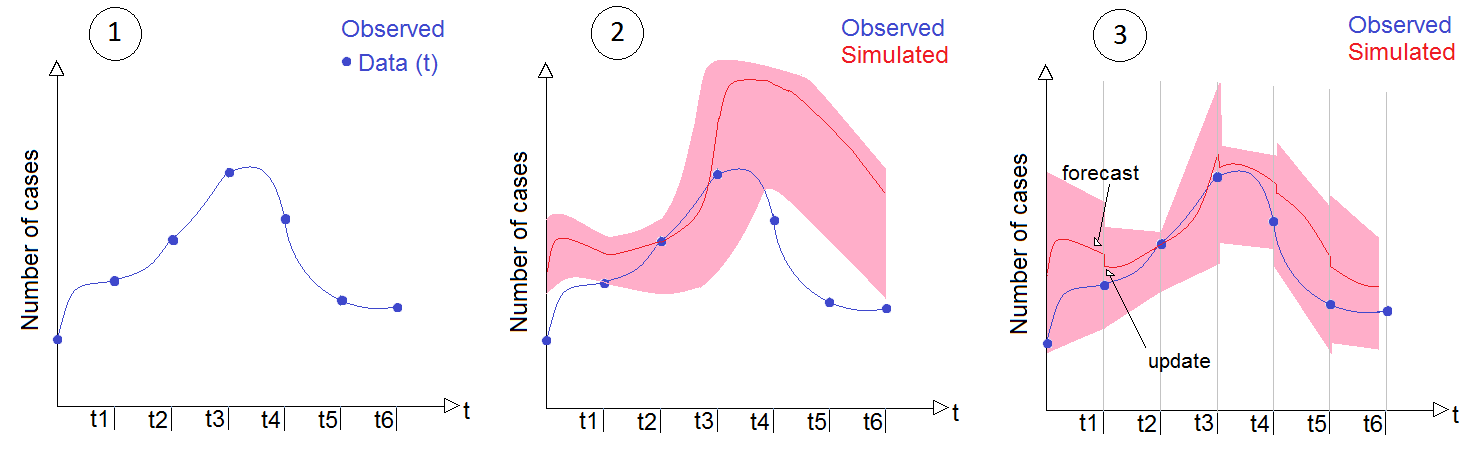}
\caption[Schematic representation of DA technique]{Schematic representation of the DA technique. For more details see Sect.\ref{enskf}.}
\label{dafig}
\end{figure}

This process can be schematically seen through Fig. \ref{dafig}. On step (1), an environmental process as a cholera epidemic, blue line, and its measurements, dots, are represented. On step (2), the mathematical model tries to mimic the real phenomena. Uncertainties, in pink, define a confidence interval for the model results, that with time becomes larger, and the mean value of the ensemble, in red, detaches from the real mechanism, leading to wrong evaluations. On step (3), DA techniques, as the Kalman Filter and EnKF, at each time step correct the model evaluations, updating state variables forecast. Uncertainties reduce in time, and the mean value approaches the observed phenomenon. 

\par As \citet{pasetto2016} reported, one disadvantage of EnKF is the so-called filter ``inbreeding": parameter distribution rapidly converges toward one value, underestimating uncertainties. In this work, we have tried to overcome this problem due to the high uncertainty associated with epidemiological records, whose variance is unknown. As we have seen, the analysis step is obtained by the Kalman gain of the ensemble, containing information on errors variance and correlations between the observations and the simulated states. We control convergence using the Kalman gain: we gradually increase the measurement error variance $\sigma^{a}_{\xi}$ until the parameter variances $\sigma_{\Theta^{a}_\xi}$ are higher than a desired tolerance. On each repetition $i$, we set a value $c^{i}_1>1$ to increase the measurement error. The update is accepted if $\sigma_{\Theta^{a}_\xi}>c_2\sigma_{\Theta^{f}_\xi}$, as to say the ratio between the parameter variances forecast and the parameter variances of the analysis step, is greater than a value $c^{i}_2$ defined between zero and one.

In this way we can control how the parameter variances decrease in time, avoiding the rapid reduction of the probability space explored by the ensemble. Moreover, we avoid drastic change in parameters values in one analysis step.

%-----------------------------------------------------------------------%
%                          	CHAPTER 4: MODEL SET UP
%-----------------------------------------------------------------------%
\chapter{Model setup and epidemiological data}
\label{chap4}

The modeling approach described in Chapter 3, together with DA calibration technique, need in order to be performed, specific inputs.

The epidemiological records, containing daily records of suspected cholera infected individuals, are required to perform the Data Assimilation, accumulating cholera cases during each week. Further details on the temporal and spatial behavior of the disease given in the records were useful to define the network in which the model performs. As explained earlier, the nodes of the network are representative of distributed communities and, therefore, information about the population distribution and size are required to define them. These requirements are the same as the gravity model of mobility, in which, we recall, the attractiveness of the nodes is represented by their population size. 
Moreover, rainfall  measurements during the period of analysis in the area of study are required to perform SIRB.
Last, but not least, initial conditions have to be chosen in order to solve the state variables $I_i(t)$,$R_i(t)$, $B_i(t)$, $C_i(t)$ in each community.

Data were processed creating appropriate MATLAB scripts and using and creating spatial references on GIS.

Further details follow.
%-----------------------------------------------------------------------%
%                          	Epidemiological records
%-----------------------------------------------------------------------%
\section{Epidemiological records}
Epidemiological records of both years 2014 and 2015 were provided by the SSMoH. Records contain data regarding the date of onset of the symptoms, the date of arrival at the health facility and the provenience of the patients. 
All the cases were reordered and timed using the date of onset of the symptoms. Georeference of the cases at the level of the Payam was possible using the information on provenience, and overlapping data of the shapefiles, provided by the SSMoH, SSNBS and the WHO, and other spatial information obtained by \citet{openstreetmap}. This step was the most delicate one.
%-----------------------------------------------------------------------%
%                          Framework
%-----------------------------------------------------------------------%
\section{Framework}
\label{framework}
Analyzing the hydrography (Fig. \ref{hydrography}), the topography (Fig. \ref{elevation}) and the spatial distribution of the disease during the two years, we decided not to consider hydrological connectivity between communities (which is present in SIRB through the parameter $l$) since there is not a clear hydrographic connection between the nodes that recorded cholera. Therefore, in order to define the spatial domain for the simulations, we used the distribution of recorded suspected cases. In both years, as we showed in Sect.\vref{colerass}, the right part of the country has been affected by the disease. 
The number of cases and the places in which the infection appeared are the reasons why we delineate the model within the borders of the states Central Equatoria, Eastern Equatoria, Jonglei and Upper Nile (for geo-political reference please see figs.\vref{administration}). As anticipated, we chose to neglect the three cases that appeared in the state Western Equatoria. From a mathematical point of view, the structure of the model does not allows for zeroes infected in the nodes, and a certain response, even if small, is always expected allover the network. Therefore, to include these cases and the node in which these were recorded in our domain is a completely arbitrary choice, that would not strongly affect the output, yet it would have increased the computational cost due to an higher number of nodes. 
\\ \par
\begin{figure}[htp]
\centering
\includegraphics[width=0.8\textwidth]{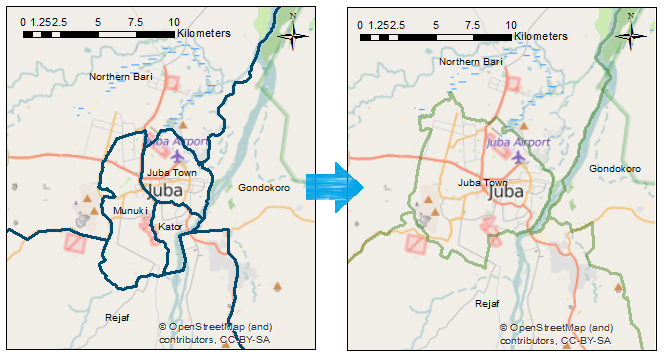}
\caption[Modifications for the Payam of Juba]{Modifications on the geo-political assessment of Juba to optimize the model. Names of the Payams are given.}
\label{Jubapayam}
\end{figure}

 To define the communities  distribution, we used the administrative division of the country. The population in each node consists in the inhabitants of the \textit{Payams}. To optimize the performance of the model, we modified the spatial information regarding this geo-political division. The mathematical structure, in fact, gains stability when the spaces in which we define the communities have homogeneous extensions. This means that we should prefer division in areas having, more or less, the same size. As Fig.\vref{administration} shows, payams sizes are very irregular and for most inhabited zones, the number of divisions is higher and irregularity greater. Especially in the capital area, Juba, this geo-political structure creates confusion and, to the purpose of this work, is not necessary. 
Based on \citep{JubaAdmin2005, openstreetmap} and the information provided by the Souther Sudanese government, we decided to merge together in a unique Payam the small payams named Muniki and Kator, located in less than $10 km$ around the city center of Juba and that can be considered as big districts of the capital. The same kind of incorporation has been done for Malakal, whose administration is actually subdivided in Central, Eastern, Northern and Southern Malakal. Fig.\vref{Jubapayam} shows the modifications for the case of Juba. We limited our modifications to most populated areas to not upset the administrative division. 

\newpage
In the domain of this work, a total of $236$ Payams, hence nodes, are considered. Nodes position is computed using the population distribution, as it will be described in sect. \ref{sectcentroids}. The model simulates the epidemics in all the nodes. The calibration, that will be discussed in the following sections, runs at the spatial level of the Counties in which cholera cases were recorded. In total, 10 Counties were severely affected by the disease between the two years: Juba, Yei, Kajo-Keji, Torit, Magwi, Ikotos, Lopa, Kapoeta North, Bor South and Malakal. The output of the model will be shown, for comparison with the real observed cases, in these areas. 

\begin{figure}[h!]
\centering
\includegraphics[width=0.9\textwidth]{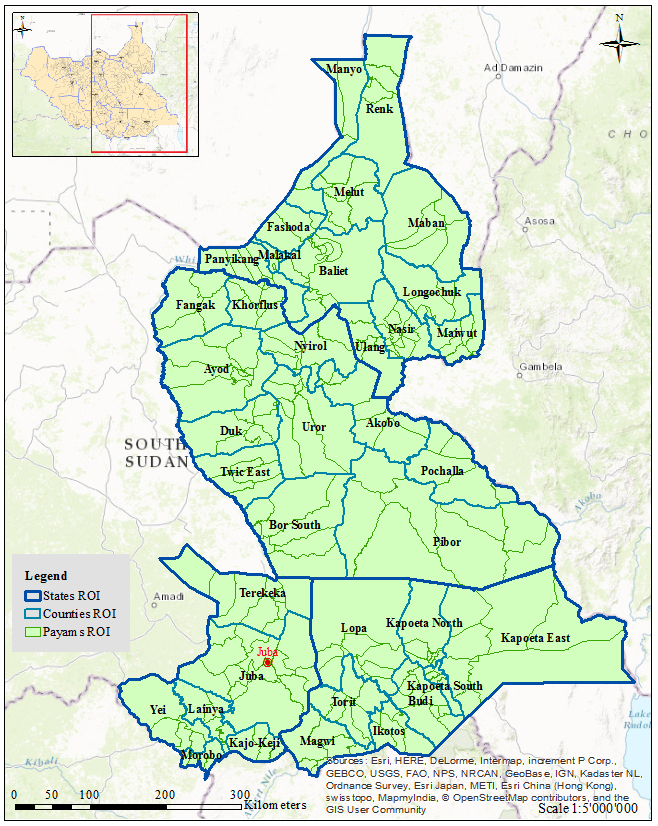}
\caption[Spatial domain of the model]{Map of the spatial domain of the model. We defined a Region Of Interest - \textit{ROI} to run the simulations. Each payam, whose borders are define by thin green lines, consists in a node in our network. Results will be shown at the level of the Counties, borders in light blue lines, whose names are reported in the map. State division is highlighted in dark blue.}
\label{modeldomain}
\end{figure}

Fig. \vref{modeldomain} redefines the Region Of Interest - \textit{ROI} with which we worked.

%-----------------------------------------------------------------------%
%                          	Centroids
%-----------------------------------------------------------------------%
\section{Centroids}
\label{sectcentroids}
The position of the nodes in the spatial domain is computed using the population distribution in each payam. We define as 'centroid' the point in which falls the center of the area of the Payam weighted on the population distribution in the same area. The concept is similar to the one of the "center of mass", where the masses are replaced by the number of inhabitant in each cell of the image.

For the population distribution we used the map provided by \citet{afripop}, in which each pixel value represents the number of people per grid, estimated using remotely sensed map. The raster image has spatial resolution equal to $0.000833333$ decimal degrees, approximately $100 m$ at the equator. The evaluation is based on land covers and remote sensing. Chosen between the four maps available, the version that we propose estimates the number of people per grid square, with national totals adjusted to match UN population division estimates and it is updated at 2013 (for further details http://esa.un.org/wpp/ and reference). 

\begin{figure}[htp]
\centering
\includegraphics[width=0.9\textwidth]{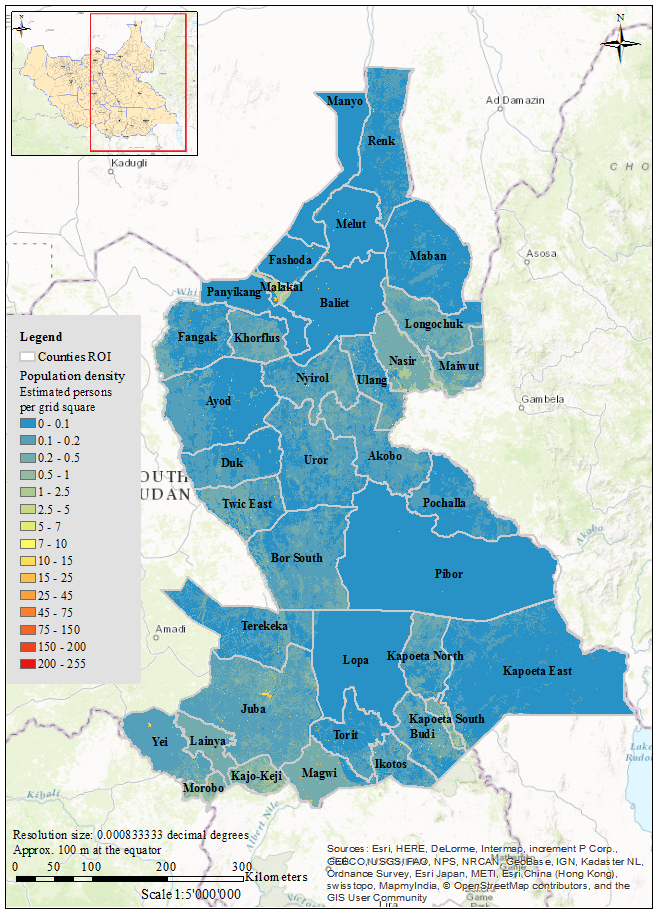}
\caption[Population distribution from WordPop]{Estimated numbers of people per grid square, with national totals adjusted to match UN population division estimates. Cell size $0.000833333$ decimal degrees, approximately $100 m$ at the equator. We highlight the administrative division at the level of Counties. Source \citet{afripop}.}
\label{poporiginal}
\end{figure}

\begin{figure}[htp]
\centering
\includegraphics[width=0.9\textwidth]{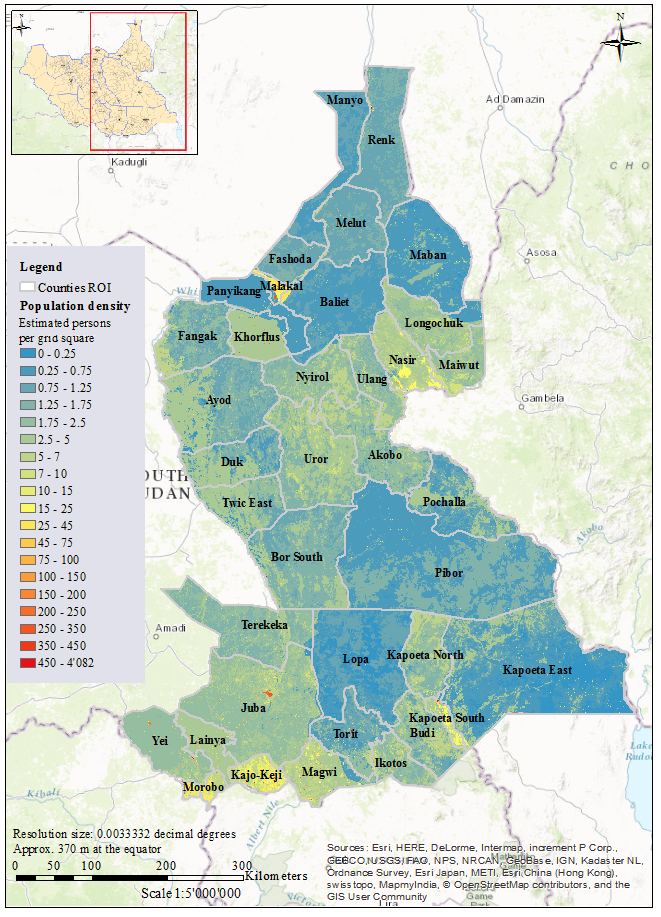}
\caption[Population distribution, re-sized raster]{Estimated numbers of people per grid square, re-sized to compute the centroids. Cell size $0.0033332$ decimal degrees, approximately $370 m$ at the equator. We highlight the administrative division at the level of Counties. The map shows that the country in the domain of study has really low population density, exceptions are the big cities as Juba, and the County Malakal.}
\label{modelpop}
\end{figure}

In order to reduce the computational cost related to the number of cell treated in the algorithm, we re-sized the original raster seen in Fig. \ref{poporiginal}, enlarging the cell size. Each cell of the new raster contains the information of 16 cells of the original one. The new raster for the population distribution has spatial resolution equal to $0.0033332$ decimal degree, approximately $370 m$ at the equator (Fig. \ref{modelpop}).

First, the algorithm aims to calculate the total population inside the areas of a single Payam. From the shapefiles provided, we create a raster image of the Payam division having resolution $0.0033332$ decimal degree. Each Payam is identified by an "ID number" from 1 to 263: these values fill the cells of the raster, locating the position of the payams.

\begin{figure}[htp]
\centering
\includegraphics[width=0.9\textwidth]{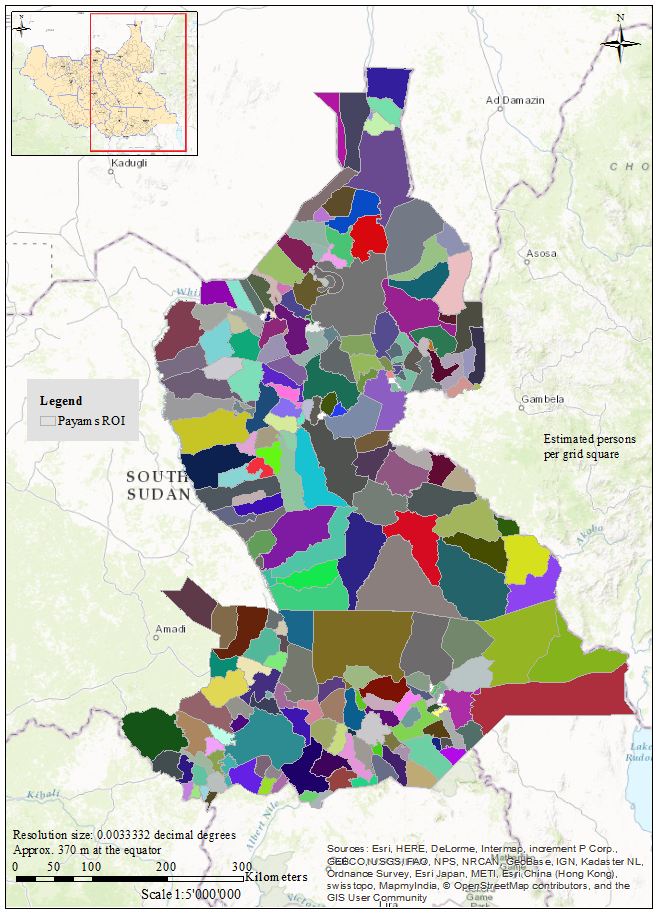}
\caption[Payam raster image for centroid computation]{Raster image of the division in Payam, created to compute the centroids. Cell size $0.0033332$ decimal degrees, approximately $370 m$ at the equator. We highlight the administrative division at the level of Payam: each one is represented using a different color.}
\label{modelpay}
\end{figure}

\begin{figure}[h!]
\centering
\includegraphics[width=0.9\textwidth]{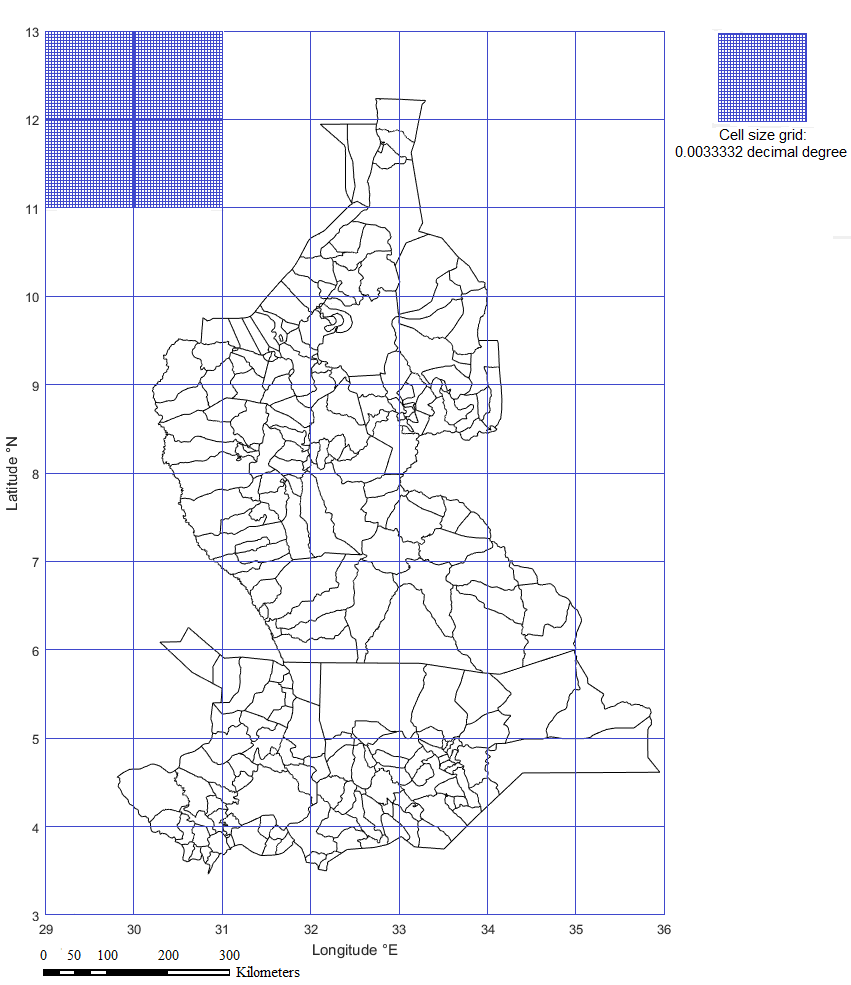}
\caption[Grid to compute centroids]{Grid in the algorithm that scans raster images of the Payams and of the population distribution to compute the centroids.}
\label{grid}
\end{figure}

\newpage
Both raster images are delimited in longitude and latitude $(\ang{29} - \ang{36} E; \ang{3} - \ang{12})$, having the same number of cells. The computation scans all the pixels located in this grid, from left to right and from the bottom to the top, comparing the ID numbers with values in the cells of the Payam raster image. When one ID number is found in the image, the position of the cell is stored and used in the Population raster image, to evaluate the total population in the area. Fig. \vref{grid} shows the grid through which the raster images are scanned.

\begin{figure}[h!]
\centering
\includegraphics[width=0.9\textwidth]{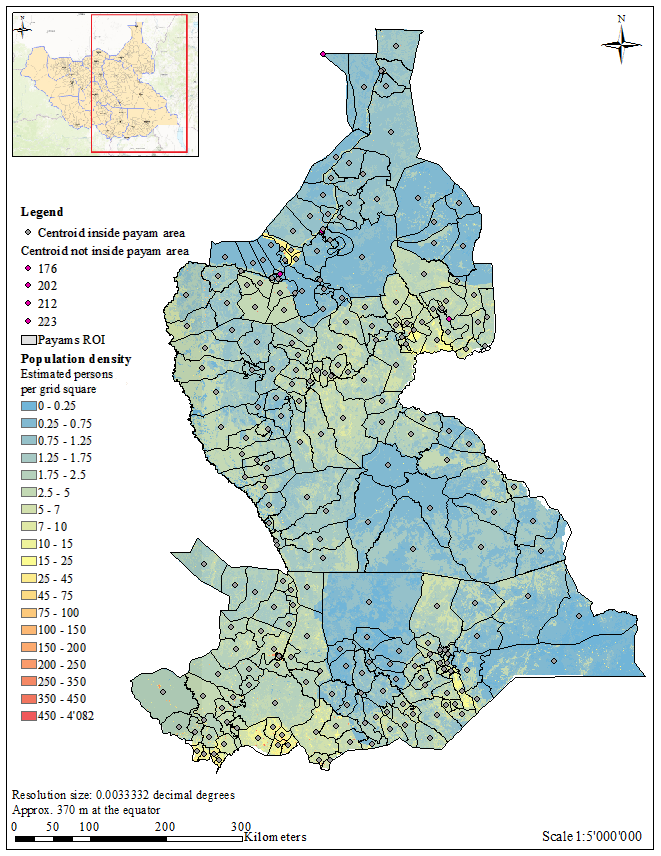}
\caption[Position of the Payams' centroids]{Position of the nodes weighted on the population distribution. 'Not inside' points are highlighted in purple.}
\label{centroid}
\end{figure}

%\begin{figure}[h!]
%\centering
%\includegraphics[width=0.5\textwidth]{modelmap/scan_centroid}
%\caption[Scan algorithm]{Procedure of scanning the images to compute the centroids.}
%\label{scan}
%\end{figure}
The centroid of each area is the result of the weighted position of each cell size in the grid considered. Due to the irregularity of the shape of the payams, in some cases the centroid did not fall into its belonging area. We forced these points to stay into the borders of the Payam in analysis, using a build-in function of MATLAB. Fig.\vref{centroid} shows the results of this algorithm, i.e. the position of the nodes in latitude and longitude coordinates.

\newpage

%The pseudo-code of this algorithm follows.
%\input{codes/centroid}

%-----------------------------------------------------------------------%
%                          	Population
%-----------------------------------------------------------------------%
\section{Population}
The map offered by \citet{afripop} is a useful source of information for population distribution. Although computing the total estimated number of people in each payam we realized that these values do not match the numbers of the 2008 last census \citep{census2008tables} and the rate of population growth provided by \citet{CIA2015}. In fact, as the website reports, population maps are updated to newest versions when improved census or other input data become available. 
Due to the unstable political conditions discussed in Chapter 2, new certain information are not available.  Reasonable population values for each payam, therefore each node, were found in official projections provided by \citep{payam_projectionSSNBS,county_projectionSSNBS}. 

These projections match the growth rate stated by the \citet{CIA2015}: comparing the data of the 2008 census with the projections for the year 2015, the average growth rate computed as:
\[
\frac{H_i(2015)-H_i(2008)}{H_i(2008)}
\]
results approximately $+0.3\%$, that approaches the values of the annual growth rate of 4\% in 7 years. These values appear in the model as $H_i$ in each node $i$.

%-----------------------------------------------------------------------%
%                          	Distance
%-----------------------------------------------------------------------%
\newpage
\section{Distances}
In order to derive the mobility of individuals around the network using the gravity model (eq. \eqref{mobility}), we compute distances $d_{ij}$ between the centroids using the \textit{haversine formula} (for further information see \citet{haversinebook}). The equation has importance in navigation, as it is useful to compute distances between two points on a sphere using longitudes and latitudes. 
For any two points on a sphere, the haversine of the central angle between them is given by:
\begin{equation}
\label{haversine}
hav \left(\frac{d}{R}\right) = hav (\varphi_2 - \varphi_1) + \cos(\varphi_1)\cos(\varphi_2)hav(\lambda_2-\lambda_1)
\end{equation}
in which $hav$ is the \textit{haversine function} defined as:
\[
hav(\theta) = \sin^{2} \left(\frac{\theta}{2}\right) = \frac{1-\cos\theta}{2}
\]
where $\theta$ is any angle and
\begin{itemize}
\item \textit{d} is the distance between the two points along a great circle of the sphere, i.e. d is the spherical distance in which we are interested;
\item \textit{R} is the radius of the sphere;
\item $\varphi_1$, $\varphi_2$ are the latitude values of the points expressed in radians;
\item $\lambda_1$, $\lambda_2$ are the longitude values of the points expressed in radians.
\end{itemize}

Solving for d, equation \eqref{haversine} becomes:
\begin{equation}
\label{disthaversine}
d = 2R \arcsin \sqrt[]{\sin^{2} \left(\frac{\varphi_2-\varphi_1}{2}\right) + \cos(\varphi_1)\cos(\varphi_2) \sin^{2} \left(\frac{\lambda_2-\lambda_1}{2}\right)}
\end{equation}
that can be easily be compacted \eqref{compactedhaversine} through the $\arctan2$ with two elements and defining the value $a$:
\[
a = \sin^{2} \left(\frac{\varphi_2-\varphi_1}{2}\right) + \cos(\varphi_1)\cos(\varphi_2) \sin^{2} \left(\frac{\lambda_2-\lambda_1}{2}\right)
\]
\begin{equation}
\label{compactedhaversine}
d = 2R \arctan2 \left(\sqrt{a},\sqrt{1-a}\right)
\end{equation}

The formula applies to our model using as points on the sphere the position of the nodes in our network and their coordinates. $R$ is the radius of the earth, whose value used is $R = 6371 km$. 

%We give a pseudo-code for the algorithm used \vref{codedistance}

%\input{codes/distance}
%-----------------------------------------------------------------------%
%                          Rainfalls
%-----------------------------------------------------------------------%
\section{Rainfall data}
Rainfall data were downloaded from the Climate Data Library \citep{rain}. Information provided regard the daily estimated precipitation, expressed in $mm$ $d^{-1}$, evaluated by the \textit{NOAA} - National Oceanic and Atmospheric Administration - in the African continent. In order to analyze both years, we took the data during the period January 1, 2014 - December 31, 2015 within the range $(\ang{29.5} - \ang{36} E; \ang{3} - \ang{12.5} N)$. The spatial resolution of the data is \ang{0.1}N, approximately $10$ km at the equator. The format in which data are provided is a table containing a vector of the time $T$, time step of a day, and a matrix $X$ x $Y$, where $X$ is the longitude and $Y$ is the latitude, in which compare the values of daily precipitation in that coordinates.

%We get the data using a MATLAB code, whose pseudo - code follows \vref{coderainurl}

\begin{figure}[h!]
\centering
\includegraphics[width=0.9\textwidth]{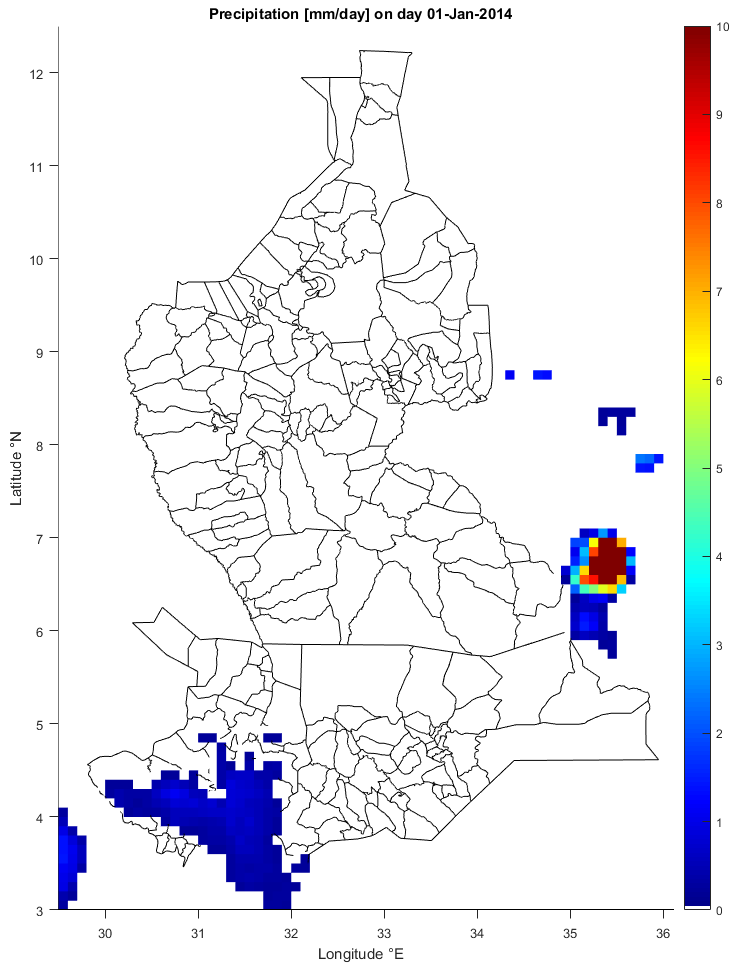}
\caption[Rain analysis on January $1^{st}$, 2014]{Rainfall estimation during the first day of analysis in our spatial domain, January $1^{st}$, 2014}
\label{rain1}
\end{figure}

%\input{codes/rainurl}
Fig. \ref{rain1} shows the rainfall estimation during the first day of analysis in our region of interest, data processed using MATLAB 2015.

\begin{figure}[h!]
\centering
\includegraphics[width=0.9\textwidth]{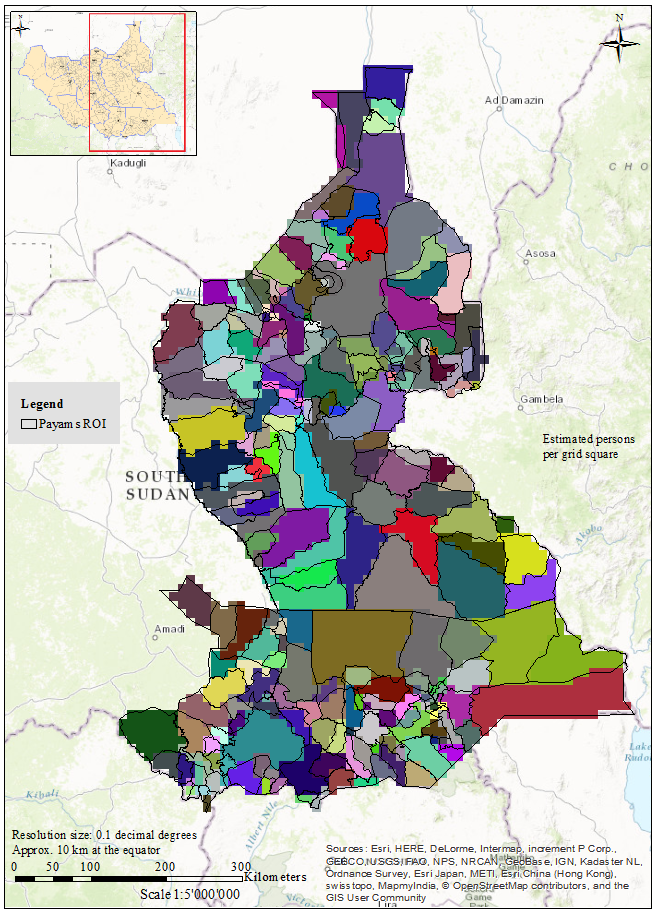}
\caption[Payam raster image for rainfall analysis]{Raster image of the division in Payam, created to compute the mean values of rain in each community. Cell size $0.1$ decimal degrees, approximately $10$ $km$ at the equator. We highlight the administrative division at the level of Payam. The color bar is the same used in fig.\vref{modelpay} for comparison.}
\label{payrain}
\end{figure}

To each single node of the network we assign the mean daily rainfall fallen in the area of the Payam, evaluating the position in each cell similarly to what was conducted to compute the total population in the area. In order to solve this algorithm we create a raster image of the payam administration having the same spatial resolution of the rainfall data, i.e. $0.1$ decimal degree. Fig. \vref{payrain} shows the new raster image for payams representation.

\begin{figure}[h!]
\centering
\includegraphics[width=0.7\textwidth]{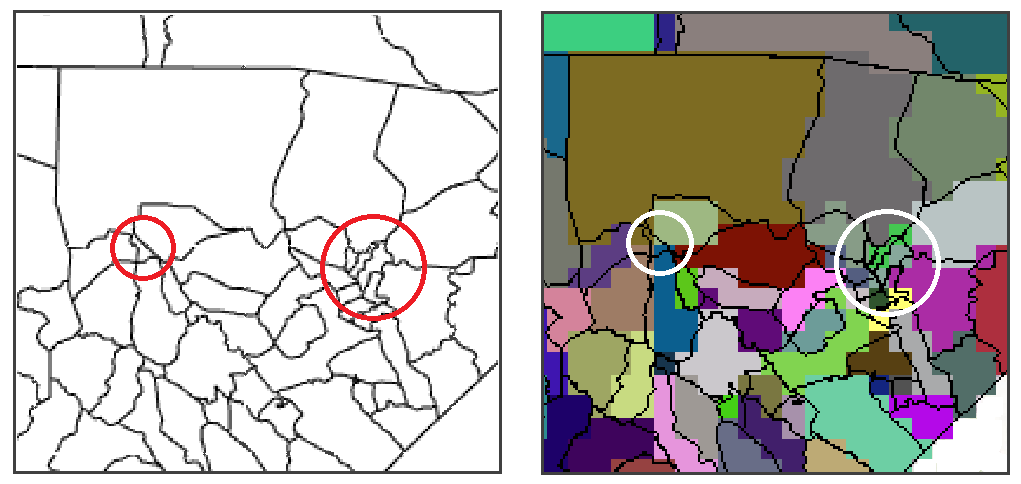}
\caption[Detail on the comparison between Payam vector and raster representation]{Comparison between the Payam vector image and its raster image, focusing on a specific area}
\label{detailcell}
\end{figure}

As the map shows and as one can notice comparing this map with the previous raster image in Fig. \ref{modelpay}, to increase the cell size compromises the resolution of the image, and borders of the areas are less detailed. Moreover, it is possible that the cell size is bigger than the dimension of the payams and that, in these cases, these are included in nearby bigger payams. Fig. \ref{detailcell} focuses in a specific area of the vector, on the left, and raster image, on the right, of the Payam representation. Circles highlight areas characterized by various small regions. Not all of these regions are represented in the raster image on the right, yet are included in other regions. 

To avoid wrong evaluation, we compare the values of the pixels in the raster image with the ID numbers. Whether some ID did not appear in the raster, the mean daily rain in the area could not be computed. In this case we assign to this payam, hence node, the daily rainfall data interpolated in the coordinates of the corresponding centroid. For this step, we used the 2D interpolation function available in MATLAB.
%
%The pseudo code that follows will give a better idea of the algorithm used.
%\input{codes/raincentr}
%-----------------------------------------------------------------------%
%                          Adapting the model
%-----------------------------------------------------------------------%
\section{Adapting the model}
The analysis on the context and on the epidemiological records let us develop an appropriate mathematical approach for this particular case study.

For the reason explained in Sect.\vref{sirb}, we consider the population in each node to be at demographic equilibrium. Therefore the equation for the susceptible class in each node $i$ is eq.\eqref{Scost}.

Since we are neglecting the hydrological transport of the bacteria, the parameter $l$ that stands for this phenomenon will be set to the constant value zero in all the simulations. The relative eq.\eqref{B} for the environmental concentration of bacteria in this approach becomes:

\begin{equation}
\frac{dB_i}{dt}=-\mu_B B_i+\frac{p}{W_i}\left[1+\lambda J_i(t)\right] I_i
\label{Bmodel}
\end{equation}

We recall that, the value $p/W_i$ is replaced by the parameter on water and sanitation $\theta$ (see Sect.\vref{calibration}).

Therefore, calibration is required for the following nine parameters, 
\[
\Theta = \{\theta, m, D, \lambda, \beta_0, \mu_{\beta}, \rho, \sigma, k\}.
\]
while $l$ is kept to constant null value.

The parameters $\mu$, $\gamma$ and $\alpha$, respectively human mortality, rate of recovery and mortality due to cholera, can be estimated from literature (see Sect. \vref{calibration}). Table \ref{fixedpars} shows the assigned values. The mortality rate is taken from the statistic provided by \citet{CIA2015}. For the recovery rate, we use the value as in Haiti model \citep{Rinaldo2014}, value that is comparable with the duration of the symptoms arise from the epidemiological records. For what concerns the mortality rate due to cholera, this is computed using the epidemiological records of 2014 as follows:

\[
\alpha = \frac{Recorded~~ deaths~~ 2014}{Tot~ cases~ 2014~ *~ 2014~ epidemic~ duration~ period}
\]

\begin{table}[h t b p]
\centering 
\caption{Fixed parameters}
\label{fixedpars}
\begin{tabular}{c c l l}\hline
\multicolumn{1}{c}{\textbf{Parameters}} & \textbf{Units} & \textbf{Value} & \textbf{Evaluation}\\
\hline
$\mu$ & $d^{-1}$ & $2.25*10^{-05}$ & \cite{CIA2015}\\
$\gamma$ & $d^{-1}$ &  0.2 & \cite{Rinaldo2014}\\
$\alpha$ & $d^{-1}$ & $1.145*10^{-04}$  & From epidemiological data 2014\\
\hline\end{tabular}
\end{table}

We decide to use the value $\alpha$ computed from 2014 into 2015 simulations, since the magnitude of this number for 2015 epidemic is the same.
%-----------------------------------------------------------------------%
%                          Calibration set-up
%-----------------------------------------------------------------------%~
\section{Calibration set-up}
\label{setupcal}
The requirements of the calibration are the size of the ensemble $N$ and the values of the constants $c_1$ and $c_2$ to control the inbreeding problem, as seen in Sect.\vref{enskf}.
Based on preliminary sensitivity analyses, we have seen that stable results are obtained with $N ~ = 3000$, $c_1 ~ = 5$, and $c_2 ~ = 0.8$. 

Outliers,i.e. updated values of the parameters outside the prior, are forced to their prior boundaries using a 'reflection technique', that projects, as a mirror on the borders, these values within the boundaries.   

The time step of the simulations is weekly. We use the "epidemiological week", whose starting day is Sunday. Weekly output are therefore given on Saturdays. This works for both observed and simulated state variables.

For each realization during DA, the update of the state variables is accepted whether the following condition is accepted:

\[
S_i = H_i-I_i-R_i ~>0
\]
as to say, it is not possible to have more susceptible than the population in the $i-th$ node. 

As anticipated in Sect.\ref{framework}, the calibration runs at the spatial level of 10 Counties. This implies that the observation function $\mathbf{H}$ in eq.\eqref{observed}, transfers the cumulative weekly cases $C_i$ simulated in each node $i$, i.e in each Payam, to the Counties, in which we define the observation points. Therefore in our case, the dimension of $\mathbf{H}$, the vector of weekly recorded cases $\mathbf{y}_k$, and the vector of the measurements error $\mathbf{\xi}_k$ are $\mathbf{H}$ $\in$ $\mathbb{R}^{10x236}$, $\mathbf{y}_k$ $\in$ $\mathbb{R}^{10}$, $\mathbf{\xi}_k$ $\in$ $\mathbb{R}^{10}$.

The goodness of each realization is computed using the Root Mean Squared Errors (RMSE). Per each realization $j$ the errors at each time $k$ are defined as:

\begin{equation}
Err = \mathbf{Hx}^{j}_k ~ - ~ \mathbf{y}^{j}_k
\end{equation}

Therefore the RMSE is:

\begin{equation}
RMSE = \sqrt[]{\frac{\sum_{j=1}^{N} (\mathbf{Hx}^{j}_k - \mathbf{y}^{j}_k)^2}{N}}
\end{equation}
that can be mediated both in time and on number of realization. To compare simulations, we will use a single value RMSE, mediated first on realizations, and in time afterwards.
%-----------------------------------------------------------------------%
%                          Initial conditions
%-----------------------------------------------------------------------%
\section{Initial conditions and time of the simulations}
In order to perform simulations and calibration, initial conditions - \textit{IC} and time of the simulations have to be set for both the two years of investigation. We apply the initial conditions to the first day of simulation $(t=0)$, that changes in the two years. The susceptible at time $t=0$ are given by the equilibrium between the population, the initial symptomatic infected and the recovered:
\[
S_(i) = H_i-I_i(0)-R_i(0)
\]
where 
\begin{center}
$R_i(0) = (1-\sigma) I_i(0)$
\end{center}
We assume $B_i(0)$ to be in equilibrium with the local number of infected individuals, therefore:
\[
B_i^{*}(0)=\frac{\theta I_i(0)}{H_i\mu_{\beta}}
\]

Details on the initial conditions for both years follows.

\begin{description}
\item{\textbf{2014 epidemic}} The 2014 epidemic started on April 23, 2014 and last until October 29, 2014. We assign as initial conditions the cumulated recorded cases until May 17, 2014. During this month, cases were recorded in the capital Juba and in four adjacent Payams. All these nodes belong to the County of Juba. Simulation last until the last day of the epidemic. 
we apply the initial conditions to the first day that is simulated, as to say May 18,2015. The simulation starts from a total number of infected equal to 168.
\item{\textbf{2015 epidemic}} The 2015 epidemic started on May 18, 2015 and last until September 23, 2015. We assign as initial conditions the cumulated recorded cases until June 20, 2015. As 2014 epidemic, during 2015 one first cases were recorded in the capital Juba and in nearest payams, so that all nodes belong to Juba County. We apply the initial conditions to the first day simulated, June 21, 2015. The simulation starts from a total number of infected equal to 172.
Simulation last until the last day of the epidemic.
\end{description}

Table \vref{ic} resumes what just stated.
\begin{table}[htbp]
\centering 
\caption{Initial conditions and times}
\label{ic}
\begin{tabularx}{\textwidth}{lX X X cX cX cX}\hline
\multicolumn{1}{c}{\textbf{Year}} & \textbf{IC Period}        	 & \textbf{Simulated} 			& \textbf{County IC} & \textbf{Nodes IC}  & \textbf{Cum. Cases IC}\\
\hline
\textbf{2014}  & $23/04/2014$~-~$17/05/2014$   & $18/05/2014$~-~$29/10/2014$	& 1; Juba County	 & 4; 7; 9; 235; 236  & 169\\
\hline
\textbf{2015}  & $18/05/2015$~-~$20/06/2015$   & $20/06/2015$~-~$26/09/2015$	& 1; Juba County	 & 7; 9; 235; 236     & 172\\
\hline\end{tabularx}
\end{table}

%-----------------------------------------------------------------------%
%                         PSEUDO CODE
%-----------------------------------------------------------------------%
\newpage
\section{Pseudo-code}
We would like to give an idea on how EnKF and SIRB model work together via the simple pseudo-code that follows.

\begin{lstlisting}[frame=trBL][\caption={Pseudo - code for EnKF + DA and SIRB},label=codemodel][firstnumber=1]
%
% Pseudo Code for EnKF and SIRB model. 
% Given for year 2014, updating both states and parameters
% We show the initialization of the model and the first step of the DA method, 
% avoiding loops for the time in which the simulation runs and the Number of the
% realization used.
%

% ================================ INPUT DATA =============================== %
% load the input files
J = load rain;					% rainfall data in each node, mean or interpolation
Pop = load population;			% population size in each node
Obs = load weekly cases;		% recorded cases in each node
Centr = load centroid;			% position of the centroids and ID number
dij = load distances;			% distances between nodes
H = load upscale matrix; 		% matrix to upscale from payams to counties: 
								  objective function
nodes = 236;					% number of nodes

Input = [J, Pop, Obs, Centr, dij, H, nodes];

% ================================== TIMING ================================= %
% definition of the time of the simulation
start_time = 18/05/2014;
end_time = 29/10/2014;

% definition of the time of the model in day
time = start_time :1: end_time;
% definition of the time of the model in weeks
week time = start_time: 7 :end_time;

% ================================ INITIALIZE =============================== %
% definition of the parameters: l is null l=0;
pars = [theta, m, D, lambda, beta0, muB, rho, sigma, K];
% for all the parameters: Domain in which the parameter can change
p_range = [low_bound : up_bound];  

% Initializing the EnKF
N = 3000; 					   % size of the ensemble
c1 = 5;						   % constant to increase the measurements errors
c2 = 0.8;					   % constant to accept the update

% Initialize the parameter ensemble: 
% definition of the random sampling in the ranges for each realization
pars = rand(p_range, Nre);

% definition of the IC for year 2014
I0 = cumsum(Obs(23/04/2014:17/05/2014));		% Infected at time t=0
R0 = (1-sigma)/(sigma*I0);						% Recovered at time t=0
S0 = H0 - I0 - R0;								% Susceptible at time t=0
B0 = (theta*I0)/(H0*muB);						% Concentration at time t=0
nodesIC = find(I0 ~= 0);						% nodes of IC
nodesIC = H*nodesIC;							% upscale to counties
												  using obj function
IC = [I0,R0,S0,B0,nodesIC];

% ============================ STATE EVALUATION ============================= %
% simulation to forecast the states at the k-th week and solve the SIRB model
a = H*exp(-dij/D);		  %mobility nominator
% term for mobility denominator djk
% need to remove the zero distance on the diagonal
b = a*(1-eye(263));		  
Qij = a/sum(b);			  %mobility probability from node i to node j

% exposure rate
beta = beta0*exp(-cumsum(Obs(1:=t))

%------------------------------- SIRB FUNCTION ------------------------------%
% function containing the differential equations of the SIRB to solve 
% t contains the time, y the actual state of the system 
% from which we solve dy per each day k+1
function dy=sirb(t,y,pars);
	% Susceptibles at k+1
    S = H - I - R;
	% force of infection
    F = ((B*(1-m))/(1+B)+m)*S*beta;
    % solve for the infected dy(1) = I(k+1)
    dy(1) = (1-sigma)*F - (gamma+mu+rho)*I;
    % solve for the recovered dy(2) = R(k+1)
    dy(2) = (1-sigma)*F + gamma*I - (mu+rho)*R;
    % solve for the concentration dy(3) = B(k+1)
    dy(3) = -muB*B + (theta/H)*(1+lambda*R)*((1-m)*(Qij)*I);
    % solve for the cumcases dy(4) = cumcases(k+1)
    dy(4) = sigma*F;
%-----------------------------------------------------------------------------%  
% solve the differential equations at daily time step. states contain the daily model 
% response in each node: as to say I,R,B,cumulative I for each realization Nre
[t,States] = ode45(@sirb, time, IC, pars, options);

% evaluate the weekly cases using a function @dafun that repeats the SIRB weekly
%the value sim_cases is what we defined as Hx, cumulative number of infected in the week
[sim_cases, y] = feval(@dafun,pars,Inputs,states);

% ============================= DA + EnKF ============================= %
% we want to update both the states and the parameters of the system 
% using the state augmentation
% We control the inbreeding, 10 times trial to update the states

trial = 0;
for i=1:10
	trial = trial + 1;
	old_pars = pars;
	old_states = y;

	Prior = [old_pars,old_states];

	% variance of the prior needed to control the inbreeding
	var_prior = var (Prior);

	% define a value for the simulated cases standard deviation
	% variation coefficient
	CV = 0.1;
	sim_std = c1*CV;

	% formula for the kalman gain K = P*H^T*(H*P*H^T + R)^-1
	% we decompose the value P, covariance matrix of the ensemble as 1/N-1*(X*X^T)
	% in which X contains the deviation from the mean of the states
	% and define as Y the product H*X
	mean_prior = mean(Prior);
	% deviation from the mean values
	X = Prior - mean_prior;

	% The value Y*Y^T=(H*X*X^T*H^T)/N-1 is simply the covariance matrix of the response of the model
	mean_sim_cases = mean(sim_cases);
	% deviation from the mean per each obs
	dev_sim = sim_cases - mean_sim_cases;
	YYT = 1/(N-1) * dev_sim*dev_sim';
	% add the value R, diagonal matrix of the errors, that completes the term (H*P*H^T + R)
	YYT_R = YYT + diag (sim_std.^2);

	% complete the terms of the kalman gain
	XY = 1/(N-1) * X * dev_sim;
	K= YYT_R \ XY;
	
	% compute normally distributed error, supposed to be independent, associated to the simulations
	pert = sim_cases + randn;
	% store the generated added error that we associate to the measurements s.t. err = (yk - Hxk)
	err= pert - sim_cases;
	
	% Update proposed
	[Update_prop] = Prior+K*err;  
    
    % compute the variance of the update to control the inbreeding 
    var_up = var (Update);
    
    ratio = sqrt (var_up/var_prior);
    
	if ratio < c2
    	if trial < 10
        	Update = Prior;
		else
        	Update = Update_prop;
        end
     end
end

% back to the steps of the model and the data assimilation

Update(1) = States;
Update(2) = pars;

[t,States] = ode45(@sirb, time, States, pars, options);

\end{lstlisting}

%-----------------------------------------------------------------------%
%                          	CHAPTER 5: RESULTS
%-----------------------------------------------------------------------%
\chapter{Results and discussion}
The input data and the set-up seen in the previous chapter are used to run calibration and simulations. Our aim would be to find constant values in time for the parameters, adequate to perform the cholera model in both years. Therefore, we start in our analysis from the first year of the epidemic, the most critical, seeking for these values. This allows to understand whether the dynamics of the infection were similar among the two years. 

As mentioned before, Sect.\ref{calibration}, during first attempts the DREAM method was used to evaluate the parameters. With the aim of reducing the discussion, we do not show these results as they were not satisfactory. We show the results obtained using DA technique, EnKF. In order to understand the response of the model, we started to perform simulations using the same prior pdfs of the Haiti model \citep{Bertuzzo2014,Mari2012}. These let us understood that appropriate ranges for this case study were required.

For all the simulations, we show the simulated epidemiological curve compared to the weekly observations, the cumulative simulated and the observed cases, the simulated curves in all the ten counties of our interest and the changes in the parameter evaluation. %In the figures regarding the parameters values, we refer to these using their Greek phonetic names, hence:

%\[\theta = theta;~\ \lambda = ~lambda;~\ \beta_0 = beta;~\ \mu_\beta = muB;~\ \rho = rho;~\ \sigma = sigma;~\ k = kappa\]

%-----------------------------------------------------------------------%
%                          	   Prior HAITI
%-----------------------------------------------------------------------%
\newpage
\section{Prior definition}

\subsection{First attempts for year 2014}
The first simulations that we performed, starting from 2014, had as prior pdfs the ones used to calibrate the model for the Haiti cholera events, shown in table \vref{priorhaiti} \citep{Bertuzzo2014}. The hydrological dispersion $l$ is set to zero, as we described earlier.
\begin{table}[htbp]
\centering 
\caption{Range of the parameter prior distribution from Haiti case study}
\label{priorhaiti}
\begin{tabular}{c c l}\hline
\multicolumn{1}{c}{\textbf{Parameters}} & \textbf{Units} & \textbf{Prior}\\
\hline
$\theta$ & $d^{-1}$ 	 		 & [ 0.01     -      10 ] \\
\co{l} & $d^{-1}$ 				 & [ 0        -       0 ] \\
\co{m} & $-$	    			 & [ 0.0001   -       1 ] \\
\co{D} & $km$ 					 & [ 1        -     300 ] \\
$\lambda$ & \textit{d} $mm^{-1}$ & [ 0.01     -      10 ] \\
$\beta_{0}$ & $d^{-1}$	 	 	 & [ 0.01     -      10 ] \\
$\mu_{\beta}$ & $d^{-1}$		 & [ 0.01     -       1 ] \\
$\rho$ & $d^{-1}$ 				 & [ 0.0005   -    0.02 ] \\
$\sigma$ & $-$ 					 & [ 0.03     -     0.2 ] \\
\co{k} & $d^{-1}$				 & [ 1        -    1000 ] \\
\hline\end{tabular}
\end{table}
The parameter $\theta$, containing all the information related to water, contamination and sanitation (eq. \eqref{teta}), represents the rate for shedding excreta and water contamination rate, and it can vary between $0.1$ days and $100$ days. The node-independent probability of travel $m$, changes between the unitary value and zero. The range definition for shape factor $D$ of the gravity model, is based on analysis of the distances considered as reachable in the Haitian territory. $\lambda$, that defines the aggravating effect of the rain, has a large domain, as the initial exposure rate to bacteria $\beta_0$ and as the other parameters values showed in the table.  

\begin{figure}[!h]
\centering
\includegraphics[width=1\textwidth,height=0.4\textheight]{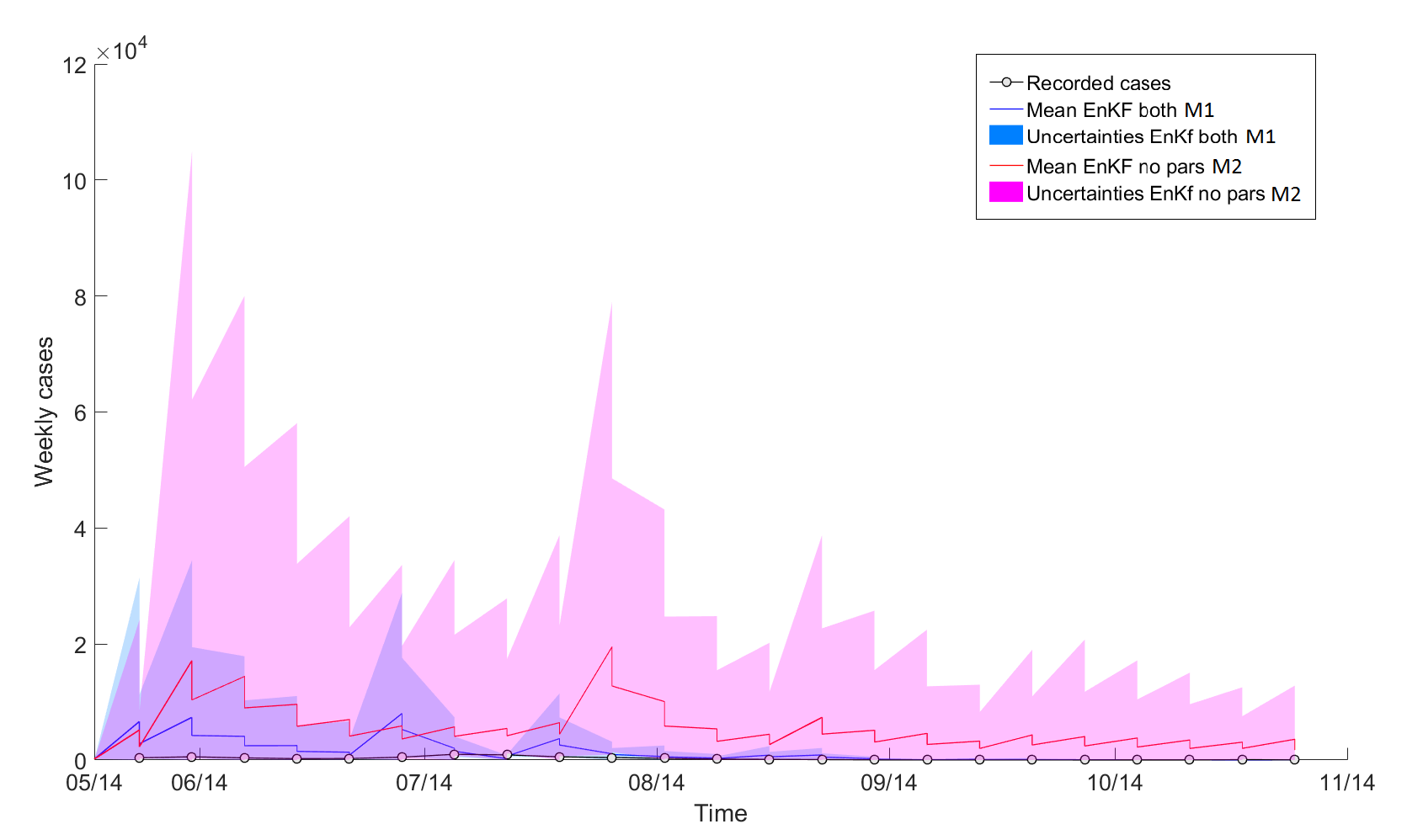}
\caption[EnKF, simulations for 2014 using prior Haiti, global behavior]{\textbf{First scenario for year 2014} using the prior pdfs for the Haitian case study. Epidemiological curve in our domain. The model overrates the process whether the parameters pdfs are updated or not. In the last case, uncertainties are higher.}
\label{sum_cases_prior14}
\end{figure}

\par As figure \ref{sum_cases_prior14} shows, using this prior, the model overestimates the number of cases recorded in our spatial domain. We compare the results of two simulations: one using the state augmentation, i.e. both update of states and parameters (blue palette), to which we will refer to as "Simulation M1" for simplicity; the other using constant values for the parameters (magenta palette), that we will call "Simulation M2". Mean values of the ensemble are represented by the lines, credible interval of 90\% - or uncertainties -  by the shadows. The black dots represent the weekly recorded cases.

\begin{figure}[!h]
\centering
\includegraphics[width=1\textwidth,height=0.4\textheight]{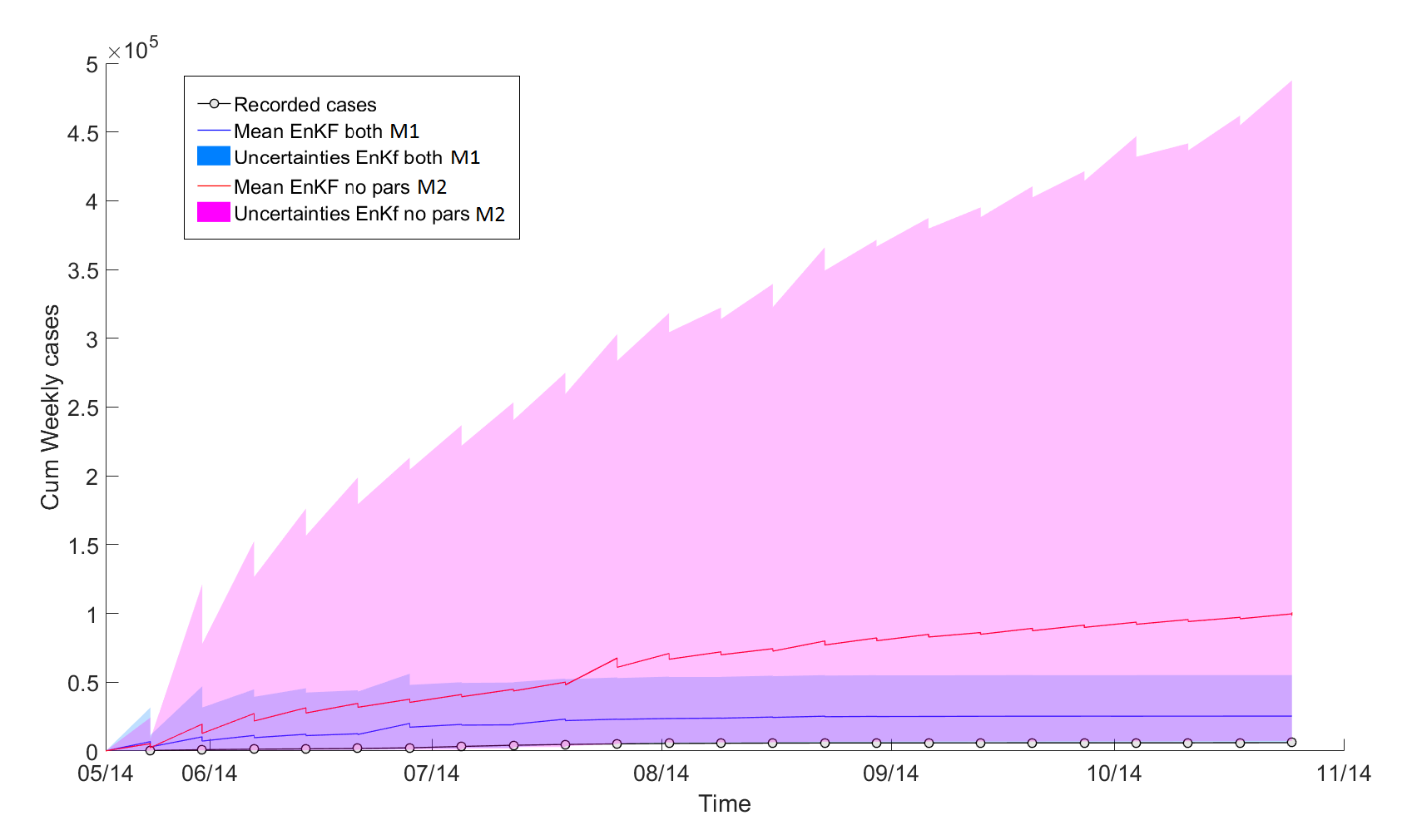}
\caption[EnKF, simulations for 2014 using prior Haiti, cumulative curve]{\textbf{First scenario for year 2014} using the prior pdfs for the Haitian case study. Cumulative curve of weekly cases. The model overrates the process whether the parameters pdfs are updated or not. In the last case, relative uncertainties are higher.}
\label{sum_cum_cases_prior14}
\end{figure}

\begin{figure}[!h]
\centering
\includegraphics[width=1\textwidth,height=0.5\textheight]{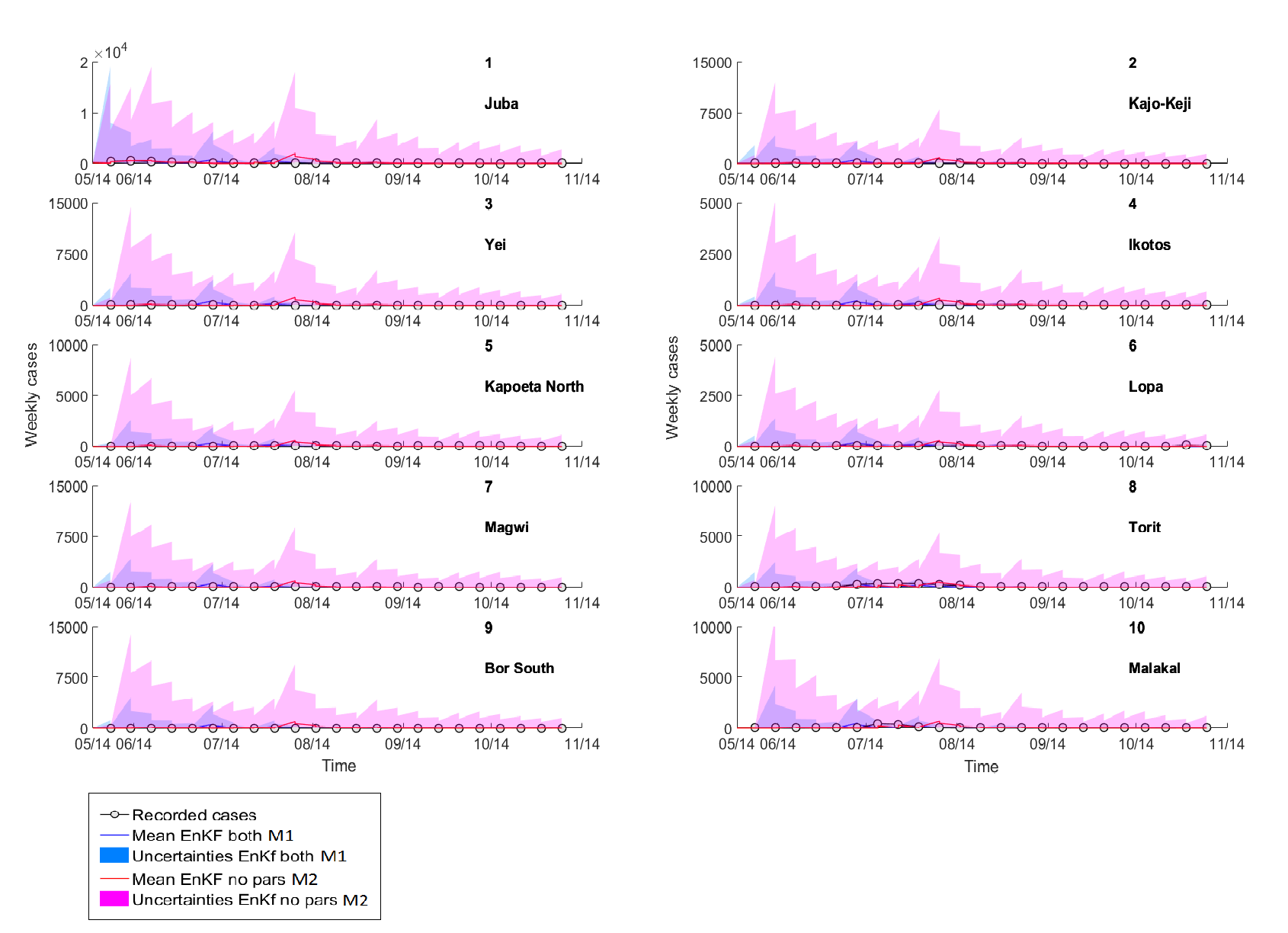}
\caption[EnKF, simulations for 2014 using prior Haiti, County-level cases]{\textbf{First scenario for year 2014} using the prior pdfs for the Haitian case study. Epidemiological curves in the Counties in which cholera was recorded. Magnitudes of the uncertainties are high and the mean values are far away from the real observed phenomena.}
\label{comm_cases_prior14}
\end{figure}

At the beginning, as it is possible to see from the cumulative curves in Fig. \ref{sum_cum_cases_prior14}, and from the response of the model at the County - level in Fig. \ref{comm_cases_prior14}, the model starts immediately to overrate the actual process. At the first step of the data assimilation, the update of both states and parameters tries to reduce the mean and the uncertainties, but there are still some wrong values that do not allow a good performance of the model. In the case in which the values for the parameters are constant, the corrections at each time step very slightly reduce the uncertainties, whose magnitudes are higher than the observations and than the results of Simulation M1. 

\begin{figure}[!h]
\centering
\includegraphics[width=1\textwidth,height=0.5\textheight]{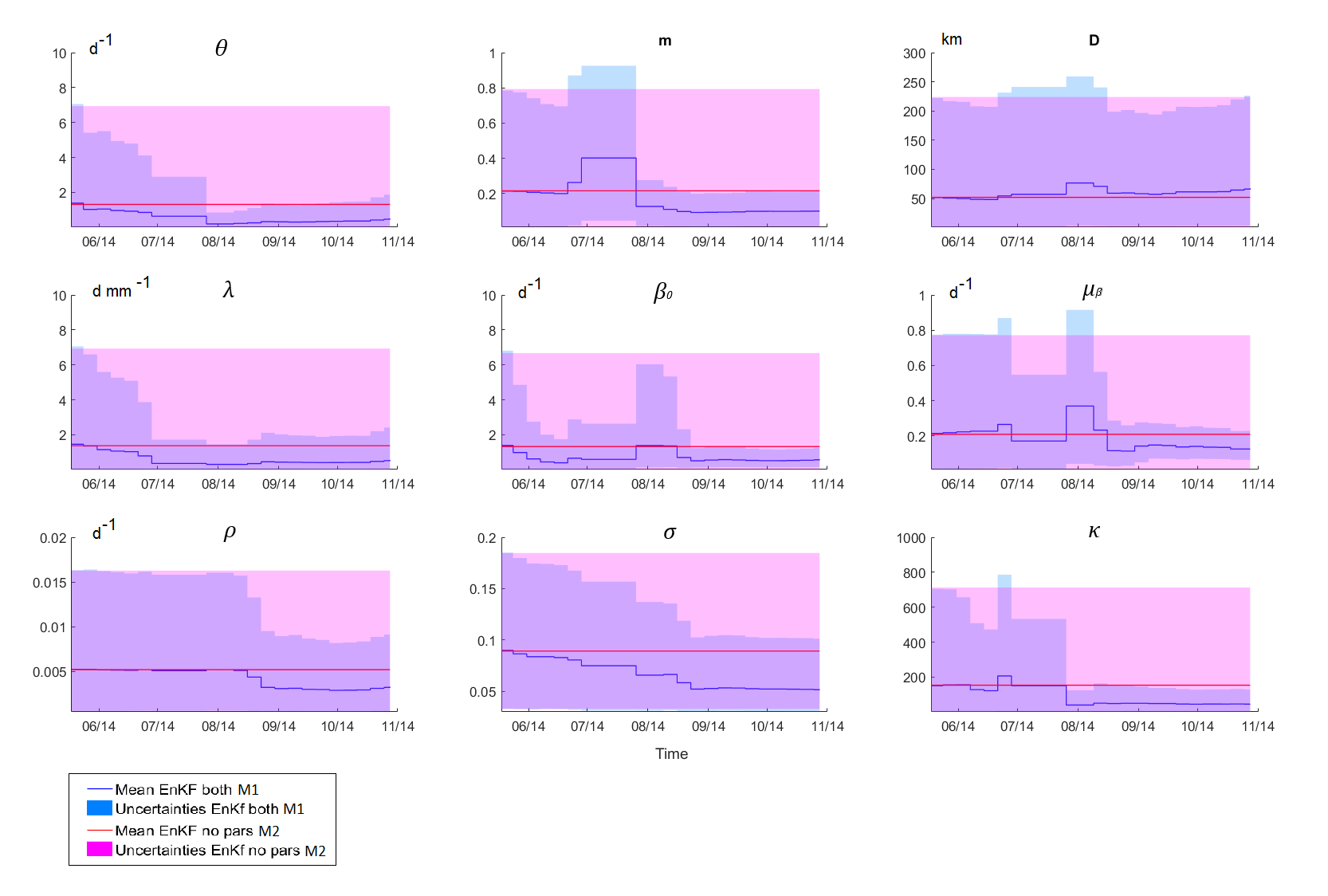}
\caption[EnKF, simulations for 2014 using prior Haiti, parameters behavior]{\textbf{First scenario for year 2014} using the prior pdfs for the Haitian case study. Parameters values change in time for Simulation M1, while remains constant during Simulation M2.}
\label{parm_val_prior14}
\end{figure}

\newpage
In order to correct this evaluation, we have tried to figure out which were the parameters that mostly affected the model, focusing on the values chosen by the algorithm on the first four weeks of simulations. Figure \ref{parm_val_prior14} shows the behavior of the parameter in time.
\par The trends in this figure made us understand that the wrong response maybe possibly due to high values of the parameters and their combinations.

Focusing on Simulation M1, parameters values change a lot during the update, restricting their prior values and moving away from the bounds.
The combined parameter $\theta$, from the prior whose maximum value is $10$ $d^-1$, reduces from a starting maximum value of almost $7$ $d^-1$ associated to the uncertainties, to a maximum value of $2$ $d^-1$ at the end of the simulation. Regarding the probability of travel $m$, the maximum value for uncertainties reduces from $0.9$ to $0.3$. The shape factor $D$ do not detach too much from the prior distribution. On the opposite, the coefficient $\lambda$ for the rainfall enhancement effect, reduces its maximum value to $2$ $d ~ mm^-1$, against its prior $10$ $d ~ mm^-1$. The same trend has the exposure rate $\beta_0$ that, by the time the data assimilation helps in decreasing the uncertainties, assess to a maximum value of $2$ $d^-1$. The same happens with the mortality rate of bacteria $\mu_\beta$, assessing to approximately $0.4$ $d^-1$; the fraction $\sigma$ of symptomatic infected, slightly restricts its prior range, having as upper bound the value $0.15$ $d^-1$. The same happens for loss of immunity rate $\rho$, whose upper bound value reduces to $0.015$ $d^-1$. The maximum value awareness parameter $k$ instead, settles during the data assimilation to $200$.

%-----------------------------------------------------------------------%
%                          	     NEW PRIOR
%-----------------------------------------------------------------------%
\subsection{Prior for South Sudan}

Based on the trends of the parameters during the Data Assimilation conducted using the prior of Haiti, we have tried to restrict the prior ranges with the aim of reducing the uncertainties of the model at the beginning of the simulations. Focusing on the first four weeks in various scenarios, we have found new prior ranges that can work well for this case study (table \ref{priorss}) and that were used to perform the EnKF and SIRB model for both years and whose results will follow.

\begin{table}[htbp]
\centering 
\caption{Range of the parameter prior distribution for South Sudan case study}
\label{priorss}
\begin{tabular}{c c l}\hline
\multicolumn{1}{c}{\textbf{Parameters}} & \textbf{Units} & \textbf{Prior}\\
\hline
$\theta$ & $d^{-1}$ 	 & [ 0.1      -     0.5 ] \\
\co{l} & $d^{-1}$ 		 & [ 0        -       0 ] \\
\co{m} & $-$	    	 & [ 0.1      -    0.35 ] \\
\co{D} & $km$ 			 & [ 50       -     250 ] \\
$\lambda$ & \textit{d} $mm^{-1}$ & [ 0.2      -     0.8 ] \\
$\beta_{0}$ & $d^{-1}$	 & [ 0.2      -     0.5 ] \\
$\mu_{\beta}$ & $d^{-1}$ & [ 0.02     -    0.07 ] \\
$\rho$ & $d^{-1}$ 		 & [ 0.0015   -   0.015 ] \\
$\sigma$ & $-$ 			 & [ 0.05     -    0.15 ] \\
\co{k} & $d^{-1}$		 & [ 100      -     450 ] \\
\hline\end{tabular}
\end{table}

As the table shows, we redesigned both upper and lower bounds. We have seen that too high values of the parameters lead to overrate of the infected in our network. From the trials done to reduce the uncertainties, we understood that too low values instead, drift the model to underrate them. Comparing this table with the one seen before, \ref{priorhaiti}, the values than adapt to South Sudan case are lower that the one used for the Haiti model. The Haitian cholera events were more severe than the ones in this case study. Weekly recorded cases in that country reached numbers as $3$x$10^4$. It is though reasonable to have,  e.g., high values of $\theta$, indicator of the worsening of the sanitation condition, and $\beta_0$, indicator of the amount of bacteria daily shed. The epidemics we are dealing with, for what we know, did not show similar attack rate.
%-----------------------------------------------------------------------%
%                          	     YEAR 2014
%-----------------------------------------------------------------------%
\newpage
\section{Year 2014}

We compare the responses of different simulations performed using the new prior. We used the Ensemble Kalman Filter algorithm twice, similarly to what has been done testing the prior of Haiti case study: once updating in one case both states and parameters using the state augmentation (blue palette in the figures), subsequently updating only the states and using constant values of the parameters in time (magenta palette in the figures). For simplicity, as before, we will refer to these two respectively as "Simulation M1" and "Simulation M2". Moreover, we tested this new prior in an "Open Loop" simulation, that performs the SIRB model without Data Assimilation and uses random values for the parameter into the defined ranges.

\subsection*{Simulations using EnKF}
\begin{figure}[!h]
\centering
\includegraphics[width=1\textwidth,height=0.4\textheight]{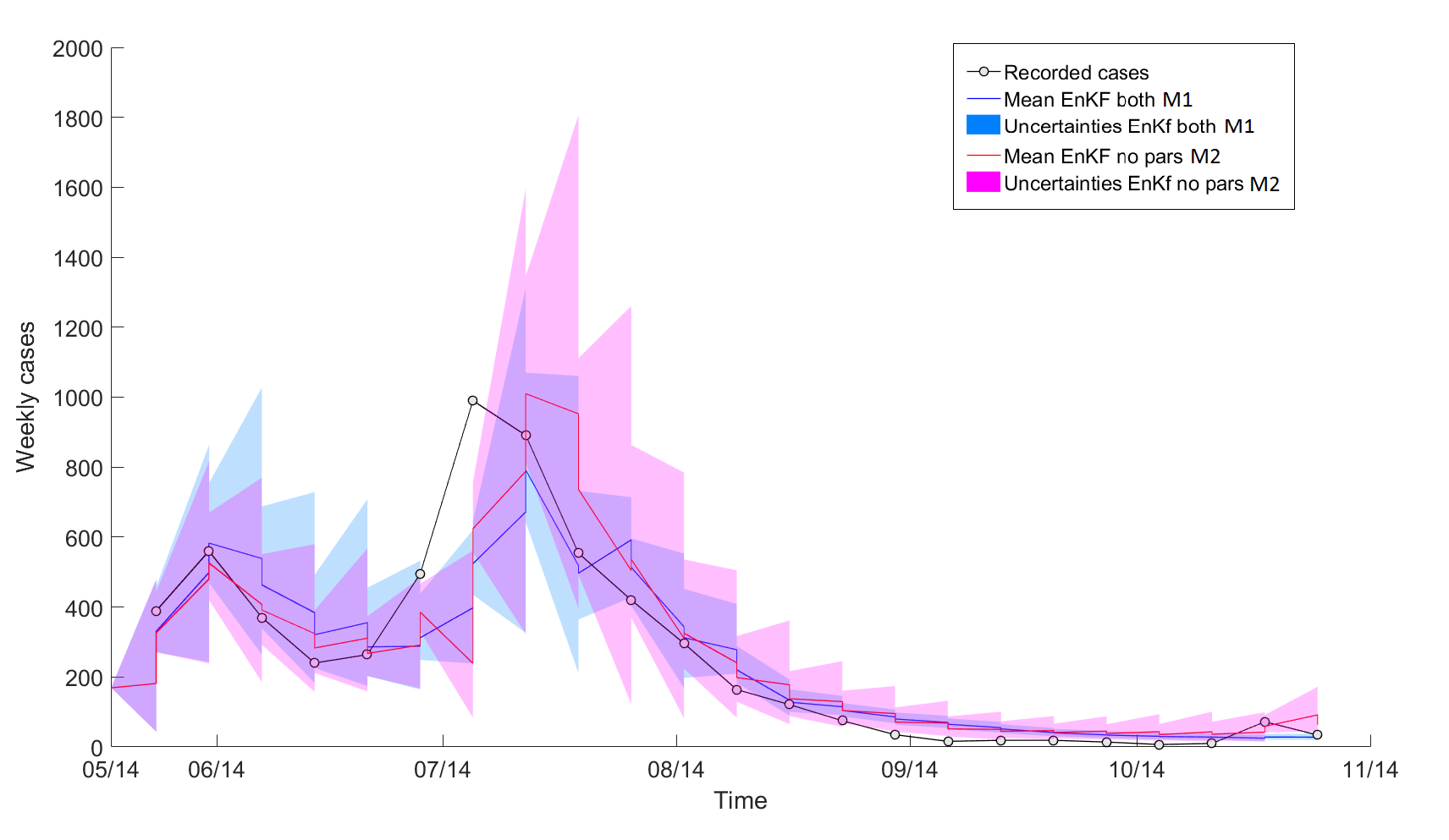}
\caption[EnKF, simulations for 2014 using the new prior, global behavior]{\textbf{Results for 2014 using the new prior in EnKF procedures.} Epidemiological curve in our domain. Mean values of both simulations approach the real observations.}
\label{sum_cases14}
\end{figure}

The model shows an overall good agreement with the data. 
As Fig.\ref{sum_cases14} shows, the mean of the ensemble of both simulations approaches the real behavior of the epidemic in our domain of study. Uncertainties, that for simulations M2 are higher than the other one, decrease in time. The peak of the epidemic, that was recorded during the second week of July 2014, is shifted of one week by the model in both simulations, but its magnitude is almost captured. 

\begin{figure}[!h]
\centering
\includegraphics[width=1\textwidth,height=0.4\textheight]{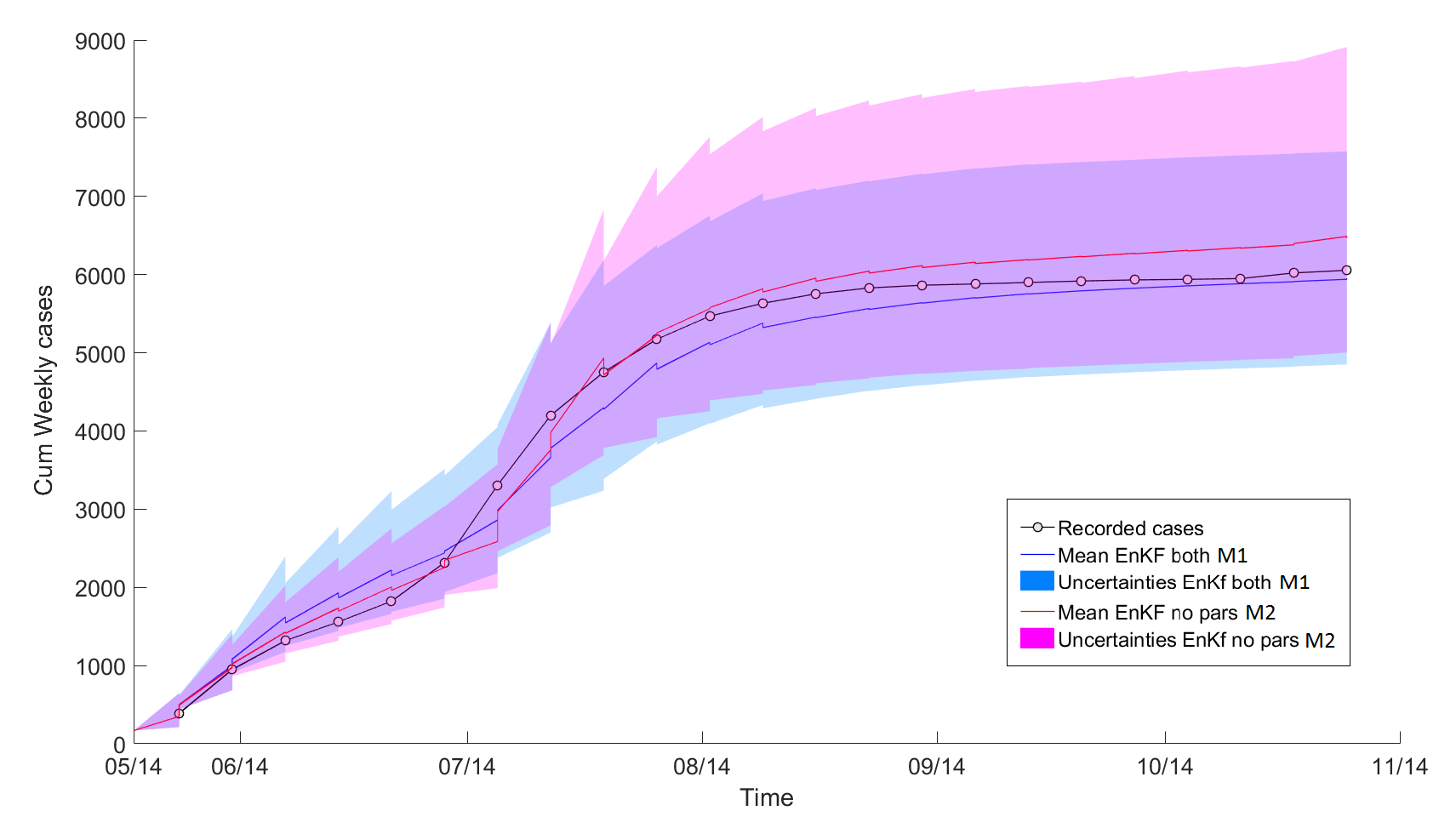}
\caption[EnKF, simulations for 2014 using the new prior, cumulative cases]{\textbf{Results for 2014 using the new prior in EnKF procedures.} Cumulative curve of weekly cases. The three curves of EnKF calibration procedure and the observation almost overlap.}
\label{cum_sum14}
\end{figure}

It is interesting to see through the cumulative response in Fig. \ref{cum_sum14}, that the model offers a good approximation of the real curve, that at the end of simulations is found in between the results of Simulation M1 and M2. As the figure shows, the confidence interval when parameters pdfs are not updated is larger.

\begin{figure}[!h]
\centering
\includegraphics[width=1\textwidth,height=0.5\textheight]{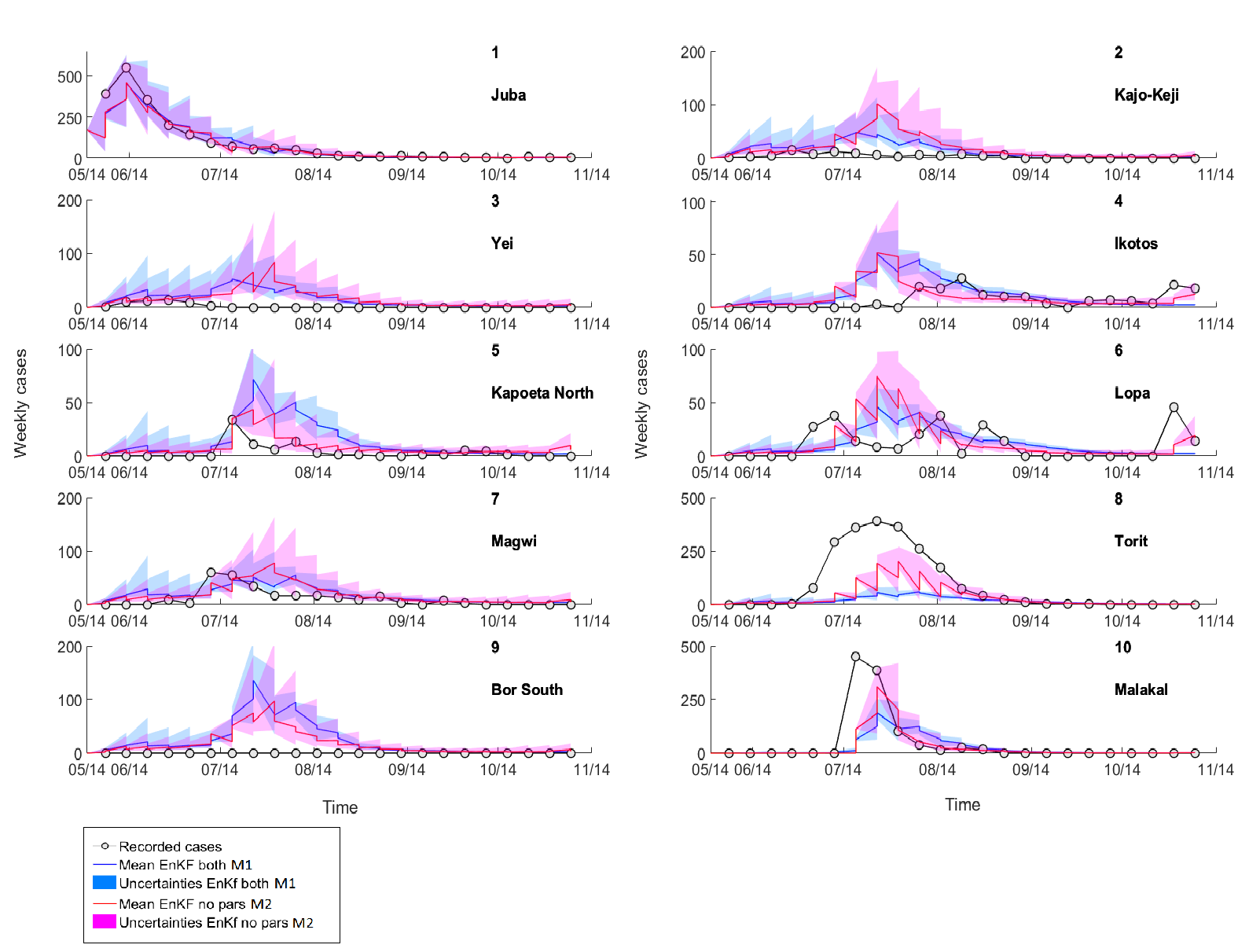}
\caption[EnKF, simulations for 2014 using the new prior, County level cases]{\textbf{Results for 2014 using the new prior in EnKF procedures.} Epidemiological curves in the Counties in which cholera was recorded. The simulations offer a good representation of the dynamic of the epidemic, overall in Juba.}
\label{comm_cases14}
\end{figure}

At the level of the Counties and in both simulations, as shown in Fig. \ref{comm_cases14}, our mathematical framework is able to reproduce almost perfectly the dynamic of the epidemic in Juba, one of the County hit more severely by the infection and where it begun. Actually, the two simulations in here almost have the same mean and confidence interval. The response of the simulations is not as perfect in the other counties. In terms of numbers, the mean and the variances are not so distant from the real number of cases recorded, and the magnitude of the simulated and real phenomena are comparable, overall for what concerns simulation M1. Some overestimation are present in the Counties of Kajo-Keji, Yei, Ikotos. As we said, a certain response is always expected in all the nodes of the network: therefore we can justify the output of the model in Bor South County, in which simulated cases are present even though there were no recorded cases in 2014 year. Regarding Torit and Malakal, it is very difficult to peak this two events, despite the mean values of simulation M2 approximate better the real values of weekly recorded cases. Further details on these behaviors will be given in Sect. \vref{discussion}.

\newpage
\begin{figure}[t]
\centering
\includegraphics[width=1\textwidth,height=0.55\textheight]{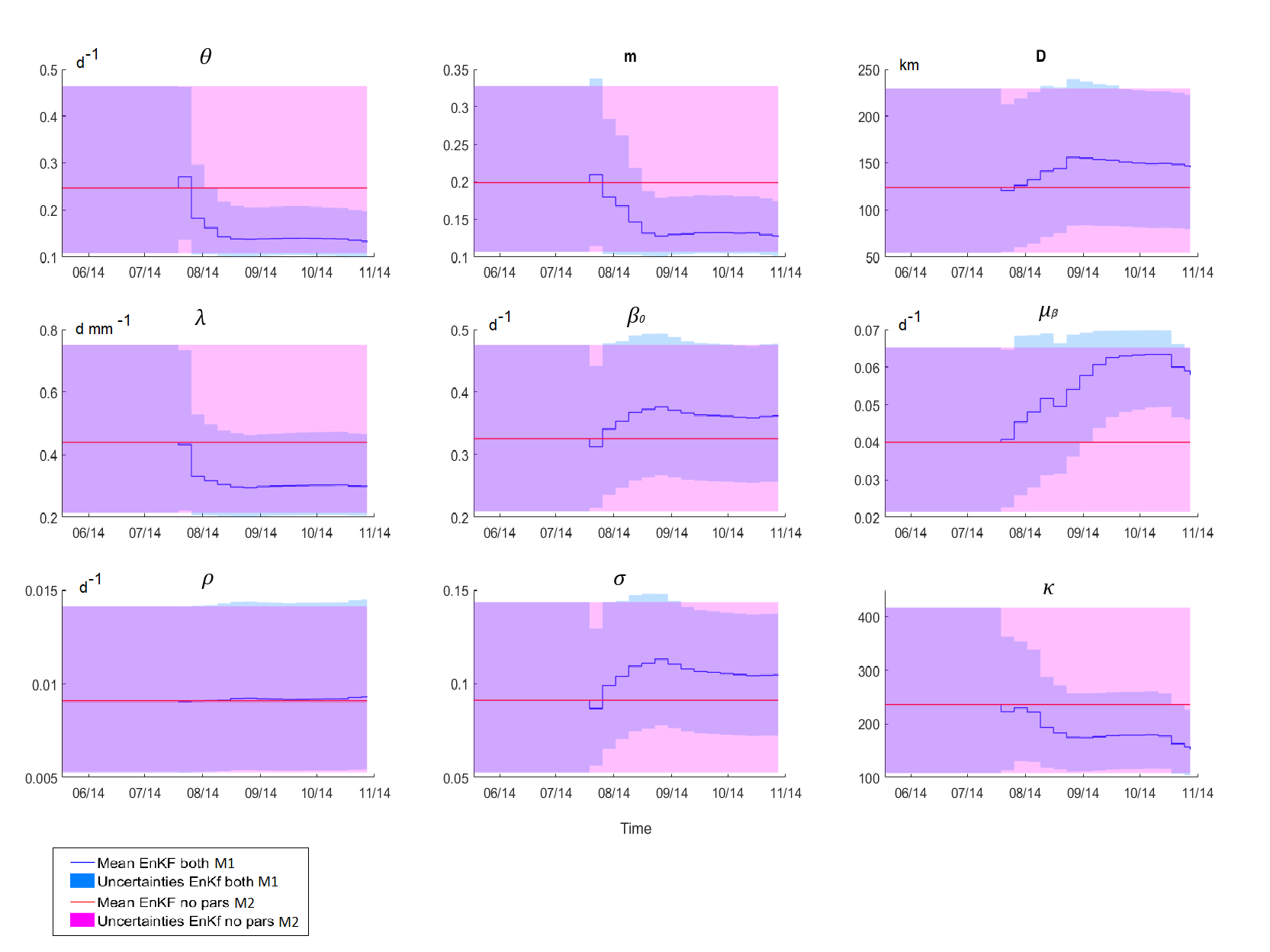}
\caption[EnKF, simulations for 2014 using the new prior, parameters behavior]{\textbf{Results for 2014 using the new prior in EnKF procedures.} Parameters values change in time for Simulation M1, while remain constant during Simulation B.}
\label{parm_val14}
\end{figure}

Fig.\ref{parm_val14} shows how parameters change during the Data Assimilation and the values assumed whether constant. Focusing on the trends of time-dependent parameters, Simulation M1, it is possible to see that even though the credible intervals restrict in time, the variability of these makes difficult to find constant values that can fit the model. We can read this figure recalling the equations of the model \eqref{S} - \eqref{B} and the meaning of the parameters. As the epidemic decreases in time, after the peak of the simulation at the end of July 2014, the mean value of $\theta$, indicator of the sanitation conditions, drastically decreases from $0.26$ $d^-1$ to $0.17$ $d^-1$, possibly mimicking some kind of strategies that acted to reduce the risk of infection. From the point of view of the phenomena described, at the beginning the simulation is reproducing a rate of contamination equal to $1/4$ days, value that changes to $1/6$ days after the peak. Complementary, the simulated mortality rate of the bacteria $\mu_\beta$, increases in time toward its upper bound in order to reduce the bacteria concentration in the water, hence the risk of the disease, with the effect of reducing the number of simulated cases in time, approaching the real infected. The lifetime values of bacteria that the model is using during the simulations are comparable to the values stated by clinical observations (e.g. \citep{Kaper1995.,Nelson2009} of few months. In our case, lifetime of bacteria is about one month ($\mu_\beta$ = $0.04$ $d^-1$). The exposure rate $\beta_0$ slightly increases after the simulated peak. Hence the interval between ingestion of contaminated food or water changes from $3$ days to $2$ days ($\beta_0$ = $0.03$ - $0.4$ $d^-1$). Following this trend, $k$, the inverse of the parameters $\psi$ that controls the decrease of the exposure rate $\beta$, decreases, meaning that the awareness of the population as regards the illness increases. 
It is interesting to see that the loss of immunity rate $\rho$ settles to a quasi-constant average value of $0.009$ $d^-1$, i.e. on average people that recovered from the disease lose their immunity after $112$ days. We can assume this value to be valid: in fact, our rate takes into account both symptomatic and asymptomatic infected that, as various authors showed (see e.g. \citep{Koelle2005,King2008}), have different immunity duration, with symptomatic infected gaining on average multi-year immunity, while asymptomatic ones few months or less. Regarding the fraction of symptomatic infected $\sigma$, it shows similar trend to the global epidemiological curve that is simulated. The effect of rainfalls, whose coefficient is $\lambda$, reduces in time following the seasonality of rain and the global trend of the epidemic. 
Looking at the mobility indicators, $m$ and $D$, the first one slowly shifts to lower values after the peak in July, while $D$ does not show big changes during the period of analysis, stating that distances do not affect in time movements of people. Further discussion on the role of these two will follow.

For what concerns the constant values, these coincides with the values of the parameters at the beginning of Simulation M1. Again, in this simulation the confidence is higher.

%======================================================================================%
%------------------------------Open Loop-------------------------------%
%======================================================================================%
\subsection*{Simulation using Open Loop} In order to test the validity of the prior, we performed an "Open Loop" simulation, in which the output of the model is given using randomly sampled parameters, without any kind of correction of state variables.

\begin{figure}[!h]
\centering
\includegraphics[width=1\textwidth,height=0.4\textheight]{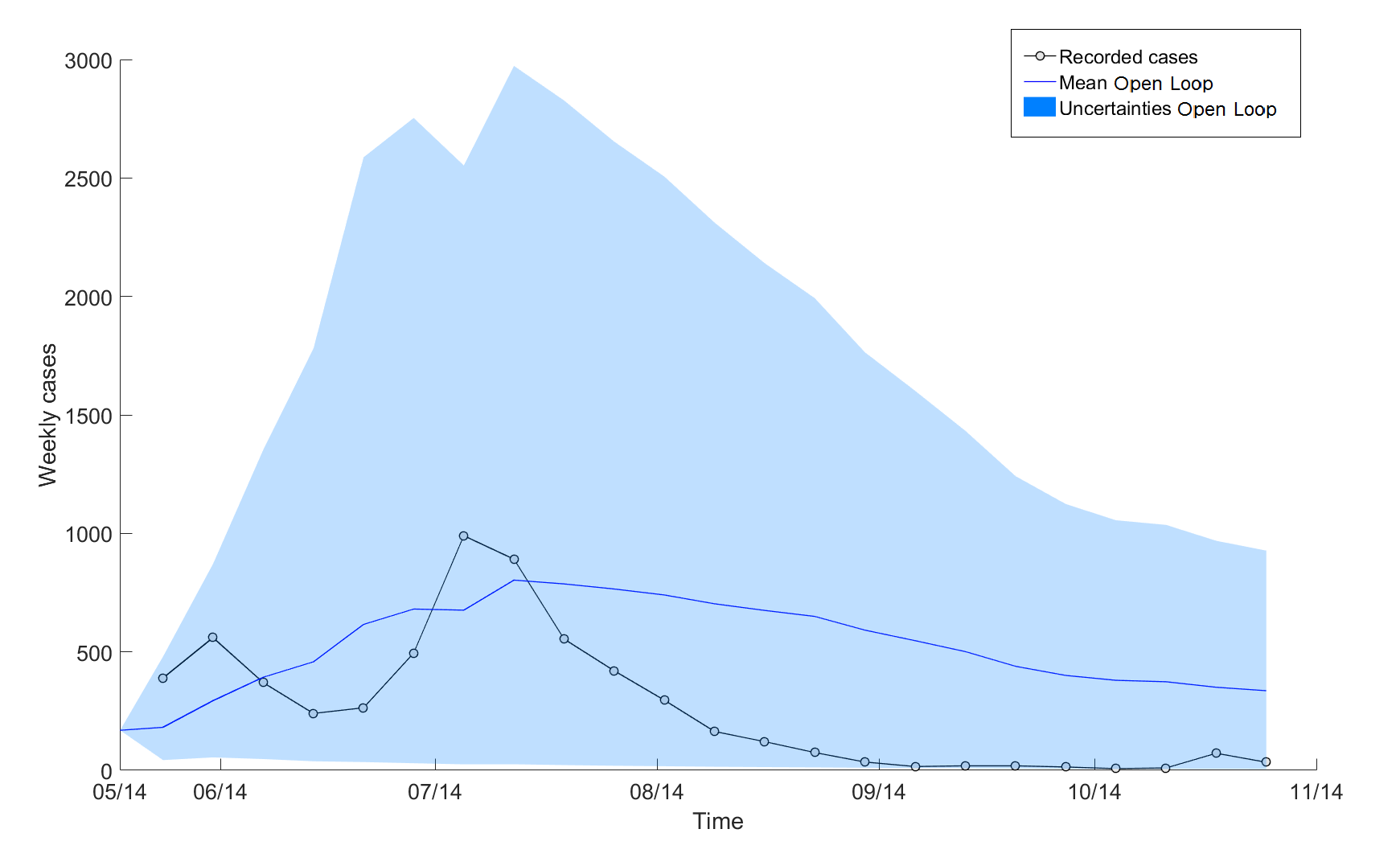}
\caption[Open Loop, simulation for 2014 using the new prior, global behavior]{\textbf{Results for 2014 using the new prior in Open Loop simulation.} We use random sampling of the parameters to perform the model without DA. Observations fall into the confidence interval, yet the mean does not approach them.}
\label{open14}
\end{figure}

As the Fig. \ref{open14} show, the confidence interval of the model contains the real curve, meaning that the prior that we have chosen fit the model and the dynamic to mimic, yet the mean does not approach the observations. This let us understand that Data Assimilation is required for a good behavior of the model and correct the state variables. 

\paragraph{}
We can compare the goodness of the simulations just described using the Root Mean Squared Errors of the four (see Sect. \vref{setupcal}). As the table \ref{rmse14} shows, using the new prior adapted for this case study, we definitely reduce the mean and the maximum RMSE with respect to the simulations performed using the prior for Haiti. This means that we are reducing the errors associated to the output of the model, approaching the real observations. As expected from the figures, the mean RMSE for the Open Loop is higher than the ones for the simulations in which the Data Assimilation technique has been developed. Among these two, the one in which parameters are kept constant, has a lower mean RMSE, therefore preferable. Maximum values of RMSE among these two are practically the same.

\begin{table}[htbp]
\centering 
\caption{RMSE comparison for simulations of year 2014}
\label{rmse14}
\begin{tabularx}{\textwidth}{X c c}\hline
\multicolumn{1}{c}{\textbf{Simulation}} & \textbf{Mean RMSE} & \textbf{Max RMSE}\\
\hline
EnKF: parameters update; prior Haiti		& 175.6385 			 & 1.1570e+03 \\
EnKF: no parameters update; prior Haiti		& 633.5467 			 & 2.2280e+03 \\
EnKF: parameters update; new prior			& 39.0639			 &	177.2061  \\
EnKF: no parameters update; new prior		& 33.9874			 &	177.9327  \\
Open Loop: new prior						& 80.5312			 &	199.6132  \\
\hline\end{tabularx}
\end{table}
%======================================================================================%
%======================================================================================%
%-----------------------------------------------------------------------%
%                          	     YEAR 2015
%-----------------------------------------------------------------------%
\newpage
\section{Year 2015}
Seen that the prior used in the Haiti case study was not successful in our domain, we chose to directly perform simulations using the new prior, whose results are quite good for the previous year. As before we performed EnKF twice and an Open Loop simulation to test the prior. We will use the definition "Simulation M1", "Simulation M2" to distinguish among the two EnKF procedures.  

\begin{figure}[!h]
\centering
\includegraphics[width=1\textwidth,height=0.4\textheight]{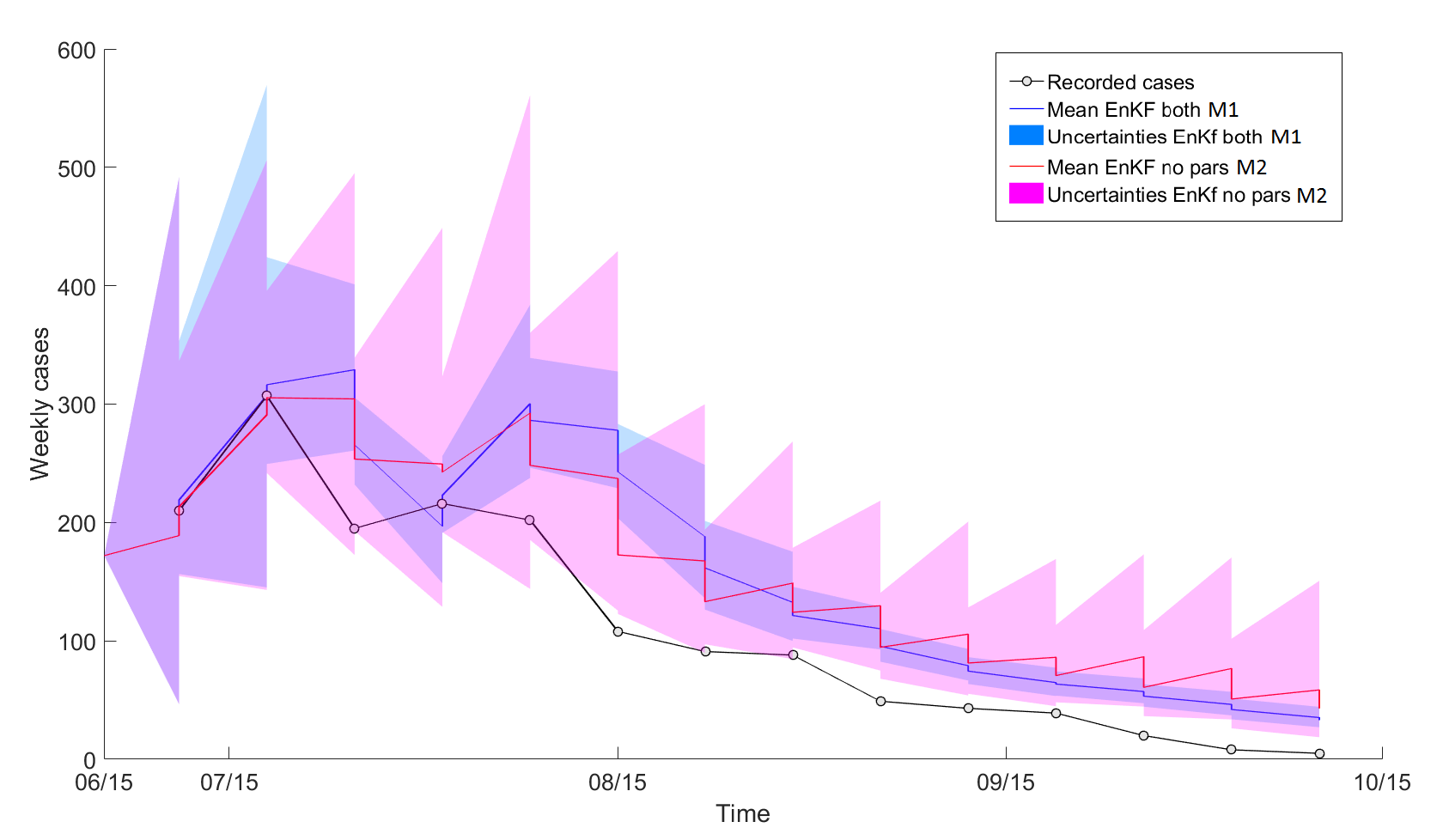}
\caption[EnKF, simulations for 2015 using the new prior, global behavior]{\textbf{Results for 2015 using the new prior in EnKF procedure} Epidemiological curve in our domain. Trends of the mean values are similar to the real observed phenomena, yet overestimation is present.}
\label{sum_cases15}
\end{figure}

As Fig. \ref{sum_cases15} shows, the model tends to overrate the global number of cases in our spatial domain. At the beginning of the simulations, as the uncertainties show, DA corrects the state variables, and updates the parameters in one case, reducing the simulated number of infected. Even though the trends are similar, the confidence interval for the simulation M2 when parameters are constant in time, is larger. 

The cumulative curves confirm the overestimation of the model in both simulations.

\begin{figure}[!h]
\centering
\includegraphics[width=1\textwidth,height=0.4\textheight]{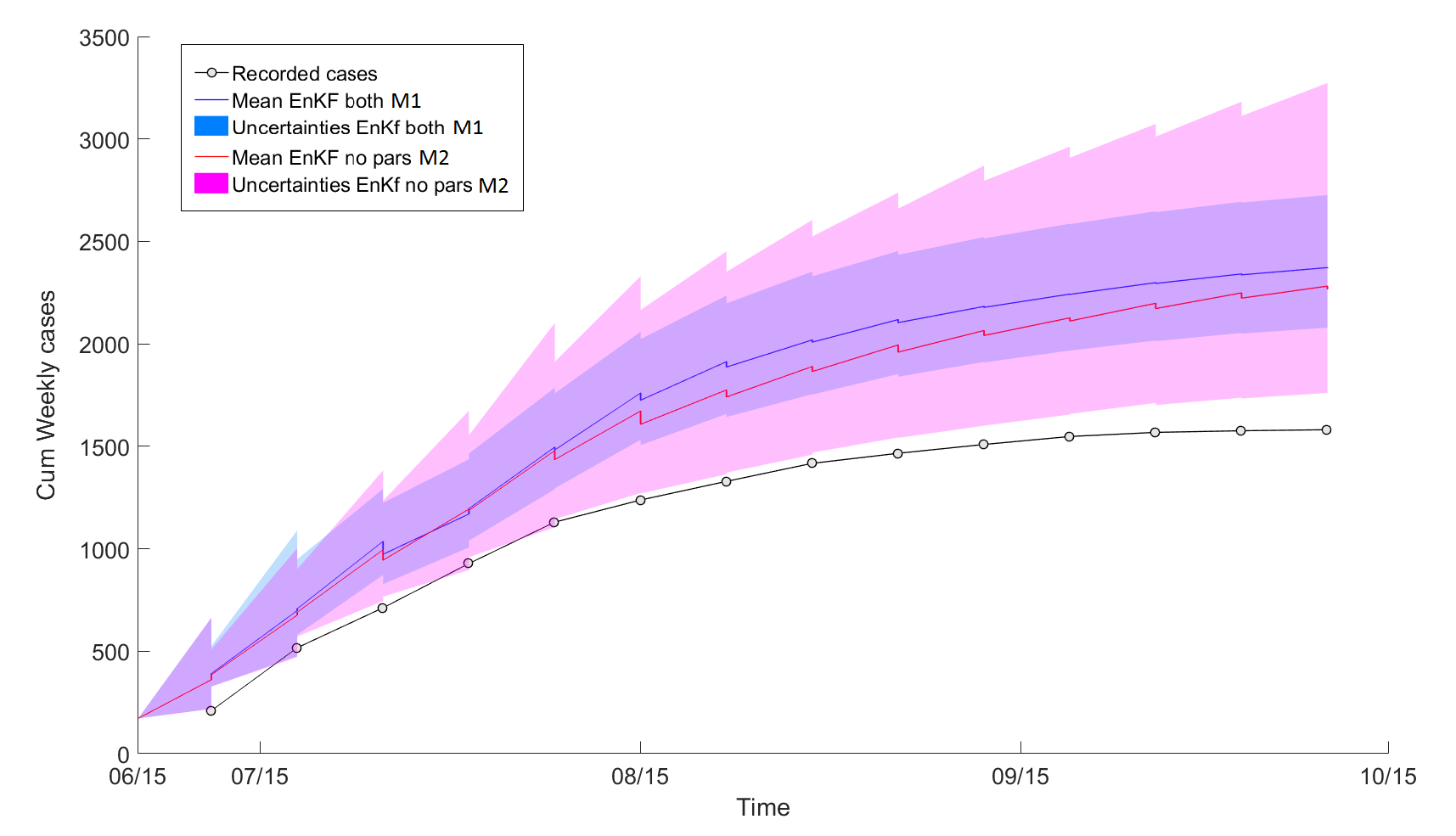}
\caption[EnKF, simulations for 2015 using the new prior, cumulative cases]{\textbf{Results for 2015 using the new prior in EnKF procedures.} Cumulative curve of weekly cases. The curves of the EnKF calibration procedure detach each other towards the end of the simulations, yet the observed cumulative curve stays below them}
\label{cum_sum15}
\end{figure}

\newpage
\begin{figure}[!h]
\centering
\includegraphics[width=1\textwidth,height=0.55\textheight]{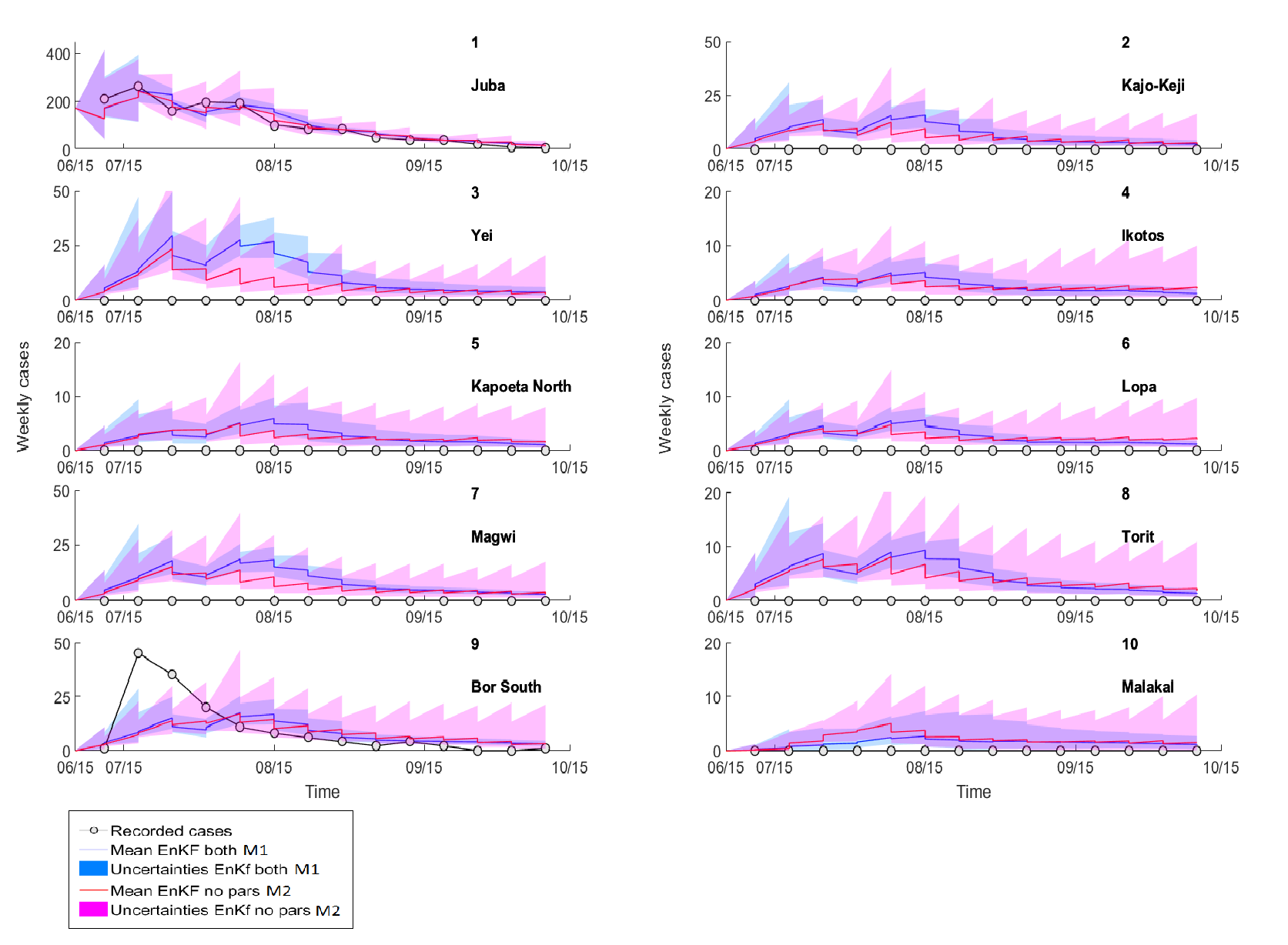}
\caption[EnKF, simulations for 2015 using the new prior, County level cases]{\textbf{Results for 2015 using the new prior in EnKF procedures.} Epidemiological curves in the Counties in which cholera was recorded. The simulations offer a good representation of the dynamic of the epidemic in Juba, and a relative low response in cholera-free Counties.}
\label{comm_cases15}
\end{figure}

At the level of the County the model captures, almost perfectly, timing and magnitude of the epidemic in Juba, whether parameters are kept constant or subject to update. As we have stated in the previous lines, the mathematical approach used does not allows for zeroes. Therefore, as we can see from Fig. \ref{comm_cases15}, the model is simulating cases in the other Counties in which there were no recorded cases, even if the magnitude of these numbers is low. We can consider this as an explication to the global overestimation seen in the epidemiological curve in our domain for 2015 (Figs \ref{sum_cases15} - \ref{cum_sum_cases15}. There are no great differences between the two EnKF simulations, that are quite the same allover the Counties. The response in Bor South, whose peak is difficult to capture, needs some further details that will be given in Sect. \ref{discussion}.

\newpage
\begin{figure}[t]
\centering
\includegraphics[width=1\textwidth,height=0.55\textheight]{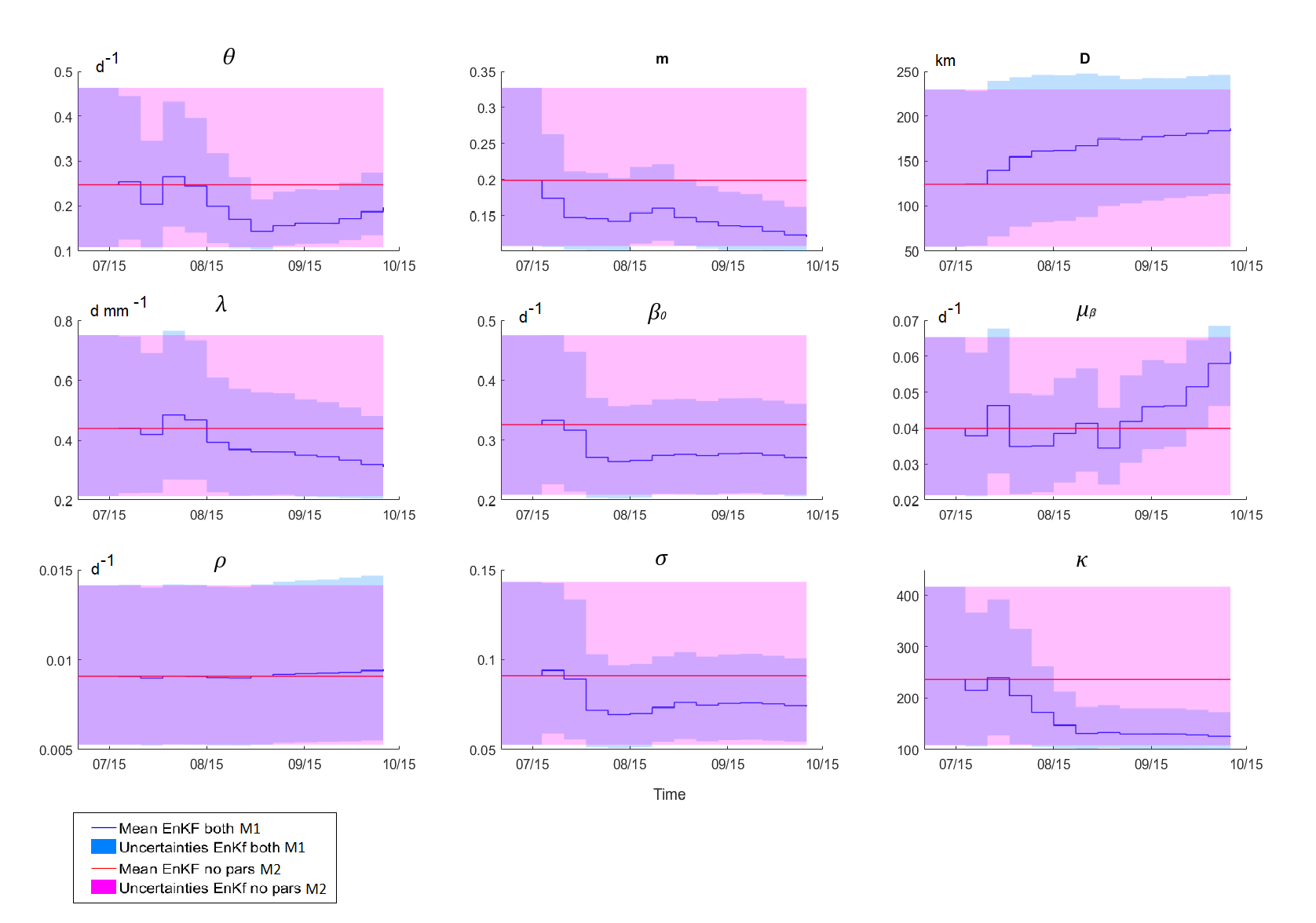}
\caption[EnKF, simulations for 2015 using the new prior, parameters behavior]{\textbf{Results for 2015 using the new prior in EnKF procedures.} Parameters values change in time for Simulation A, while remain constant during Simulation B.}
\label{parm_val15}
\end{figure}

As regards the parameters, focusing on Simulation M1, their variability and uncertainties are higher compared to the trends seen during 2014 simulations. $\theta$, for the sanitation condition and the contamination rate, firstly decreases to a lower frequency $1/7$ day, while at the end of the simulation it becomes $1/5$ days. Despite the worsening of the sanitation condition through the parameter $\theta$, for consistency of the simulation with the data, the awareness of the population regarding the risk of contracting the illness increases. This is represented by the parameter $k$, whose value rapidly decreases towards its lower bound $100$. 
Complementary, the mortality rate of bacteria $\mu_\beta$, subsequently to some oscillations, increases from $0.04$ $d^-1$ to $0.06$ $d^-1$ at the end of the simulation; the initial exposure rate, $\beta_0$ whose starting value is $1/3$ days, changes to $1/4$ days, leading to a lower possibility of contract the infection via ingestion of contaminated food and water. The influence of the rain reduces through the coefficient $\lambda$, whose values change from $0.4$ $d^mm^-1$ to $0.3$ $d~mm^-1$. Again, the mean value of the loss of immunity rate $\rho$ assesses around $0.009$ $d^-1$, while $\sigma$ defining the fraction of symptomatic infected follows the reduction of infected, and reduces, after the peak of the epidemic in July, to the mean value $0.07$, i.e. the 7\% of the people that get in touch with the bacteria showed symptoms. For what concerns the mobility parameters, $m$ decreases in time from $0.2$ to $0.12$, trying to contain the spatial spreading of the epidemic and let the simulation to be consistent with the data. This behavior of the model is confirmed by the shape factor of mobility $D$ that increases in time, giving more importance to distances as deterrence factors for mobility.

To what concerns the constant values, these coincides with the values of the parameters at the beginning of Simulation M1. Again, in this case, the confidence interval is higher.

\subsection*{Open Loop} 
\begin{figure}[!h]
\centering
\includegraphics[width=1\textwidth,height=0.4\textheight]{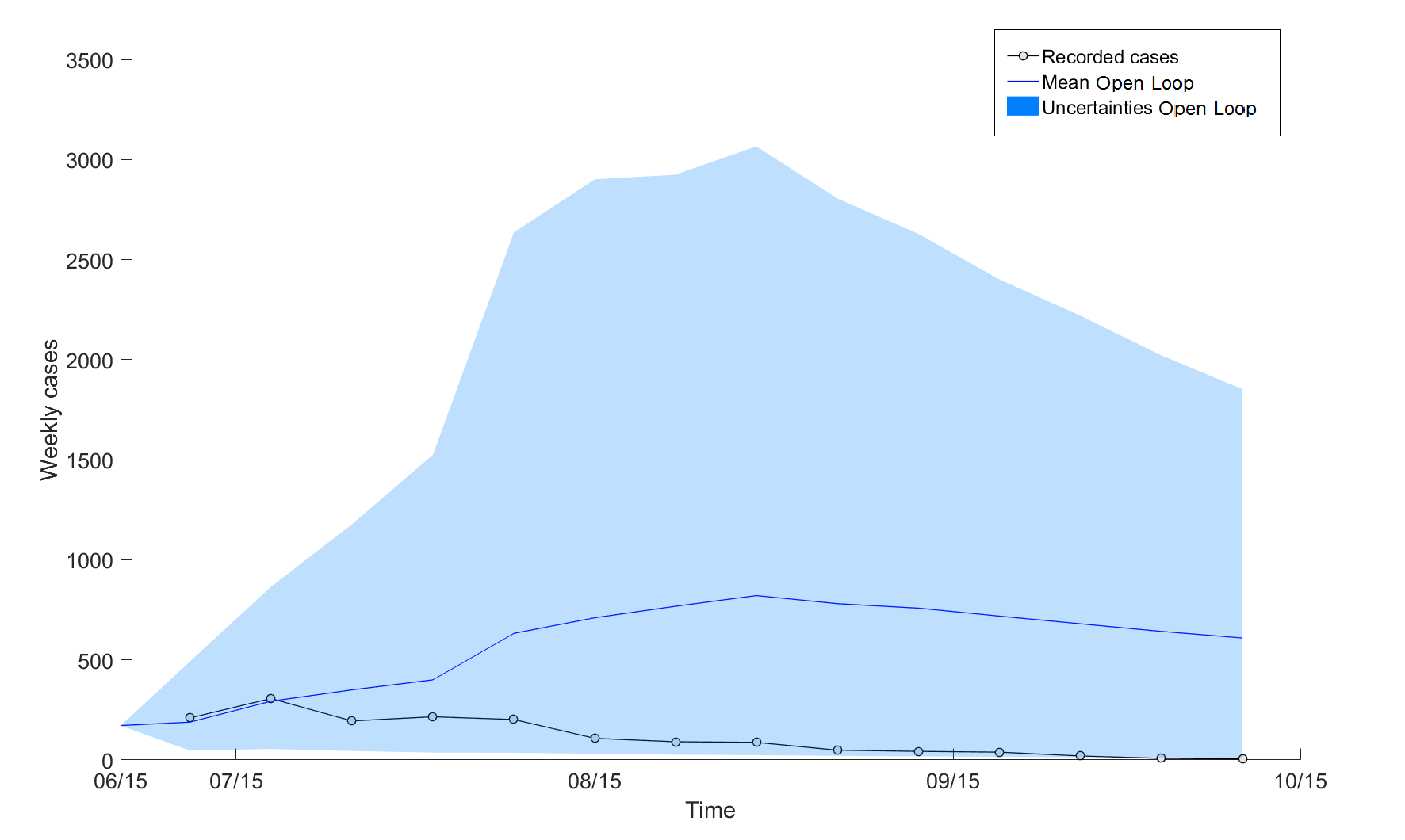}
\caption[Open Loop, simulation for 2015 using the new prior, global behavior]{\textbf{Results for 2015 using the new prior in Open Loop simulation.} We use random sampling of the parameters to perform the model without the DA. Observations are into the confidence interval, yet the mean do not approaches them.}
\label{open15}
\end{figure}

Similarly to what we have stated for 2014, the Open Loop in Fig. \ref{open15} for year 2015 let us understand that, even though the confidence interval of the simulation contains the curve, DA is required for a good agreement of the model with the data. 

What stated is confirmed by the RMSE computation, Table \ref{rmse15}. The Open Loop again results non adequate for this case study. For what concerns EnKF, values of the mean RMSE of the two simulations are really near, and whether is needed to choose, we would prefer the simulations in which parameters are subjected to update in order to reduce the uncertainties, that as we saw, are higher in case of non-update of parameters pdfs.
\begin{table}[htbp]
\centering 
\caption{RMSE comparison for simulations of year 2015}
\label{rmse15}
\begin{tabularx}{\textwidth}{X c c}\hline
\multicolumn{1}{c}{\textbf{Simulation}} & \textbf{Mean RMSE} & \textbf{Max RMSE}\\
\hline
EnKF: parameters update; new prior			& 11.8527			 &	34.8859  \\
EnKF: no parameters update; new prior		& 12.1196			 &	34.8859  \\
Open Loop: new prior						& 64.7669			 &	89.0379  \\
\hline\end{tabularx}
\end{table}

\section{Discussion}
\label{discussion}
The SIRB model, together with the calibration technique EnKF, shows an overall good agreement with the cholera cases recorded in the spatial domain under study. The results that we have previously described are sufficiently satisfactory on the global scale (Figs \ref{sum_cases14} - \ref{sum_cases15}), yet looking at the spatial level of the Counties, the model has some difficulties to capture the actual trend of the epidemics (see  Figs \ref{comm_cases14} - \ref{comm_cases15}), with the exception of Juba, whose dynamics are perfectly represented by all the simulations performed.

\begin{figure}[!h]
\centering
\includegraphics[width=1\textwidth,height=0.5\textheight]{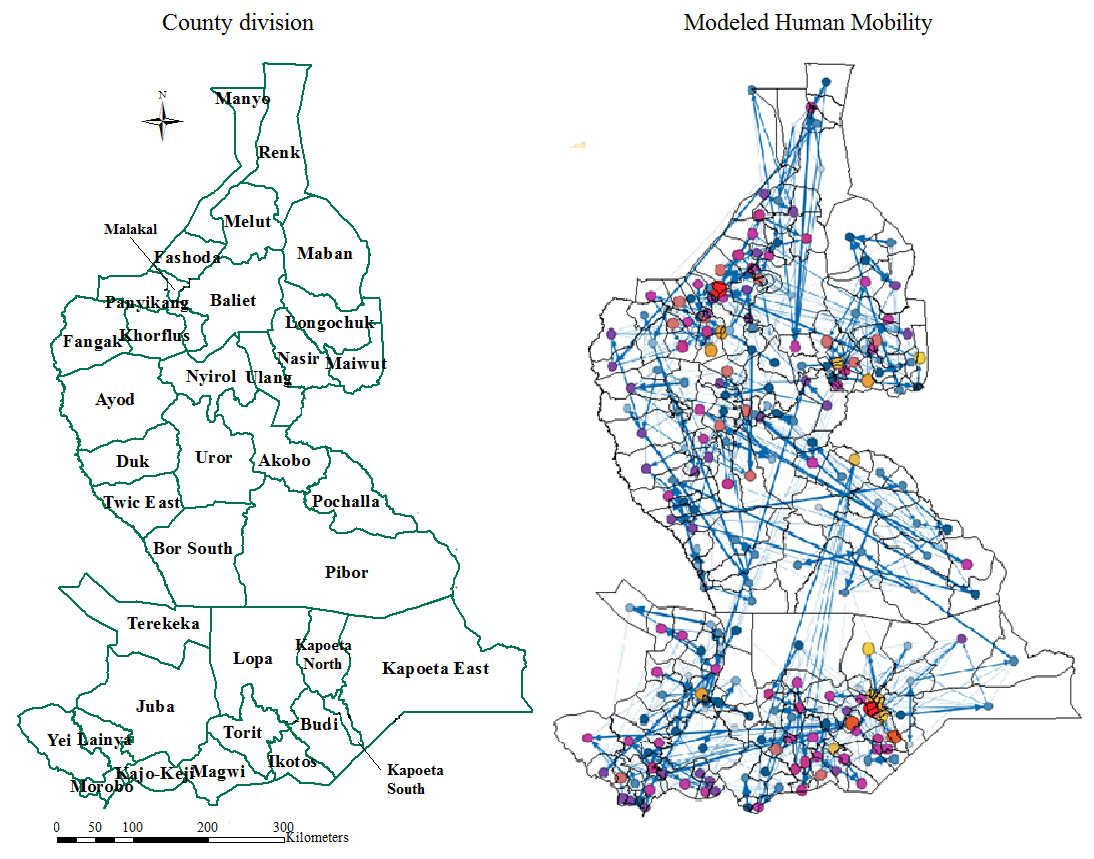}
\caption[Response of the gravity model]{Response of the gravity model. From each node $i$, the thickness of the arrow is proportional to the probability $P>0.01$ of choosing the node $j$ as destination. The color and magnitude of the nodes is proportional to the number of connections for each node. As nodes are representative of the Payams, the underlying figures highlights the administrative areas. For comparison, the County division used for the calibration is represented on the left.}
\label{mobilita_resp}
\end{figure}

Focusing on year 2014 at the level of the Counties (Fig.\vref{comm_cases14}), the simulated epidemiological curves deviate from the real ones especially in Malakal and Torit, in which the model does not reproduce correctly times and numbers of the infection. Looking at the spatial distribution of the recorded cases in 2014, Fig. \ref{map2014}, the infection seems to be not spatially correlated, and the possibilities that single travelers moved the epidemic away from its origin site, as for Malakal in 2014, could be high. The same works for the following year, when cases were recorded only in Juba and Bor in 2015 (Fig.\ref{map2015}). Therefore, one possible reason for this behavior of the model can be found in the mobility. As \citet{Ujjiga2015} identified through their study, travel to cholera outbreak areas is considered as a risk factor. Moreover, previous studies regarding cholera outbreak in Southern Sudan showed that the risk was associated with being a visitor to Juba \citep{CDC2007}. As we said in Chapter 2, the country is subject to important migratory fluxes that are not that easy to simulate via a simple gravity model. These fluxes move people towards the capital, in which are present PoCs sites, and to other places. In fact, the epidemic started in Juba, far away from Malakal, and that can be reached by flight connections that were not considered in this work. In our mathematical framework it is difficult to simulate cases in more distant areas, as e.g. Malakal, without spreading the infection all over the nodes between these two, as happens for example in Bor South County during 2014 simulations, located in between Juba and Malakal. The gravity model, whose response is shown in Fig. \ref{mobilita_resp}, let the disease spread more severely in places that are more connected between each other. As the figure shows, Malakal is not directly connected to Juba and the model is favoring the nodes in Yei and Kajo-Keji County, since the modeled probability of moving towards these is higher. As regards 2015 epidemics, the all zeroes observed data in the counties around Juba that the model uses to corrects the states of the system, try to contain the response in these places and affect the simulation in Bor South, despite this one is well connected to Juba through the mobility model.

Furthermore, the migrants fluxes and camps establishment change the distribution of the inhabitants in the areas considered, affecting the position of the nodes in the network, in a loop that involves the gravity model.

Last, one should include among the possible reasons for the misfitting between the simulation and the observed data, the biased case reporting or identification. Our analysis includes in fact, all the suspected cases that were recorded among the health facilities. During 2014, when the infection moved from Juba to Malakal, there were no confirmed cases in Jonglei. We should think about the possibility that the bacteria affected displaced people in this state that lived outside the vaccinated PoC camps \citep{Abubakar2015}. Hence, there are reasons to believe that there is some lack of information regarding cases distributed in diverse areas. This hypothesis can be supported by the three cases recorded in Western Equatoria and that were not included in this analysis (see Sect. \ref{colerass}): there is no assurance that no other infected were present in that area, considering the fact that, for what we know, no vaccination campaign took place over there. To support this hypothesis, different studies show that education and age can influence people in going to health care infrastructure. As an example, \citet{Luby2004} reported in their studies, children are less likely to consult a health care practitioner for symptoms of diarrhoeal diseases as cholera.

\chapter{Conclusions}
Improving global access to water, sanitation and hygiene is a critical step to reducing Africa's cholera burden. Vaccinations campaigns are helping people all over the continent and their validity is tested everyday more (e.g. \citep{Luquero2014,Abubakar2015}), unfortunately a lot of people still do not have access to them. We argue that epidemiological model can help in understand risk factors and dynamics of transmission and even more, forecast possible new events. Therefore we would like to make the point on our work, suggesting future possibilities and further studies that can help in treating cholera, especially in South Sudan, where the unstable social conditions have lead to another cholera outbreak at the moment of writing this manuscript.
%%..................%%
\\ \par The mathematical model that we proposed was adapted to analyze and to represent the epidemics that affected South Sudan in 2014 and 2015. This deterministic mathematical approach is not exempt from limitations and uncertainties, yet we would like to highlight its potentiality.

The use of DA technique for the calibration of the parameters is an interesting choice that can be supported by its advantages. First, it allows to accelerate the calibration procedure reducing the computational cost compared with other calibration method (e.g. DREAM) \citep{pasetto2016}. Second, the method admits the parameters to change in time, sequentially updating their values via incorporation of new observations. Looking at their physical meaning, this is preferable in cholera modeling. The exposure rate $\beta_0$, the population awareness on cholera risk factors $k$ and the sanitation conditions $\theta$ can actually improve during the time of the disease, acting on the transmission routes and changing the probability of contracting the infection. Especially in this case study, as shown by the results in the previous chapter, the dynamic of the epidemics are better retrieved using the state augmentation technique (see Sect. \ref{enskf}). In fact, the confidence intervals in all the simulations M2, in which parameters were kept constant, were larger than the ones in simulations M1, in which uncertainties reduced in time. Of course, in this procedure the basic point was to find appropriate domains of variability, resulted definitely smaller than the Haitian prior pdfs first used.

Despite the difficult context and the lack of information that favored the uncertainties, the SIRB model was able to retrieve the trends of the epidemics during both years, capturing numbers and timing. The best agreement with the data was showed for the global curves and for Juba during both years, having the model able to reproduce the epidemiological curves almost perfectly. 

As regards the response in the other counties, we already discussed about possible reasons and limitations for what concerns the model for human mobility, which is not able to reproduce the migratory fluxes all over the spatial framework considered. For further understanding of the epidemics, one possibility to go over this problematic can be the incorporation of mobile phone data to understand the role of mass gathering in spreading the infection, as \citet{Finger2016} in the last formulation of the SIRB model. In fact, mobile phone data sets are an important new source of information about the dynamics of populations overall in Africa and provide an opportunity to measure human mobility directly for entire populations \citep{mobilityafrica}. By the way, according to \cite{keyindicators}, only the 15\% of the Southern Sudanese population posses a phone, and further investigations would be needed to understand whether these data are reliable or have changed.

Another methodology would be, whether data are available, to adapt and modify the gravity model. On one hand, we could include into the formulation an indicator for the "safety" of the site. In literature, example of the validity of socio-political indicators in modeling mobility to understand spread of disease can be found, as \citet{adaptgrav} studied for Influenza. On another hand, we could introduce meta-population dynamics to better understand spatial heterogeneity of the disease, as in \citet{Xia2015}.

Looking at the limitation of the model, one should include the strong assumptions used to simplify the mathematical framework. As we described in Chapter 3, we consider the population to be at demographic equilibrium. This assumption deletes further stochasticity of host demography, yet it is not realistic in our context. In fact, the high growth rate in the country (see Sect. \ref{popolazione}) should be considered when modeling the year 2015, yet this would required certain data on population size and distribution.

Other assumption that can compromise the results of our approach is the hypothesis of non pre-existing immunity. We are not considering in fact the vaccination campaigns that took place at the beginning of 2014 and during the period in study. Therefore, all the population is considered all the time as susceptible and in risk of contracting the disease. The same works for 2015, in which we are not considering neither the vaccinations, nor the previous year epidemic that create immunity in both symptomatic and asymptomatic infected. Hence, we would like to explore the response of the model introducing either as initial condition during calibration procedure, or as a new equation in the epidemiological model, the recovered individuals that gained immunity and the individuals that in each node obtained immunity through vaccinations. Moreover, as \citet{Azman2016} reported, cholera showed both inside and outside the camp. In order to understand the transmission routes, additional information could be acquired by the conduction of surveys and questionnaires among refugees, together with environmental assessments, to evaluate the use of safe drinking water, personal food and hygiene and the use of improved sanitation facilities. These could be useful tools for understanding risk of disease, as showed by \citet{Ujjiga2015, Swerdlow1997, Cummings2012}. Moreover, surveys can be helpful in modeling the mobility and understand as well how movements are affected by the rain season, that as we discussed in this work, impede connections between the cities and that we did not took into account. 

Finally, another possibility would be to focus on the role of Juba, trying to understand whether there are some phenomenon that enhance the risk of infection in the capital, associated to the risk of spreading the infection allover the Country.
\\ \par We hope with our work to be helpful and that new and more precise information will be soon available to improve it.

%\input{greetings}
%-----------------------------------------------------------------------%
%                          	 BIBLIOGRAFIA
%-----------------------------------------------------------------------%
\cleardoublepage
\bibliographystyle{apalike}
\bibliography{references}

\begin{thebibliography}{}

\bibitem[Abubakar et~al., 2015]{Abubakar2015}
Abubakar, A., Azman, A.~S., Rumunu, J., Ciglenecki, I., Helderman, T., West,
  H., Lessler, J., Sack, D.~A., Martin, S., Perea, W., Legros, D., and Luquero,
  F.~J. (2015).
\newblock The first use of the global oral cholera vaccine emergency stockpile:
  Lessons from south sudan.
\newblock {\em PLoS Medicine}.
\newblock doi:10.1371/journal.pmed.1001901.

\bibitem[Andrews and Basu, 2011]{Andrews2011}
Andrews, J. and Basu, S. (2011).
\newblock {Transmission dynamics and control of cholera in Haiti: an epidemic
  model}.
\newblock {\em Lancet}, 377:1248--1255.
\newblock doi:10.1016/S0140-6736(11)60273-0.

\bibitem[Azman et~al., 2016]{Azman2016}
Azman, A., Rumunu, J., Abubakar, A., West, H., Ciglenecki, I., and Helderman,
  T. e.~a. (2016).
\newblock {Population-Level Effect of Cholera Vaccine on Displaced Populations,
  South Sudan, 2014}.
\newblock {\em Emerg Infect Dis.}, 22(6).
\newblock doi:10.3201/eid2206.151592.

\bibitem[Barab{\'a}si, 2016]{Barabasi}
Barab{\'a}si, A.-L. (2016).
\newblock {\em Network science}.
\newblock Cambridge university press.

\bibitem[Baratti, 2014]{Baratti}
Baratti, E. (2014).
\newblock {\em {Stima dei parametri di modelli idrologici mediante
  ottimizzazione dell'utilita}}.
\newblock PhD thesis, Alma Mater Studiorum - Università di Bologna.

\bibitem[BBC, 2016]{bbc}
BBC (2016).
\newblock \url{http://www.bbc.com/news/world/africa}.
\newblock News update available online.

\bibitem[Bertuzzo et~al., 2008]{Bertuzzo2008}
Bertuzzo, E., Azaele, S., Maritan, A., Gatto, M., Rodriguez-Iturbe, I., and
  Rinaldo, A. (2008).
\newblock On the space-time evolution of a cholera epidemic.
\newblock {\em Water Resources Research}, 44.
\newblock doi:10.1029/2007WR006211.

\bibitem[Bertuzzo et~al., 2010]{Bertuzzo2010}
Bertuzzo, E., Casagrandi, R., Gatto, M., Rodriguez-Iturbe, I., and Rinaldo, A.
  (2010).
\newblock {On spatially explicit models of cholera epidemics}.
\newblock {\em Interface}, 7(43):321--33.
\newblock doi:10.1098/rsif.2009.0204.

\bibitem[Bertuzzo et~al., 2014]{Bertuzzo2014}
Bertuzzo, E., Finger, F., Mari, L., Gatto, M., and Rinaldo, A. (2014).
\newblock {On the probability of extinction of the Haiti cholera epidemic}.
\newblock {\em Stoch Environ Res Risk Assess}, pages 1--13.
\newblock doi:10.1007/s00477-014-0906-3.

\bibitem[Bertuzzo et~al., 2012]{Bertuzzo2012}
Bertuzzo, E., Mari, L., Righetto, L., Casagrandi, R., Gatto, M.,
  Rodriguez-Iturbe, I., and Rinaldo, A. (2012).
\newblock {\em An epidemic model for the future progression of the current
  Haiti cholera epidemic}, volume~14.
\newblock Provided by the SAO/NASA Astrophysics Data System.

\bibitem[Bertuzzo et~al., 2011]{Bertuzzo2011}
Bertuzzo, E., Mari, L., Righetto, L., Gatto, M., Casagrandi, R., Blokesch, M.,
  Rodriguez-Iturbe, I., and Rinaldo, A. (2011).
\newblock Prediction of the spatial evolution and effects of control measures
  for the unfolding haiti cholera outbreak.
\newblock {\em Geophysical Research Letters}, 38(6).
\newblock doi:10.1029/2011GL046823.

\bibitem[Beven and Binley, 1992]{oncalibuncertain}
Beven, K. and Binley, A. (1992).
\newblock {The future of distributed models: Model calibration and uncertainty
  prediction}.
\newblock {\em Hydrological Processes}, 6:279 -- 298.
\newblock doi:10.1002/hyp.3360060305.

\bibitem[Bhattacharya et~al., 2009]{Bhattacharya2009}
Bhattacharya, S., Black, R., Bourgeois, L., Clemens, J., Cravioto, A., Deen,
  J.~L., Dougan, G., Glass, R., Grais, R.~F., and Greco, M. e.~a. (2009).
\newblock {Public health. The cholera crisis in Africa.}
\newblock {\em Science}, 324(5929):885.
\newblock doi:10.1126/science.1173890.

\bibitem[Camporese et~al., 2009]{Camporese2009}
Camporese, M., Paniconi, C., Putti, M., and Salandin, P. (2009).
\newblock {Ensemble Kalman filter data assimilation for a process-based
  catchment scale model of surface and subsurface flow}.
\newblock {\em Water Resources Research}, 45.
\newblock doi:10.1029/2008WR007031.

\bibitem[Capasso and Paveri-Fontana, 1979]{Capasso1979}
Capasso, V. and Paveri-Fontana, S.~L. (1979).
\newblock {A mathematical model for the 1973 cholera epidemic in the European
  Mediterranean region}.
\newblock {\em Revue d'épidémiologie et de santé publique}, 27(2):121--32.

\bibitem[Cazelles and Chau, 1997]{hivfore}
Cazelles, B. and Chau, N. (1997).
\newblock {Using the Kalman filter and dynamic models to assess the changing
  HIV/AIDS epidemic}.
\newblock {\em Elsevier - Mathematical Biosciences}, 140:131 -- 154.
\newblock doi:10.1038/ncomms3837.

\bibitem[{CDC}, 2009]{CDC2007}
{CDC} (2009).
\newblock {Morbidity and Mortality Weekly Report: Cholera Outbreak in Southern
  Sudan 2007-2009}.
\newblock Technical report, Centers for Disease Control and Prevention.
\newblock Available online; accessed on March 2016.

\bibitem[{CIA}, 2015]{CIA2015}
{CIA} (2015).
\newblock {South Sudan}.
\newblock Central Intelligence Agency. Available online; accessed on March
  2016.

\bibitem[Code{\c{c}}o, 2001]{Codeco2001}
Code{\c{c}}o, C.~T. (2001).
\newblock {Endemic and epidemic dynamics of cholera: the role of the aquatic
  reservoir}.
\newblock {\em BMC infectious diseases}, 1:1.

\bibitem[Collins, 2015]{Collins2015}
Collins, R. (2015).
\newblock {South Sudan}.
\newblock \url{http://www.britannica.com/place/South-Sudan}.
\newblock Available online; accessed on April 2016.

\bibitem[Colwell, 1996]{Colwell1996}
Colwell, R.~R. (1996).
\newblock {Global climate and infectious disease: the cholera paradigm}.
\newblock {\em Science}, 274(5295):2025--31.

\bibitem[Cummings et~al., 2012]{Cummings2012}
Cummings, M.~J., Wamala, J.~F., Eyura, M., Malimbo, M., Omeke, M.~E., Mayer,
  D., and Lukwago, L. (2012).
\newblock {A cholera outbreak among semi-nomadic pastoralists in northeastern
  Uganda: epidemiology and interventions}.
\newblock {\em Epidemiology and Infection}, 140(08):1376--1385.
\newblock doi:10.1017/S0950268811001956.

\bibitem[De~Rochars et~al., 2011]{DeRochars2011}
De~Rochars, V., Tipret, J., Patrick, M., Jacobson, L., Barbour, K., Berendes,
  D., Bensyl, D., Frazier, C., D., J., Archer, R., Roels, T., Tappero, J., and
  Handzel, T. (2011).
\newblock Knowledge, attitudes, and practices related to treatment and
  prevention of cholera, haiti, 2010.
\newblock {\em Emerging Infectious Diseases}, 17(11).

\bibitem[Di~Molfetta and Sethi, 2012]{sethi}
Di~Molfetta, A. and Sethi, R. (2012).
\newblock {\em {Ingegneria degli acquiferi}}.
\newblock Springer Science {\&} Business Media.
\newblock doi:10.1007/978-88-470-1851-8 11.

\bibitem[Erlander and Stewart, 1990]{Erlander1990}
Erlander, S. and Stewart, N.~F. (1990).
\newblock {\em {The gravity model in transportation analysis : theory and
  extensions}}.
\newblock VSP.

\bibitem[Evensen, 1994]{evensen}
Evensen, G. (1994).
\newblock {Sequential data assimilation} with a nonlinear quasi-geostrophic
  model using monte carlo methods to forecast error statistics.
\newblock {\em Journal of Geophysical Research}, 99:10143 – 10162.
\newblock doi:10.1029/94JC00572.

\bibitem[Ewens and Grant, 2005]{Ewens2005}
Ewens, W. and Grant, G. (2005).
\newblock {\em {Statistical Methods in Bioinformatics: An Introduction}}.
\newblock Springer Science+Business Media, Inc., 2 edition.

\bibitem[Finger et~al., 2016]{Finger2016}
Finger, F., Genolet, T., Mari, L., de~Magny, G.~C., Manga, N.~M., Rinaldo, A.,
  and Bertuzzo, E. (2016).
\newblock Mobile phone data highlights the role of mass gatherings in the
  spreading of cholera outbreaks.
\newblock {\em Proceedings of the National Academy of Sciences},
  113(23):6421--6426.
\newblock doi:10.1073/pnas.1522305113.

\bibitem[Finger et~al., 2014]{Finger2014}
Finger, F., Knox, A., Bertuzzo, E., Mari, L., Bompangue, D., Gatto, M.,
  Rodriguez-Iturbe, I., and Rinaldo, A. (2014).
\newblock Cholera in the lake kivu region (drc): Integrating remote sensing and
  spatially explicit epidemiological modeling.
\newblock {\em Water Resources Research}, 50(7):5624--5637.
\newblock doi:10.1002/2014WR015521.

\bibitem[Gelman and Rubin, 1992]{Gelman1992}
Gelman, A. and Rubin, D.~B. (1992).
\newblock {Inference from Iterative Simulation Using Multiple Sequences}.
\newblock {\em Statistical Science}, 7(4):457--472.
\newblock doi:10.1214/ss/1177011136.

\bibitem[{Global Security}, 2016]{GlobalSecurity2016}
{Global Security} (2016).
\newblock {South Sudan}.
\newblock \url{www.globalsecurity.org}.
\newblock Available online; accessed on 15-May-2016.

\bibitem[Grawert, 2010]{Grawert2010}
Grawert, E. (2010).
\newblock {\em {After the Comprehensive Peace Agreement in Sudan}}.
\newblock Boydell {\&} Brewer.
\newblock isbn:1847010229, 9781847010223.

\bibitem[Gupta et~al., 2015]{enskftb}
Gupta, H., Verma, K., and Sharma, P. (2015).
\newblock {Using Data Assimilation Technique and Epidemic Model to Predict TB
  Epidemic}.
\newblock {\em International Journal of Computer Applications}, 128(9):5.
\newblock doi:10.5120/ijca2015906625.

\bibitem[Harris et~al., 2012]{Harris2012}
Harris, J.~B., LaRocque, R.~C., Qadri, F., Ryan, E.~T., Calderwood, S.~B., and
  Morris, J. (2012).
\newblock {Cholera}.
\newblock {\em The Lancet}, 379(9835):2466--2476.
\newblock doi:10.1016/S0140-6736(12)60436-X.

\bibitem[Inman, 2010]{haversinebook}
Inman, J. (2010).
\newblock {\em {Navigation and Nautical Astronomy for the Use of British Seamen
  (1849)}}.
\newblock Kessinger Publishing, LLC.

\bibitem[IRI/LDEO, 2016]{rain}
IRI/LDEO (2016).
\newblock \url{http://iridl.ldeo.columbia.edu/}.
\newblock Climate Data Library. Available online; accessed on 7-April-2016.

\bibitem[Kalman, 1960]{Kalman1960}
Kalman, R. (1960).
\newblock {A New Approach to Linear Filtering and Prediction Problems}.
\newblock {\em Journal of Basic Engineering}, 88:35 -- 45.

\bibitem[Kaper et~al., 1995]{Kaper1995.}
Kaper, J.~B., Morris, J.~G., and Levine, M.~M. (1995).
\newblock {Cholera}.
\newblock {\em Clinical microbiology reviews}, 8(1):48--86.

\bibitem[Kelly, 2016]{Kelly2016}
Kelly, M. (2016).
\newblock {Quick facts: What you need to know about the South Sudan crisis}.
\newblock {\em Mercy Corps}.
\newblock Available online, accessed on 19-May-2016.

\bibitem[Kermack and McKendrick, 1927]{KermackMc}
Kermack, W.~O. and McKendrick, A.~G. (1927).
\newblock {\em {A Contribution to the Mathematical Theory of Epidemics}},
  volume 115.
\newblock doi:10.1098/rspa.1927.0118.

\bibitem[King et~al., 2008]{King2008}
King, A., Ionides, E., Pascual, M., and Bouma, M. (2008).
\newblock {Inapparent infections and cholera dynamics}.
\newblock {\em Nature}, 454:877--880.
\newblock doi:10.1038/nature07084.

\bibitem[Koelle et~al., 2005]{Koelle2005}
Koelle, K., Rod{\'{o}}, X., Pascual, M., Yunus, M., and Mostafa, G. (2005).
\newblock {Refractory periods and climate forcing in cholera dynamics}.
\newblock {\em Nature}, 436(7051):696--700.
\newblock doi:10.1038/nature03820.

\bibitem[Leung et~al., 2012]{Leung2012}
Leung, D., Chowdhury, F., Calderwood, S., Qadri, F., and Ryan, E. (2012).
\newblock {Immune responses to cholera in children}.
\newblock {\em Expert Review of Anti-infective Therapy}, 4(10):435 – 444.
\newblock doi:10.1586/eri.12.23.

\bibitem[Li et~al., 2011]{adaptgrav}
Li, X., Tian, H., Lai, D., and Zhang, Z. (2011).
\newblock {Validation of the Gravity Model in Predicting the Global Spread of
  Influenza}.
\newblock {\em Int J Environ Res Public Health}, (8):3134--3143.
\newblock doi:10.3390/ijerph8083134.

\bibitem[Lipp et~al., 2002]{Lipp2002}
Lipp, E.~K., Huq, A., and Colwell, R.~R. (2002).
\newblock {Effects of Global Climate on Infectious Disease: the Cholera Model}.
\newblock {\em Clinic Microbiology Reviews}, 15(4):757--770.
\newblock doi:10.1128/CMR.15.4.757–770.2002.

\bibitem[Luby et~al., 2004]{Luby2004}
Luby, S., Agboatwalla, M., Painter, J., Altaf, A., Billhimer, W., and Hoekstra,
  R. (2004).
\newblock {Effect of intensive handwashing promotion on childhood diarrhea in
  high-risk communities in Pakistan}.
\newblock {\em Journal of the American Medical Association}, 29(1):2547--2554.

\bibitem[Luquero et~al., 2014]{Luquero2014}
Luquero, F.~J., Grout, L., Ciglenecki, I., Sakoba, K., Traore, B., Heile, M.,
  Diallo, A.~A., Itama, C., Page, A., Quilici, M., M., M.~A., Eiros, J.,
  Serafini, M., Legros, D., and Grais, R.~F. (2014).
\newblock {Use of Vibrio cholerae vaccine in an Outbreak in Guinea}.
\newblock {\em New England Journal of Medicine}, 370(22):2111--2120.
\newblock doi:10.1056/NEJMoa1312680.

\bibitem[Mandel, 2009]{Mandel2009}
Mandel, J. (2009).
\newblock {A Brief Tutorial on the Ensemble Kalman Filter}.
\newblock Available online under the GNU Free Documentation License.

\bibitem[Mari et~al., 2015]{Mari2015}
Mari, L., Bertuzzo, E., Finger, F., Casagrandi, R., Gatto, M., and Rinaldo, A.
  (2015).
\newblock {On the predictive ability of mechanistic models for the Haitian
  cholera epidemic}.
\newblock {\em J. R. Soc. Interface}, 12.

\bibitem[Mari et~al., 2012]{Mari2012}
Mari, L., Bertuzzo, E., Righetto, L., Casagrandi, R., Gatto, M.,
  Rodriguez-Iturbe, I., and Rinaldo, A. (2012).
\newblock {Modelling cholera epidemics: the role of waterways, human mobility
  and sanitation}.
\newblock {\em J. R. Soc. Interface}, 9:376–388.

\bibitem[{Mekalanos J. and Lehman A.}, 2012]{cholerabacter}
{Mekalanos J. and Lehman A.} (2012).
\newblock {Confronting Cholera}.
\newblock \url{www.hms.harvard.edu/news/confronting-cholera}.
\newblock Available online; accessed on 30-June-2016.

\bibitem[Miller et~al., 1985]{Miller1985}
Miller, C.~J., Feachem, R.~G., and Drasar, B.~S. (1985).
\newblock {Cholera epidemiology in developed and developing countries: new
  thoughts on transmission, seasonality, and control}.
\newblock {\em Lancet}, 1(8423):261--2.

\bibitem[Nelson et~al., 2009]{Nelson2009}
Nelson, E.~J., Harris, J.~B., Glenn~Morris, J., Calderwood, S.~B., and Camilli,
  A. (2009).
\newblock {Cholera transmission: the host, pathogen and bacteriophage dynamic}.
\newblock {\em Nature Reviews Microbiology}, 7(10):693--702.
\newblock doi:10.1038/nrmicro2204.

\bibitem[OpenStreetMap, 2015]{openstreetmap}
OpenStreetMap (2015).
\newblock \url{www.openstreetmap.org}.
\newblock Available online; accessed on April-2016.

\bibitem[Pasetto, 2013]{Damianotesi}
Pasetto, D. (2013).
\newblock {\em {Reduced order models and Data Assimilation for hydrological
  applications}}.
\newblock PhD thesis, Universit\'{a} degli Studi di Padova, Dipartimento di
  Matematica.

\bibitem[Pasetto et~al., 2016]{pasetto2016}
Pasetto, D., Finger, F., Rinaldo, A., and Bertuzzo, E. (2016).
\newblock {Real-time projections of cholera outbreaks through data assimilation
  and rainfall forecasting}.
\newblock {\em Submitted to Advances in Water Resources}.

\bibitem[Porta et~al., 2014]{vaccinationprev}
Porta, M., Lenglet, A., de~Weerdt, S., Crestani, R., Sinke, R., Frawley, M.,
  Van~Herp, M., and Zachariah, R. (2014).
\newblock {Feasibility of a preventive mass vaccination campaign with two doses
  of oral cholera vaccine during a humanitarian emergency in South Sudan}.
\newblock {\em Royal Society of Tropical Medicine and Hygiene}.
\newblock doi:10.1093/trstmh/ztu153.

\bibitem[Righetto et~al., 2011]{Righetto2011}
Righetto, L., Bertuzzo, E., Casagrandi, R., Gatto, M., Rodriguez-Iturbe, I.,
  and Rinaldo, A. (2011).
\newblock {Modelling human movement in cholera spreading along fluvial
  systems}.
\newblock {\em Ecohydrology}, 4:49--55.

\bibitem[Rinaldo et~al., 2014]{Rinaldo2014}
Rinaldo, A., Bertuzzo, E., Mari, L., Righetto, L., Blokeschc, M., Marino, G.,
  Casagrandi, R., Murraye, M., Vesenbeckhe, S.~M., and Iturbef-Rodriguez, I.
  (2014).
\newblock {Reassessment of the 2010–2011 Haiti cholera outbreak and
  rainfall-driven multiseason projections}.
\newblock {\em PNAS}, page 6602–6607.

\bibitem[Ross, 2010]{ross}
Ross, S. (2010).
\newblock {\em {A first course in probability}}.
\newblock Pearson Education, Inc., 8 edition.
\newblock Available online; accessed on 14-July-2016.

\bibitem[Shaman et~al., 2013]{influenzafore}
Shaman, J., Karspeck, A., Yang, W., Tamerius, J., and Lipsitch, M. (2013).
\newblock {Real-time influenza forecasts during the 2012-2013 season}.
\newblock {\em Nature Communication}, 4(2837).
\newblock doi:10.1038/ncomms3837.

\bibitem[SSNBS, 2010]{census2008tables}
SSNBS (2010).
\newblock {Southern Sudan Counts: Tables from the 5th Sudan Population and
  Housing Census, 2008}.
\newblock Technical report, Southern Sudan Centre for Census Statistics and
  Government of Southern Sudan, Juba, South Sudan.

\bibitem[{SSNBS}, 2010]{SSYearbook10}
{SSNBS} (2010).
\newblock {Statistical Yearbook for Southern Sudan 2010}.
\newblock Technical report, Southern Sudan Centre for Census Statistics and
  Evaluation, Juba, South Sudan.

\bibitem[SSNBS, 2012]{keyindicators}
SSNBS (2012).
\newblock Key indicators for south sudan.
\newblock Technical report, South Sudan National Bureau of Statistics.

\bibitem[{SSNBS}, 2015a]{general_projectionSSNBS}
{SSNBS} (2015a).
\newblock {Population Projections for South Sudan 2015 - 2020}.
\newblock Technical report, South Sudan National Bureau of Statistics.

\bibitem[{SSNBS}, 2015b]{county_projectionSSNBS}
{SSNBS} (2015b).
\newblock {Population projections for South Sudan by County 2015 - 2020}.
\newblock Technical report, South Sudan National Bureau of Statistics.

\bibitem[{SSNBS}, 2015c]{payam_projectionSSNBS}
{SSNBS} (2015c).
\newblock {Population Projections for South Sudan by Payam From 2015 - 2020}.
\newblock Technical report, South Sudan National Bureau of Statistics.

\bibitem[{Sudan Tribune}, 2015]{SudanTribune2015}
{Sudan Tribune} (2015).
\newblock {South Sudan’s Kiir appoints governors of 28 new states}.
\newblock \url{www.sudantribune.com}.
\newblock Available online; accessed on April 2016.

\bibitem[Swerdlow et~al., 1997]{Swerdlow1997}
Swerdlow, D.~L., Malenga, G., Begkoyian, G., Nyangulu, D., Toole, M., Waldman,
  R.~J., Puhr, D. N.~D., and Tauxe, R.~V. (1997).
\newblock {Epidemic cholera among refugees in Malawi, Africa: treatment and
  transmission}.
\newblock {\em Epidemiology and Infection}, 118(3).
\newblock doi:10.1017/S0950268896007352.

\bibitem[ter Braak, 2006]{terbraak}
ter Braak, C. (2006).
\newblock {A Markov Chain Monte Carlo version of the genetic algorithm
  Differential Evolution: easy Bayesian computing for real parameter spaces}.
\newblock {\em Springer Science {\&} Business Media}, (16):239 – 249.
\newblock doi:10.1007/s11222-006-8769-1.

\bibitem[Treccani, 2016]{treccani}
Treccani (2016).
\newblock {Sud Sudan}.
\newblock \url{http://www.treccani.it/enciclopedia/sud-sudan/}.
\newblock Available online; accessed on 15-May-2016.

\bibitem[Ujjiga et~al., 2015]{Ujjiga2015}
Ujjiga, T.~T., Wamala, J.~F., Mogga, J.~J., Othwonh, T.~O., Mutonga, D.,
  Kone-Coulibaly, A., Shaikh, M.~A., Mpairwe, A.~M., Abdinasir, A., Abdi,
  M.~A., Yoti, Z., Olushayo, O., Nyimol, P., Lul, R., Lako, R.~L., and Rumunu,
  J. (2015).
\newblock {Risk Factors for Sustained Cholera Transmission, Juba County, South
  Sudan, 2014}.
\newblock {\em Emerging Infectious Diseases}, 21(10):1849--1852.
\newblock doi:10.3201/eid2110.142051.

\bibitem[{UN}, 2015]{popUNprospect}
{UN} (2015).
\newblock {World Population Prospects: Key findings and advance tables}.
\newblock Technical report, United Nations.

\bibitem[{UNHCR}, 2016]{UNHCR2016}
{UNHCR} (2016).
\newblock {Global Focus for South Sudan}.
\newblock Technical report, United Nations High Commissioner for Refugees.

\bibitem[{UNMISS}, 2015]{UNMISS2010}
{UNMISS} (2015).
\newblock {Weekly Flight Schedule Effective June 2010}.
\newblock Technical report, United Nation Mission In South Sudan.

\bibitem[Upton and Cook, 2014]{statinf}
Upton, G. and Cook, I. (2014).
\newblock {\em {A dictionary of statistics}}.
\newblock Oxford University Press, 3 edition.
\newblock doi:10.1093/acref/9780199679188.001.0001. Available online; accessed
  on 14-July-2016.

\bibitem[USAID, 2005]{JubaAdmin2005}
USAID (2005).
\newblock {Juba Assessment: Town Planning and Administration}.
\newblock Technical report, United States Agency for International Development,
  Juba.

\bibitem[Vrugt, 2016]{Vrugt2016}
Vrugt, J.~A. (2016).
\newblock {Markov chain Monte Carlo simulation using the DREAM software
  package: Theory, concepts, and MATLAB implementation}.
\newblock {\em Environmental Modelling {\&} Software}, 75:273 -- 316.

\bibitem[Vrugt et~al., 2012]{Vrugt2012}
Vrugt, J.~A., ter Braak, C., Diks, C., and Schoups, G. (2012).
\newblock {Hydrologic data assimilation using particle Markov chain Monte Carlo
  simulation: Theory, concepts and applications}.
\newblock {\em Advances in Water Resources}, 51:457–478.

\bibitem[VUVM, 2011]{koppenclimate}
VUVM (2011).
\newblock {World Maps of K\"{o}ppen-Geiger climate classification}.
\newblock \url{www.koeppen-geiger.vu-wien.ac.at}.
\newblock Vienna University of Veterinary Medicine - Institute for Veterinary
  Public Health. Available online; accessed on 15-July-2016.

\bibitem[Wesolowski et~al., 2015]{mobilityafrica}
Wesolowski, A., Prudhomme~O’Meara, W., Eagle, N., Tatem, A., and Buckee, C.
  (2015).
\newblock {Evaluating Spatial Interaction Models for Regional Mobility in
  Sub-Saharan Africa}.
\newblock {\em PLoS Comput Biol}, 11(7).
\newblock doi:10.1371/journal.pcbi.1004267.

\bibitem[{WHO}, 2010]{WHO2010Weekly}
{WHO} (2010).
\newblock {Weekly Epidemiological Record, No. 13, 26 March 2010}.
\newblock Technical report, World Health Organization.

\bibitem[{WHO}, 2015a]{2015WHOUpdates}
{WHO} (2015a).
\newblock {South Sudan cholera outbreak updates}.
\newblock Technical report, World Health Organization.
\newblock Available online; accessed on February 2016.

\bibitem[{WHO}, 2015b]{WHO2015}
{WHO} (2015b).
\newblock {South Sudan health situation reports}.
\newblock Technical report, World Health Organization.
\newblock Available online; accessed on February 2016.

\bibitem[{WHO {\&} UNICEF}, 2015]{WHO/UNICEF2015JointAssessment}
{WHO {\&} UNICEF} (2015).
\newblock {Joint Water Supply and Sanitation Monitoring Programme: 25 years of
  progress on sanitation and drinking water. 2015 update and MDG assessment}.
\newblock Technical report, World Health Organization and UNICEF.
\newblock isbn:9789241509145.

\bibitem[{World Atlas}, 2016]{WorldAtlas2016}
{World Atlas} (2016).
\newblock {Africa - South Sudan}.
\newblock \url{www.worldatlas.com}.
\newblock Available online; accessed on 15-May-2016.

\bibitem[WorldPop, 2013]{afripop}
WorldPop (2013).
\newblock \url{http://www.worldpop.org.uk/}.
\newblock Available online; accessed on March 2016.

\bibitem[Xia et~al., 2004]{Xia2015}
Xia, Y., Bj{\o}rnstad, O., and Grenfell, B. (2004).
\newblock {Measles Metapopulation Dynamics: A Gravity Model for Epidemiological
  Coupling and Dynamics}.
\newblock {\em The American Naturalist}, 164(2).

\end{thebibliography}
\addcontentsline{toc}{chapter}{Bibliography}
\end{document}